# AI-driven imitation can increase semantic diversity in news while omitting source information

Emil Bakkensen Johansen
Digital Democracy Centre
University of Southern Denmark
emilbj@sam.sdu.dk

Oliver Baumann
Strategic Organization Design
University of Southern Denmark
oliv@sdu.dk

**Abstract**
AI agents can imitate human-generated content at scale. Will their adoption homogenize the news? We simulate imitating AI agents entering the Danish news ecosystem (104,761 articles from 2022), varying imitation strategy, AI prevalence and task specification across two information environments, and measure semantic diversity between articles within news topics. Under identical benchmark instructions, single-source imitation raises semantic diversity in the observed ecosystem, made comparatively homogeneous by news agency content, but reduces it in the more heterogeneous environment, across three LLMs. In the observed ecosystem, every task specification tested with DeepSeek-V3.2 except minimal-edit instructions diversifies coverage. Relative to human-written articles, generated articles stand apart more where existing coverage is more uniform. In a source-matched comparison, DeepSeek-V3.2 rewrites omit named entities at over twice the rate of human adaptations. The consequences of AI-driven imitation thus depend on both task specification and existing coverage, and greater semantic diversity need not imply richer information.

The rapid advancement of large language models (LLMs) has accelerated the deployment of AI agents capable of performing tasks with minimal or no human intervention[1,2]. While this technological transformation is reshaping creative and knowledge-based industries, a critical interdisciplinary challenge lies in understanding how AI-generated content might reshape information environments[3–5]. Specifically, with AI agents increasingly capable of imitating human-generated content, an important question concerns the effect AI-driven imitation might have on semantic diversity. Traditional theories posit that imitation promotes convergence[6], yet imitation can also introduce novel variation[7,8] and heterogeneity in complex systems[9], rendering the aggregate consequences of AI-driven imitation difficult to predict.

Despite growing interest in how LLMs may affect semantic diversity, there remains no consensus on the diversity of outputs they produce. Comparatively little attention has been paid to AI-driven imitation as a distinct phenomenon. Given that imitation is a defining characteristic of news production, where journalists routinely draw on competitors' coverage[10–12], the increasing use of AI systems for imitation could substantially change how news, knowledge, and public discourse are produced[13].

We refer to these systems as *imitating AI agents*, using the term 'agent' in the minimal classical sense common in artificial intelligence, namely as a system that maps inputs from an environment to actions, without implying learning, adaptation, or strategic optimization[14]. In our setting, imitating AI agents receive reference content selected according to predefined simulation rules and generate rewritten outputs via probabilistic LLMs. Once deployed, they operate autonomously but do not update their behavior over time, and may exhibit their own systematic stylistic and cultural tendencies[15].

The introduction of imitating AI agents into news ecosystems raises significant societal questions about whether AI-driven imitation will reinforce existing journalistic variation or introduce new patterns of convergence or divergence[5,16]. News diversity broadly concerns the variety of news content available within an information environment[17]. In this study, we focus on a specific but consequential dimension of news diversity: semantic diversity in how the same news topics are covered across outlets. Understanding whether imitating AI agents homogenize or diversify news coverage is therefore important for assessing their potential implications for information environments like news, but semantic diversity must be interpreted alongside changes in informational content rather than as an intrinsically beneficial outcome[18,19].

This paper addresses this gap by presenting a simulation-based study explicitly examining how imitating AI agents might affect semantic diversity and information retention at the level of a national news ecosystem. Specifically, we use an extensive dataset of 104,761 Danish digital news articles published in 2022 to simulate potential near-future scenarios of AI-driven imitation among news providers. The Danish news ecosystem provides a well-suited empirical setting for this purpose, combining high levels of digitalization with a strong central news agency (Ritzau) whose syndicated content shapes cross-outlet similarity[11], akin to the role played by agencies such as AP (United States), AFP (France), or Reuters (United Kingdom).

Our scenario simulation approach systematically explores plausible near-term system-level consequences that empirical observational methods cannot directly reveal. By simulating multiple parallel realities of the Danish news ecosystem before the widespread adoption of generative AI, we gain insights into whether imitating AI agents could reinforce existing patterns or generate novel ones. Across these simulated worlds, the underlying topic structure is held constant, while we vary how imitation is performed (single-source versus multi-source), the prevalence of AI-generated articles substituting for human-written articles (see Fig. 1), and the task specifications (see Methods for details). To isolate these dimensions, imitation targets are selected randomly within news topics. The simulations therefore represent stylized imitation regimes in which target selection is exogenous and do not reproduce the endogenous source-selection and imitation processes of actual news production.

To examine whether the consequences of AI-driven imitation depend on the existing structure of news coverage, we compare relatively homogeneous and heterogeneous information environments. Our primary condition is the observed Danish news ecosystem, which constitutes a comparatively homogeneous empirical baseline because syndicated national news agency content contributes to cross-outlet similarity. We contrast this with a more heterogeneous comparison environment constructed by excluding articles that rely on national news agency content[11]. Thus, our approach follows the logic of 'science fiction science': using experiments and simulations to examine plausible sociotechnical scenarios rather than predict the future[20].

We measure semantic diversity using pairwise cosine similarity between full-text article embeddings within the same news topic. Lower average similarity indicates greater differentiation among the articles in embedding space. The measure does not identify which changes to a text account for that differentiation, nor does it establish that the articles contain more information. We therefore examine article characteristics and the retention of named entities and numerical references alongside semantic diversity, rather than treating increased differentiation as evidence of informational enrichment.

The impact of AI-driven imitation on semantic diversity remains uncertain, as studies point to divergent effects. A growing body of work suggests LLM-generated outputs can exhibit substantial uniformity. LLMs tend to converge toward statistical norms, producing outputs within a narrow representational range, which may lead to an "algorithmic monoculture" effect[21–24]. LLMs also exert "machine bias" and produce outputs distinct from human samples[25,26] and excel at representing typical features of genres[27], which could have the side effect of systematically filtering out stylistic outliers and reinforcing typicality in content. Further, AI-generated content often exhibits lower variance than human-created outputs in settings where AI-generated texts replace human creativity[28,29], mimic human outputs[30–33], or scale collective production, leading to reductions in aggregate diversity even when average quality improves[34]. Recent evidence further shows that LLMs can produce creative outputs while exhibiting compressed semantic distance distributions relative to humans, indicating that apparent creativity may coexist with structural homogenization[35,36].

Conversely, some research suggests that LLMs can successfully mimic humans, i.e., possess "algorithmic fidelity," by reproducing variation in linguistic styles and subgroup characteristics[37–40], an ability likely stemming from their extensive accumulation of human knowledge, culture, and information[41]. Under the right conditions, LLMs can accurately capture heterogeneity in human responses and maintain stylistic differentiation across sources[42]. Techniques such as prompt engineering can further increase output variation[43,44], and generative AI can potentially generate novel cultural variation through recombination[16] that can extend beyond human imagination[45]. However, these effects may diminish over time as AI-generated content increasingly conforms to dominant structures[46]. These two bodies of work need not conflict. The system-level effect of AI-generated content depends not only on how varied that content is, but also on how it relates to the content it replaces and the content that remains. Even relatively similar AI-generated articles can increase the diversity of a collection if they differ sufficiently from the existing coverage.

AI agents can imitate existing content by rewriting individual articles or combining multiple sources. Although imitation is commonly associated with convergence[6], research on organizational search shows that imperfect copying and subsequent adaptation can sometimes improve imitators' performance and sustain differences among firms[7,8]. These findings motivate asking whether imitation necessarily produces convergence in news coverage, but they do not establish how AI-driven imitation affects semantic similarity. We find that the same form of AI-driven imitation can have opposite consequences for the semantic diversity of news coverage. Under the benchmark instructions, rewriting single articles increases semantic diversity in the comparatively homogeneous observed ecosystem but reduces it in the more heterogeneous environment. Across news topics, generated articles stand apart more from the surrounding coverage, relative to human-written articles, where existing coverage is more uniform. Together with the substantial variation across task specifications, these findings help explain why the consequences of AI-driven imitation depend on both the instructions given to the model and the material being imitated. At the same time, greater semantic diversity does not necessarily imply informational enrichment: generated articles omit source information, and in a source-matched comparison, DeepSeek-V3.2 rewrites omit named entities at more than twice the rate observed in human adaptations.

## Results

### The effect of the initial heterogeneity of the information environment

Before any AI-driven imitation, mean within-topic cosine similarity among human-written articles is higher in the homogeneous environment (0.838) than in the heterogeneous environment (0.762; vertical reference lines in Fig. 2). On the 5,338 topics covered in both environments, the homogeneous baseline is 0.797, so the gap shrinks from 0.076 to 0.035. About half of the overall difference comes from topics present only in the homogeneous environment, which are covered mainly through news agency copy.

Fig. 2 shows results across LLMs under a benchmark differentiation-oriented prompt (the "benchmark task specification"; Supplementary Text 2). Across GPT-4o, DeepSeek-V3.2, and Llama-4-Scout-17B-16E-Instruct and mixed-LLM worlds, AI-driven single-source imitation increases semantic diversity in initially homogeneous environments. In heterogeneous environments, by contrast, single-source imitation raises similarity above baseline for all three LLMs at every prevalence level (by 0.019 to 0.044 cosine points). Multi-source imitation changes similarity by at most 0.01 in the homogeneous environment and at 10% and 25% prevalence in the heterogeneous environment, with no consistent sign across LLMs. In the heterogeneous environment at 50% prevalence multi-source imitation raises similarity (by 0.010 to 0.021). Compared on the same topics, multi-source imitation leaves heterogeneous topics less similar than single-source imitation for every LLM and prevalence (by 0.019 to 0.036). On the topics represented in both environments the single-source pattern is the same, but the reduction in the homogeneous environment is smaller (0.010 to 0.021 at 50% prevalence, against 0.023 to 0.039 on topics covered mostly through news agency copy; Extended Data Fig. 1).

To further examine the conditional effects of AI-driven imitation, we estimate pooled mixed-effects regression models (details provided in Supplementary Text 3) predicting each article's mean within-topic similarity ($\mu$) and the standard deviation of those similarities ($\sigma$), including interactions between LLM and environment conditions (Fig. 3a, Extended Data Table 1). Because observations from the same topic share underlying pairwise comparisons, confidence intervals and $P$ values use standard errors clustered by topic ($n = 9{,}267$). A robustness analysis aggregating to one generated-minus-human contrast per news topic and simulated world, so that each news topic contributes one observation per condition, reproduces the reference-condition differences for all three LLMs (Supplementary Table 5).

In the benchmark reference condition (homogeneous environment, single-source imitation, benchmark task specification, and 10% AI prevalence) AI-generated articles have lower mean within-topic similarity than human-written articles (Fig. 3a). The estimated differences are −0.022 cosine points for GPT-4o (95% CI [−0.023, −0.021], $z = -31.35$), −0.018 for Llama (95% CI [−0.019, −0.016], $z = -21.01$), and −0.042 for DeepSeek-V3.2 (95% CI [−0.044, −0.039], $z = -34.84$; all $P < 0.001$, Extended Data Table 1). Thus, the DeepSeek-V3.2 difference is approximately 1.9 and 2.4 times as large as the GPT-4o and Llama differences, so LLM choice affects the size but not the direction of the difference in this reference condition.

Without controls for article characteristics that AI rewriting itself changes, the reference-condition differences are −0.034 (GPT-4o), −0.026 (Llama) and −0.048 (DeepSeek-V3.2), between one third and two thirds of the difference in baseline similarity between the two environments (0.076) and comparable to the difference on their shared topics (0.035). Both estimates compare articles within the same simulated world, whereas the effect of imitation on the environment as a whole is the change in within-topic similarity shown in Fig. 2 (Extended Data Table 1).

The magnitude of these differences changes with environment structure and AI prevalence (Fig. 3b). In heterogeneous environments, the LLM-versus-human differences are approximately 0.013 to 0.015 cosine points smaller than in homogeneous environments (Extended Data Table 1). The attenuation with AI prevalence is more pronounced. For example, in the homogeneous single-source setting, the GPT-4o difference declines from 0.022 points at 10% AI prevalence to 0.012 at 25% (GPT-4o-minus-human contrast: $\beta = -0.012$, 95% CI [−0.013, −0.011], $z = -19.02$, $P < 0.001$) and reverses at 50% (GPT-4o-minus-human contrast: $\beta = 0.011$, 95% CI [0.010, 0.012], $z = 19.00$, $P < 0.001$).

The contrast narrows at higher prevalence partly because the reference group changes. The remaining human-written articles become less similar to their topic as more of their comparison articles are AI-generated (by 0.011 to 0.024 points at 50% prevalence under single-source imitation in the homogeneous environment), although the articles themselves are unchanged. The similarity of the whole ecosystem keeps falling, as Fig. 2 shows.

Dispersion in similarity is lower for AI-generated than for human-written articles under multi-source imitation, in all 18 conditions (by 0.012 to 0.021), but does not differ consistently under single-source imitation (lower in 4, indistinguishable in 3 and higher in 11 of 18 conditions; Fig. 3b). Combining three LLMs does not uniformly increase system-level dispersion either (Supplementary Text 15 and Supplementary Table 6).

**Local homogeneity and heterogeneity within environments**

To examine where within each environment these effects arise, we ranked topics by how similar their human-written articles are before imitation. Across this ranking, AI-generated articles differ most from the existing coverage where the human-written articles are most alike. In the homogeneous environment under single-source imitation at 50% prevalence, GPT-4o's articles are more similar to the rest of the coverage than human-written articles (+0.013, 95% CI [0.012, 0.014]) in the third of topics where human coverage is least alike, and less similar (−0.017, 95% CI [−0.018, −0.017]) in the third where it is most alike (Fig. 4a).

We then compare each AI-generated article's similarity to the rest of its topic's coverage with that of its sources (see "Topic- and source-level analyses", Methods). Rewriting a single article moves it away from the rest of the coverage, and so does combining two near-identical sources. When the two sources differ, the multi-source article ends up closer to the rest of the coverage than its sources were (Fig. 4b). In the heterogeneous environment, multi-source imitation raises similarity more on political topics that involve both conventional Danish political blocs than on topics involving one bloc (by 0.025 to 0.029 across LLMs; for GPT-4o, 0.015, 95% CI [0.001, 0.028], against −0.011, 95% CI [−0.019, −0.003]; Fig. 4c).

Within topics, AI-generated articles are less similar to the human-written articles they did not rewrite than human-written articles are to each other, in all 36 conditions and under all three embedding models (0.731 against 0.769 for GPT-4o under single-source imitation at 50% prevalence in the homogeneous environment), while remaining close to the article they rewrote (0.914). Under multi-source imitation at 25% and 50% prevalence, AI-generated articles are also more similar to each other than human-written articles are (0.830 against 0.769 at 50% for GPT-4o; Extended Data Table 2 and Supplementary Tables 7 and 8). These comparisons describe how the aggregate effect arises from where AI-generated articles sit within topics, not a causal mechanism within the LLMs.

**The role of task specification**

Task specification constitutes a second source of conditionality in AI-driven imitation. Whereas LLM choice determines the generative system executing the task, variation in prompting determines the form of imitation it performs[44,47,48]. In Fig. 5, we therefore vary task specification in DeepSeek-V3.2 worlds through three ablations of the benchmark prompt and three alternative newsroom-oriented prompts (see "Varying agents' task specifications" in Methods, Supplementary Text 2 for all prompt templates, and also Supplementary Fig. 8 for mixed-effects models testing task specification effects).

In homogeneous environments, single-source imitation lowers similarity below the human baseline under every task specification except minimal editing, but the magnitude varies considerably. At 50% AI prevalence, reductions range from 0.007 cosine points when the LLM is asked to preserve sources and figures to 0.046 for the wire-agency summary, against a human baseline of 0.838, while minimal editorial adjustment changes it by less than 0.005 at every prevalence level (+0.001 at 50%, 95% CI [−0.001, 0.002]; −0.002 and −0.003 at 10% and 25%; equivalence tests with bounds of ±0.005, all $P < 0.001$). The same directional effect holds when AI-generated articles within a world are randomly assigned one of the seven prompt templates. Under multi-source imitation, task specifications diverge in direction, from a 0.026-point reduction for the wire summary to a 0.024-point increase when sources and figures are to be preserved. The differentiation instruction in the benchmark task specification accounts for roughly half of the effect. At 50% prevalence under single-source imitation in the homogeneous environment, removing it reduces the decline in similarity by 45–64% depending on the embedding model, whereas removing the outlet name enlarges it (−0.035 against −0.029 for the benchmark).

In heterogeneous environments, single-source imitation raises similarity under every task specification and at every prevalence level, by 0.012 (wire summary) to 0.065 cosine points (minimal editing) at 50% prevalence, against a human baseline of 0.762. Under multi-source imitation, task specifications again diverge in direction. Where the wire summary lowers similarity at every prevalence level, the benchmark task specification and the version without the outlet name lower it at 10% and 25%, and the other specifications raise it. At 50% prevalence every specification except the wire summary raises similarity. When different task specifications coexist within the same world, and articles are randomly assigned a prompt template during imitation, the direction of the effect again depends on whether imitation is single- or multi-source (Fig. 5).

We also compare generative imitation with human null conditions that retain the replacement selections and imitation targets of the corresponding simulations, substituting the selected human source article under single-source imitation and the embedding centroid of the two sources under multi-source imitation (Supplementary Text 12). Both nulls raise mean similarity above the corresponding human baseline at every prevalence level and in both environments, reaching 0.856 (95% CI [0.853, 0.858]) under single-source and 0.893 (95% CI [0.891, 0.895]) under multi-source replacement at 50% prevalence in the homogeneous environment (Fig. 5). Replacement and random target selection alone therefore do not produce the pattern, although any process that introduces texts systematically offset from the existing coverage could.

**Longitudinal dynamics of AI-driven imitation**

To examine whether the diversification observed after the first round of imitation persists over subsequent rounds, we extended the homogeneous, single-source DeepSeek-V3.2 simulation under the benchmark task specification to three rounds in total (Fig. 6). We focus on this condition because it provides a direct test of whether an initially clear diversification effect dissipates once generated articles accumulate and become potential imitation targets. At each round, 10%, 25%, or 50% of article slots are selected for replacement within separate simulation branches, while previously generated articles remain in the simulated world and can themselves become imitation targets in subsequent rounds. AI prevalence therefore accumulates over time (see Supplementary Text 13 for details).

In this benchmark configuration, the initial diversification does not reverse over the three rounds examined. Mean within-topic cosine similarity stays below the all-human baseline ($\mu = 0.838$) in all three branches through t+3. In the 10% and 25% branches it declines at every round, reaching 0.814 (95% CI [0.812, 0.817]) and 0.806 (95% CI [0.804, 0.809]) at t+3. However, in the 50% branch it is lowest after the first round (0.809) and partly recovers to 0.821 (95% CI [0.819, 0.823]) at t+3 (Fig. 6a). The recovery comes mainly from the changing mix of article pairs. AI–AI pairs, which are more similar to each other than human-AI pairs, grow from 23% to 77% of a topic's pairs, while similarity within pair types changes little in total. These results show that the diversifying effect observed after one round can persist as AI-generated articles accumulate and become imitation targets themselves, although with diminishing effects as AI-generated articles make up a growing share of each topic.

**AI agents' distinct writing style**

Even when explicitly prompted to represent the imitating news outlet's style, AI agents' articles differ systematically from human-written news along several observed textual characteristics, including named entities, numerical references, length, and sentiment (see Supplementary Text 5 for feature extraction details).

Relative to human-written articles, AI-generated articles contain fewer named entities (standardized differences of −0.33 to −0.49 across LLMs) and fewer numerical references (−0.43 to −0.46; all $P < 0.001$, Supplementary Table 1). Article length depends on the LLM. GPT-4o and Llama articles are shorter than human-written articles (by 0.31 and 0.06 standard deviations), whereas DeepSeek-V3.2 articles are longer (by 0.75) yet still contain fewer named entities and numerical references (Supplementary Table 1).

All three LLMs write more positively than journalists. Mean sentiment is 0.07 to 0.18 standard deviations higher, with more articles at the strongly positive end of the scale (Supplementary Table 1 and Extended Data Fig. 2).

Beyond these differences in means, AI-generated articles are more uniform within topics. For all three LLMs, the dispersion of word count, named entities, numerical references and sentiment is lower than for human-written articles (Supplementary Text 7 and Supplementary Table 2), most clearly for named entities and, for GPT-4o and Llama, article length. Yet, reduced dispersion may partly reflect standardized production constraints rather than an intrinsic property of AI.

**Information loss and gain under AI-driven imitation**

Because semantic distance in embedding space can arise from multiple changes in a text, an increase in semantic diversity cannot by itself establish that the information environment contains richer factual information. We therefore examine whether source information is retained, lost, or introduced during imitation. For each generated article, we compute the share of source elements missing from the AI-generated text (information loss) as well as the share of elements present in the AI-generated text but absent from the original (information gain).

Under the benchmark task specification, Fig. 7 shows that, across LLMs, environments and imitation strategies, AI-generated articles exhibit substantially higher rates of information loss than information gain for both named entities and numbers. Depending on the LLM, environment, and imitation strategy, they omit 24–38% of named entities from the imitated article while adding 12–27% new entities. For numbers, they omit 10–28% but add only 1–5%.

Single-source imitation generally preserves more source information than multi-source imitation. Across LLMs and environments, mean named-entity loss ranges from 24% to 30% under single-source imitation, compared with 34% to 38% under multi-source imitation. Number loss similarly ranges from 10% to 14% under single-source imitation and from 21% to 28% under multi-source imitation. This pattern persists across all task specifications (Supplementary Fig. 7), although the magnitude of information loss varies substantially. For example, information loss is particularly pronounced under the wire-summary specification, whereas prompts emphasizing minimal editorial adjustment or factual preservation attenuate information loss.

As a human reference, we gave DeepSeek-V3.2 the same news agency (Ritzau) article and target outlet as the journalist who adapted it. AI imitations omitted 25.4% of the agency article's named entities, against 11.8% for the journalist (difference 13.6 percentage points, 95% CI [12.4, 14.8]; $n$ = 1,700 articles), and slightly more of its numbers (14.8% against 11.3%). Rewriting the journalists' adaptations with each of the three LLMs gives the same pattern (Extended Data Fig. 3). Placing both outcomes side by side, within each environment and imitation strategy the rank correlation between a condition's change in similarity and the share of source items it omits is negative (Spearman's $\rho = -0.67$ to $-0.88$ across nine conditions each; Extended Data Fig. 4). In other words, conditions that increase semantic diversity more, or reduce it less, also omit more source information.

**Discussion**

Our results reveal a context-dependent impact of AI-driven imitation on semantic diversity, shaped jointly by the preexisting structure of the information environment and by how the imitation task is specified. Under the benchmark task specification, single-source imitation consistently increases semantic diversity in the comparatively homogeneous environment, whereas in the more heterogeneous environment it consistently increases similarity. Task specification mainly changes the size of the effect under single-source imitation, and within each environment, single-source imitation moves similarity in the same direction under almost every specification. Thus, instructions that encourage differentiation of individual articles do not, by themselves, determine whether the resulting collection of articles becomes more or less diverse.

These results are consistent with a distributional explanation. Generated articles that differ from an initially similar body of coverage reduce average similarity even if they resemble one another, while standardized outputs replacing articles in more varied coverage can raise it. The copying and source-averaging benchmarks do not reproduce the diversification (Fig. 5). Pair-type analyses show that generated articles remain close to their sources but differ from other human-written coverage, while their similarity to one another can partly offset this. Within each environment, AI-generated articles stand apart more from the surrounding coverage, relative to human-written articles, in news topics where human-written coverage is most homogeneous (Fig. 4a), consistent with this account.

This interpretation does not attribute differences between source articles and generated texts to random copying errors. Some changes are explicitly requested by the prompt, and others may arise unintentionally during generation. Our simulations compare the outcomes of specified forms of imitation without separately identifying the contributions of these processes.

The consequences of AI-driven imitation therefore cannot be inferred from the underlying LLM alone. Two organizations using the same LLM may produce different system-level effects if they instruct it to imitate content differently, or rewrite single sources rather than combine several. The relationship between the sources also matters. Combining relatively similar articles tends to move the generated article further from the surrounding coverage than its sources were, whereas combining dissimilar articles tends to bring it closer to that coverage than its sources were on average. Even small parts of a task specification have implications. For example, naming the imitating outlet appears to anchor the rewrite, and without it, rewrites move further from both their source and the surrounding coverage, but only when the LLM is also told to differentiate.

The semantic diversity documented here concerns differences in how the same news topics are covered, as represented in embedding space, rather than differences in which events are covered. Importantly, this differentiation should not be equated with informational enrichment. AI-generated articles frequently omit named entities and numerical references present in their sources.

Our study has several limitations. First, we document two changes that occur together: greater semantic diversity can coexist with lower retention of source information. Our measures identify omissions of named entities and numerical references rather than establishing which facts are lost or assessing their importance. Whether such diversity is beneficial, harmful, or inconsequential for citizens depends on how these changes affect exposure, comprehension, and engagement, which are outcomes that our simulation does not measure. While we do investigate information loss, we do not address the potential issues of AI-generated "hallucinations" and the proliferation of low-quality content. Even if imitating AI agents add new information, such systems may simultaneously facilitate the spread of "fake news" or unreliable content[49]. Future studies should examine how semantic diversity, information retention, factual accuracy and audience responses interact in deployed systems, including under explicitly convergence-oriented instructions, for human editors and LLMs working under identical instructions, and in newsrooms adopting AI after 2022.

Second, our analysis examines imitation under specified task instructions and source-selection rules. The agents do not learn, adapt, or strategically select targets. The resulting estimates therefore describe these particular forms of AI-driven imitation rather than an effect of imitation independent of how it is implemented. In real-world deployments, however, AI agents are likely to interact with their environments and with other systems, updating their behavior over time. Recent work on interacting LLM populations further shows that collective outcomes can differ qualitatively from individual-level behavior and vary with population size[50].

Third, the imitating AI agents in our simulations select imitation targets at random within topics. In real-world settings, imitation decisions are often shaped by social cues, reputational hierarchies, audience metrics, and search heuristics. Such preferential selection might amplify or potentially reverse the system-level effects documented here depending on which parts of the existing information distribution are disproportionately sampled. Similarly, our design conditions on the observed set of news topics and does not model endogenous decisions about which events are selected for imitation in the first place. Exploring how non-random or heuristic target and event selection influences semantic diversity is an important direction for future research.

Fourth, our construction of the more heterogeneous environments also warrants qualification. Excluding news-agency-dependent content lowers baseline semantic similarity, but it does not manipulate similarity in isolation, as it also reduces topic coverage and the number of articles and outlets represented within topics. On topics covered in both environments, the reduction in similarity in the homogeneous environment is smaller than on topics covered mostly through agency copy. Nevertheless, the environment comparison should be interpreted as a comparison between empirically distinct information structures rather than as a controlled causal estimate of baseline similarity alone.

Lastly, our longitudinal analysis captures recursive imitation across three successive rounds in one benchmark condition, but does not establish longer-run dynamics across the full set of conditions examined in the main simulations. Longer time horizons may generate additional feedback processes. For example, as human language becomes more similar to that of LLMs[5,51], and as AI agents become more prevalent, imitating AI agents may also increasingly imitate AI-generated content, which could pollute the training data used for forthcoming LLMs[52] and increase the risk of model collapse[52,53]. Relatedly, human-AI feedback loops can alter human perceptual, emotional, and social judgments, implying that changes in information distributions may propagate through exposure and interaction over time[54]. Furthermore, because AI prevalence increases while generated articles also become eligible imitation targets, this extension does not separate the contribution of AI-to-AI rewriting from that of increasing the share of generated content.

Overall, our simulations show that AI-driven imitation need not homogenize news coverage. In the more homogeneous environment, single-source imitation increased semantic diversity across three LLMs and six of seven task specifications tested with DeepSeek-V3.2, with a near-zero effect under minimal-editing instructions. Under the same benchmark instructions, single-source imitation increased similarity in the more heterogeneous environment. Greater semantic diversity also coexisted with the omission of source entities and numbers, with DeepSeek rewrites omitting named entities at more than twice the rate observed in human adaptations of matched agency material. Whether AI-driven imitation diversifies or homogenizes news therefore depends on both the coverage it enters and how the imitation task is specified. Neither diversification nor homogenization alone establishes whether the news environment becomes more informative. Assessing the consequences of AI-driven imitation thus requires attention to both how it changes the relationships among articles and what information is retained in the process.

## Methods

This research complies with all relevant ethical regulations. The only component involving human participants is the survey reported in Supplementary Text 11. All other analyses reported in the manuscript concern news articles and computational simulations and did not involve human participants. The survey was administered by the survey agency Epinion as a module within an existing survey fielded for the authors’ home institution. Under the applicable Danish regulations, ethics committee approval was not required for this social-science questionnaire study. The authors' home institution Research Ethics Committee provides voluntary rather than mandatory review for research outside the statutory ethics committee system. Participants provided informed consent prior to participation. The authors received survey responses in a form that did not allow individual respondents to be identified. Participants were compensated by Epinion according to its standard panel procedures. Respondents rated the similarity of two short news article snippets and were asked which snippet they believed to be AI-generated. None of the studies reported in this manuscript were preregistered.

We investigate whether AI-driven imitation reinforces existing structures or reshapes semantic diversity, and how these effects interact with the preexisting heterogeneity of the information environment. To do so, we modeled a full news ecosystem by utilizing a comprehensive dataset of full-text news articles published online by Danish news organizations in 2022 ($n$ = 104,761). No statistical methods were used to pre-determine sample sizes. The simulations use all 104,761 articles published online by Danish news organizations in 2022 in our data, and the survey sample size ($n$ = 1,201) was set by the survey module in which it was fielded. In the simulations, the articles replaced and the imitation targets were drawn at random within topics with fixed seeds. In the survey, respondents were randomly assigned to one of five stimulus pairs and were not told which snippet was AI-generated. The dataset included relevant metadata on the source of each article and its date of publication. We used a BERTopic[55] modeling approach to identify and segment news topics from 2022 (see Supplementary Text 1 for details), ensuring that estimates of semantic diversity reflect differences in how the same news topic ($n$ = 9,267) is covered. We then constructed a multi-world simulation setup.

**Imitating AI agents setup**

Following recent work that instantiates LLM-based simulation units via controlled prompting to generate context-conditioned responses under fixed rules[56,57], we operationalize imitating AI agents as prompt-specified generative systems that execute predefined imitation strategies without learning or strategic target selection.

Articles chosen as imitation targets are selected randomly within topics. We use this stylized target-selection rule so that target selection does not depend on article attractiveness, isolating the consequences of AI generation, imitation strategy, task specification, and environment structure from endogenous processes determining which articles attract imitation, such as outlet status[12]. The resulting estimates should therefore be interpreted as effects under a random-target imitation regime, rather than as estimates of how imitation would unfold under empirically observed patterns of source prominence or AI agents' actual internal selection heuristics.

We developed imitating AI agents to pursue one of two imitation strategies: single-source imitation and multi-source imitation (Fig. 1). In single-source imitation, the agent receives one randomly selected reference article from the same topic and generates a new article under the assigned task specification (Fig. 1b). Changes to the source text may reflect the requested rewriting as well as unintended departures from the instructions. We do not operationalize these changes as copying errors, but instead examine how the resulting articles affect semantic diversity in the simulated world.

In multi-source imitation, the agent receives two randomly selected reference articles from the same topic and is instructed to produce a single article using their content (Fig. 1c). Providing multiple sources does not by itself imply greater semantic diversity. We compare generative multi-source imitation with a non-generative source-averaging benchmark and examine how its consequences vary across task specifications and information environments.

As a sensitivity check, we used two reasoning scaffolds to examine how the scaffold affected semantic diversity: zero-shot (ZeroShot) and chain-of-thought (CoT). Both strategies optimized for differentiation by rewriting the article in the imitating outlet's style (see Supplementary Text 2 for prompt templates and Supplementary Text 9, Supplementary Figs 3 and 4 for temperature setting sensitivity checks). Drawing on the idea of "silicon sampling"[37] – the notion that an LLM implicitly encodes stylistic and topical features from its training data – we explicitly named the imitating news outlet in our prompts, enabling the LLM to generate output that is consistent with the outlet's historical characteristics. When using internal CoT reasoning, we instructed the LLM to internally "think aloud" by articulating its reasoning process during article generation.

In the single-source imitation scenario, the LLM was presented with one reference article and instructed to write a verbose, restructured, and expansive version of it in the imitating outlet's usual style, matching the original article's word count (Supplementary Text 2). In the multi-source imitation scenario, the LLM was presented with two reference articles from the same topic and instructed to write a verbose, restructured, and expansive article using rewritten elements of both, again in the imitating outlet's usual style and matching the original word count. These instructions specified the requested degree of rewriting and stylistic adaptation, but did not instruct the LLM to alter thematic content or to blend the sources' perspectives, and any resulting differences in themes, perspectives, or factual content are assessed empirically rather than assumed from the task specification (see Supplementary Text 5).

**Multi-world simulations**

To examine the effects of AI-driven imitation on semantic diversity under various conditions, we constructed a multi-world setup, simulating distinct "worlds" that each represented different scenarios and combinations of original and AI-generated news articles. With this approach, we explored the potential near-future impact of imitating AI agents at varying levels of prevalence and types of imitation strategies.

In World Original, we established a baseline using only original human-written news articles, organized into specific news topics using the BERTopic modeling approach (see Supplementary Text 1 and Fig. 1a). World Original served as a control group, representing the empirical baseline against which we compared alternative worlds, i.e., the state of diversity in news articles without any AI involvement during 2022 in Denmark. To simulate a reality with greater heterogeneity in human-written news articles, World Original* excluded news articles that were identified as relying on the Danish national news agency, Ritzau[1]. Excluding news agency content creates a baseline environment with systematically lower semantic similarity, allowing us to examine whether the effects of AI-driven imitation depend on initial similarity structures rather than assuming uniform impacts across contexts. Yet, the exclusion also reduces topic coverage and within-topic density in heterogeneous information environments (see Supplementary Text 14). We therefore interpret World Original and World Original* as empirically distinct information environments rather than as differing exclusively in baseline similarity.

[1] Articles identified as relying on the news agency are those whose body text has a cosine similarity of at least 0.65 to an article from the national news agency, Ritzau, published the same day or the day before or after.

In World Single-Source, we simulated the reality in which AI-generated articles were created through single-source imitation. For each topic, we randomly selected 10%, 25%, or 50% of the original articles and replaced them with AI-generated versions by prompting the LLM to imitate a specific original article from the same topic, but in the imitating outlet's usual style (Fig. 1b). AI-generated articles were then assigned to their original topic, replacing the original article previously in the topic from the imitating outlet. We replaced original articles with their AI-generated counterparts, rather than merely adding AI-generated articles to the existing pool, to isolate the effect of AI-driven imitation from confounding factors such as volume increases. Furthermore, substitution approximates organizational AI adoption decisions under fixed publishing budgets and newsroom capacity constraints.

We applied the same procedure in World Multi-Source, using two source articles instead of one (Fig. 1c). When analyzing World Single-Source and World Multi-Source, we modeled outcomes at varying prevalence levels of AI-driven imitation, replacing between 10–50% of original articles with AI-generated versions. In doing so, we explored the incremental impacts of AI-driven imitation on the resulting semantic diversity.

Imitation targets were drawn from the same topic as the focal article being replaced, and the focal article itself was excluded from this pool, so an AI-generated article could never imitate the human article it replaced. For multi-source imitation, the two source articles could originate from the same or different media outlets. The number of AI-generated articles per topic was $max(1, round(p \cdot n))$, where n is the number of articles in the topic and $p$ the AI prevalence. Topics without an eligible source received none, i.e., single-article topics and, for multi-source imitation, topics with fewer than four articles. Requested generations that failed were left as the original human-written article and labelled human (0.37% of intended AI-generated articles in the Fig. 2 worlds and 0.63% in the Fig. 5 worlds). Prevalence levels are nominal, and realised shares are close to them, except under multi-source imitation in the heterogeneous environment, where topics with two or three articles receive no AI-generated article (47% at a nominal 50%). A fixed random seed was reused across prevalence levels applied to the same article pool, so the 10% replacement set is a subset of the 25% set, which is a subset of the 50% set, within each topic group.

Worlds that randomly mix the seven task specifications (the benchmark task specification, three ablations, and three alternative newsroom prompts) assigned each replaced article to one specification using the same hash-based procedure, giving each specification approximately equal (1/7) probability per replacement article slot.

**LLM details**

For content generation, we used OpenAI's batch processing API with the GPT-4o LLM (gpt-4o-2024-08-06, default temperature = 0.7) due to its large context window and relative cost-effectiveness (see Supplementary Text 9 for sensitivity checks on temperature settings, and Supplementary Fig. 6 for sensitivity check to prompting technique). We also ran simulations of imitating AI agents via DeepInfra API using DeepSeek-V3.2 and Llama-4-Scout-17B-16E-Instruct under matched prompting templates and prevalence levels. At 50% prevalence, all LLMs replace the same articles, while at 10% and 25% the replaced articles were not the same for all LLMs. Rebuilding these worlds with identical replacement sets changes no estimate by more than 0.005, leaves the direction of every single-source effect unchanged and preserves the position of DeepSeek-V3.2 relative to the other LLMs (Supplementary Text 15). This multi-LLM comparison allows us to explore whether the effect of imitating AI agents on semantic diversity varies depending on their LLM basis. All inputs given and outputs returned from the OpenAI and DeepInfra APIs were saved locally for the authors' reproducibility purposes.

**Varying agents' task specifications**

Each generation request specified an editorial role, the full text of one reference article for single-source imitation or two reference articles for multi-source imitation, and instructions for producing a Danish-language news article. The reference articles were selected by the simulation procedure, and not chosen by the LLMs. Requests comprised a system message defining the editorial role and a user message containing the reference material, task instructions, and a structured output template. The output template separated the generated article text from metadata identifying the imitating outlet and reference articles. In the benchmark specification, LLMs were instructed to produce a verbose, restructured, and expansive article in the named outlet's usual style, while matching the original article's word count. These were generation instructions, but not any guarantees of factual preservation, stylistic fidelity, or length compliance. This specification represents a plausible newsroom use of AI, as outlets likely seek to adapt reporting from other sources while maintaining a distinctive editorial identity and differentiating their coverage from competitors.

To examine how agent objectives shape the consequences of AI-driven imitation, we compared this specification with tasks varying the requested degree and form of transformation. Three ablation conditions removed the explicit differentiation-oriented instructions, removed the named imitating-outlet cue from the prose instructions, or removed both. The differentiation ablation replaced instructions emphasizing expansion, restructuring, and stylistic adaptation with a more generic rewriting task. The outlet-cue ablation removed the named publication from the prose instructions. Three additional specifications represented alternative newsroom objectives: 1) reproducing source material with only minor editorial adjustments, reflecting an emphasis on minimizing editorial effort, 2) summarizing its key facts in approximately 200 words as neutral wire-service copy, reflecting an emphasis on concise redistribution, and 3) rewriting while explicitly preserving named sources, figures, chronology, and factual claims, reflecting an emphasis on factual fidelity.

We implemented these task variants using DeepSeek-V3.2 with zero-shot prompting and temperature 0.7, for both single-source and multi-source imitation in homogeneous and heterogeneous information environments, at 10%, 25%, and 50% AI prevalence. Alternative requests were constructed from the corresponding original requests, retaining the reference article texts and identifying metadata while changing the editorial instructions. The ablations isolate components of the benchmark task specification, whereas the alternative tasks examine broader combinations of editorial priorities in the agents' task specification. Full system and user prompt templates are provided in Supplementary Text 2.

**Estimating semantic diversity**

Because our interest lies in semantic differentiation within shared news topics, we measure semantic diversity by estimating the average semantic similarity among articles covering the same news topic on a given day. We obtained article embeddings with the sentence-transformer distiluse-base-multilingual-cased. Because the embedding model reads at most 128 tokens and human-written texts are longer (median 523 tokens), we split each article into consecutive 126-token windows, embedded each window and represented the article by the length-weighted mean of its window embeddings, normalized to unit length. As a robustness check, we re-estimated all embedding-based results with two multilingual long-context models, BGE-M3[58] and Qwen3-Embedding-4B[59], which encode each article in a single pass (Supplementary Text 10). We then calculated the mean within-topic pairwise cosine similarity across all original and generated articles in each world (see Supplementary Text 1 for details on similarity estimation and Supplementary Text 10 for text embedding sensitivity checks).

In world-level analyses, intervals are 95% intervals across news topics (equal topic weights), conditional on one stored simulation per condition. Whether a change differs from zero is assessed from the paired per-topic differences between a simulated world and the human baseline, and error bars in the figures show the interval of each world's mean. The pooled article-level models (Fig. 3, Supplementary Text 3) are linear mixed models fitted by restricted maximum likelihood with a random intercept for topic, and inference uses $z$ statistics with standard errors clustered by topic (CR2; 9,267 clusters) and 95% intervals of ±1.96 standard errors. All tests are two-sided. We report article-level models with and without the article-level covariates (word count, named entities, numbers and sentiment). Because AI rewriting changes these textual characteristics, neither estimate is the total effect of imitation, and their difference is not an estimate of mediation. Importantly, cosine similarity captures overall differentiation in embedding space and does not independently identify whether observed differences arise from factual content, style, framing, language, or other textual characteristics.

Because our analyses rely on embedding-based cosine similarity to capture differences between news articles, we conducted a limited post-hoc sanity check to examine whether this measure covaries in the expected direction with readers' overall perceptions of textual similarity. Rather than a full-scale calibration, the goal of this exercise was to verify that variation in cosine similarity corresponds in the expected direction to how human readers perceive similarity between human-written and AI-generated texts, and not whether that difference is attributed to any factual, stylistic or informational dimension. We fielded a survey to a representative sample of Danish respondents ($n$ = 1,201; self-reported gender: 594 women, 605 men, 2 other gender identity), in which participants were randomly assigned to evaluate one of five pairs of short news article snippets drawn from the corpus. Each pair consisted of one human-written article and one AI-generated imitation from the same topic.

Respondents rated the similarity of the two snippets on a 0–10 scale. For each stimulus pair, we computed the mean perceived similarity and compared these ratings to the corresponding embedding cosine similarity scores used in the simulations. Across the five stimulus pairs, higher cosine similarity corresponded to higher mean perceived similarity (Pearson's $r = 0.95$, exact permutation $P = 0.008$, see Supplementary Fig. 9a and Supplementary Text 11). Given the small number of stimulus pairs (n = 5), we interpret this result as a limited directional sanity check rather than as a comprehensive validation of the similarity measure. Specifically, it indicates that embedding similarity and overall human similarity judgments vary in the same direction across these stimuli, but does not establish that this relationship generalizes beyond the five pairs or which factual, stylistic, framing, or other textual dimensions account for either measure. Further, these findings are consistent with prior work validating cosine similarity of SBERT-style sentence embeddings against human similarity judgments in semantic textual similarity tasks[60,61] and with evidence that similarity judgments for news are driven primarily by article body text[62].

In addition, respondents were asked to identify which snippet they believed to be AI-generated, choosing among four options (AI article, human article, both, or neither). As shown in Supplementary Fig. 9b, respondents selected incorrect options more often than the correct one, most frequently misidentifying the human-written snippet as AI-generated. This pattern suggests that the AI-generated stimuli are not trivially distinguishable from human-written text.

**Topic- and source-level analyses**

For each topic we computed the mean similarity among its human-written articles in the all-human baseline and ranked topics by this value within each environment (percentile ranks). In each simulated world, we compare how similar the AI-generated and the human-written articles on a topic are to the other articles on that topic. For each article, we compute its average similarity to all other articles on the topic, as in Fig. 3. The AI-human contrast is the average of this value for the topic's AI-generated articles minus the average for its human-written articles. To check for regression to the mean, we also split each topic's human-written articles at random into two halves, ranked topics using one half, and computed the contrast using only the other half. The contrast falls from the least to the most similar third of topics in all 36 conditions in both versions (in the split-half version the intervals of the two thirds are separate in 35 of 36).

The source-relative shift compares an AI-generated article with the articles it was written from. We compute the average similarity of the AI-generated article to the other human-written articles on its topic in the original coverage, and do the same for its source articles. The shift is the difference between the two. Because both are compared with the same set of articles, a positive shift means that the AI-generated article is closer to the existing coverage than its sources were. Negative values mean that the AI-generated article is further from the rest of the coverage than the material it was written from. Single-source rewrites have negative shifts in every condition (−0.035 to −0.055). For multi-source articles, we ranked articles by the similarity between their two sources. Here, the shift is negative when the two sources are nearly identical (−0.041 to −0.075 in the most similar third of source pairs) and positive when they differ (+0.021 to +0.040 in the least similar third). As a check for regression to the mean, we compared multi- and single-source rewrites whose source articles are equally typical of their topic (fifths of the sources' similarity to the topic's coverage): in the least typical fifth, multi-source articles sit 0.043 to 0.055 cosine points closer to the coverage than single-source rewrites, per LLM and environment. These analyses describe how the aggregate effect arises from changes in individual articles; the comparisons across topics are descriptive and do not identify a causal mechanism.

Topics were classified as having no political mentions, political with parties or politicians from one bloc, or political with both blocs, according to their mentions of Danish parties and politicians, identified from party-letter tags after names, the full names of members of the Folketing in 2019–2026 (from Wikidata), and written-out party names. The detector was validated by hand. All 100 sampled detected mentions named a Danish party or politician (95% CI 96.4–100%), and none of 50 randomly sampled articles without a detected mention, drawn from topics with other articles containing political coverage, named one (95% CI 92.9–100%). The classification describes mentions, does not analyze a topic's subject or framing, and the comparison between topic types is descriptive.

**Human benchmark for information loss**

To compare AI rewriting with journalists' editing of the same input, we drew 1,700 Ritzau articles at random from those adapted by a Danish outlet and gave each to DeepSeek-V3.2 with the benchmark task specification, naming the same outlet as the journalist's adaptation. For the agency article, the journalist's adaptation and the AI rewrite, we extracted named entities and numbers with the same procedure (Supplementary Text 5) and computed the share of the agency article's items missing from each. As a supporting comparison, each of the three LLMs also rewrote the journalists' adaptations of a Ritzau article (Extended Data Fig. 3, chains). Intervals for the direct comparison are across articles, whereas intervals for the Ritzau-human-AI chains are clustered by Ritzau article.

The comparison is matched on the source article and the target outlet, not on the instructions, as the journalists worked under their own editorial conditions rather than under our task specification. It therefore places the AI omission rates in context but does not establish that AI is generally worse than journalists at preserving source information under identical tasks. The shares are source-relative omissions and do not measure factual errors.

**Robustness checks**

We conducted additional robustness analyses to assess whether our main findings depend on specific modeling choices. First, we examined sensitivity to the temperature parameter used during text generation. Repeating the analysis of semantic diversity in a homogeneous world at lower (0.2, 0.5) and higher (1.2) temperatures yields substantively similar patterns (Supplementary Figs 3 and 4). While the magnitude of the effects varies modestly, higher temperatures amplify and lower temperatures attenuate the diversifying effect, as expected given differences in output randomness.

Second, we compared embedding pipelines in one benchmark world: homogeneous, GPT-4o, single-source imitation, zero-shot prompting, and 50% AI prevalence. For each of 9,267 date-specific topics, we calculated the mean within-topic pairwise cosine similarity separately using the main measure and OpenAI's text-embedding-3-large. These paired topic-level means were strongly correlated (Pearson's $r = 0.954$, 95% CI [0.952, 0.956], $t(9{,}265) = 307.32$, $P < 0.001$, Supplementary Fig. 5 and Supplementary Text 10). This comparison shows substantial agreement in the relative similarity of topics, but does not establish that the pipelines yield the same baseline-relative effects or ordering across all simulated conditions. All embedding-based results were also estimated with BGE-M3 and Qwen3-Embedding-4B (see 'Estimating semantic diversity'). The main results are the same under all three representations. They differ mainly in the size of some effects and, in a few conditions, in sign (Supplementary Text 10 and Extended Data Fig. 5).

Third, we examined the extent to which homogeneous and heterogeneous environments differ in addition to their baseline similarity levels (Supplementary Text 14). Measured article characteristics were closely balanced across the shared topic support, with all absolute standardized differences below 0.04, while the total-variation distance between the overall outlet distributions was 0.034 (Supplementary Table 4C). Further, we replicated Fig. 2 using only the 5,338 topics represented in both environments (Extended Data Fig. 1): the direction of the single-source effect is the same in both environments, while the reduction in the homogeneous environment is smaller than on all topics.

**Data availability statement**

This study uses a proprietary dataset obtained via the API from Infomedia A/S, which cannot be shared publicly due to legal constraints and copyrighted material. Access to the data can be purchased from Infomedia (https://infomedia.dk) with written consent from the media organizations included in the sample. With access to the dataset and our materials, the study can be replicated in full.

**Code availability statement**

Custom code was developed for data processing, large-scale simulation, text generation via large language models, and the estimation of semantic similarity metrics that are central to the results reported in this study. Text processing was conducted using external APIs (OpenAI, DeepInfra) under terms specifying that submitted data are not used for training. Due to the use of proprietary data and API-based content generation, the full computational pipeline cannot be executed without authorized access to the underlying data sources.

All custom code and non-proprietary derived data are made public upon publication in a Zenodo repository.

# Figures

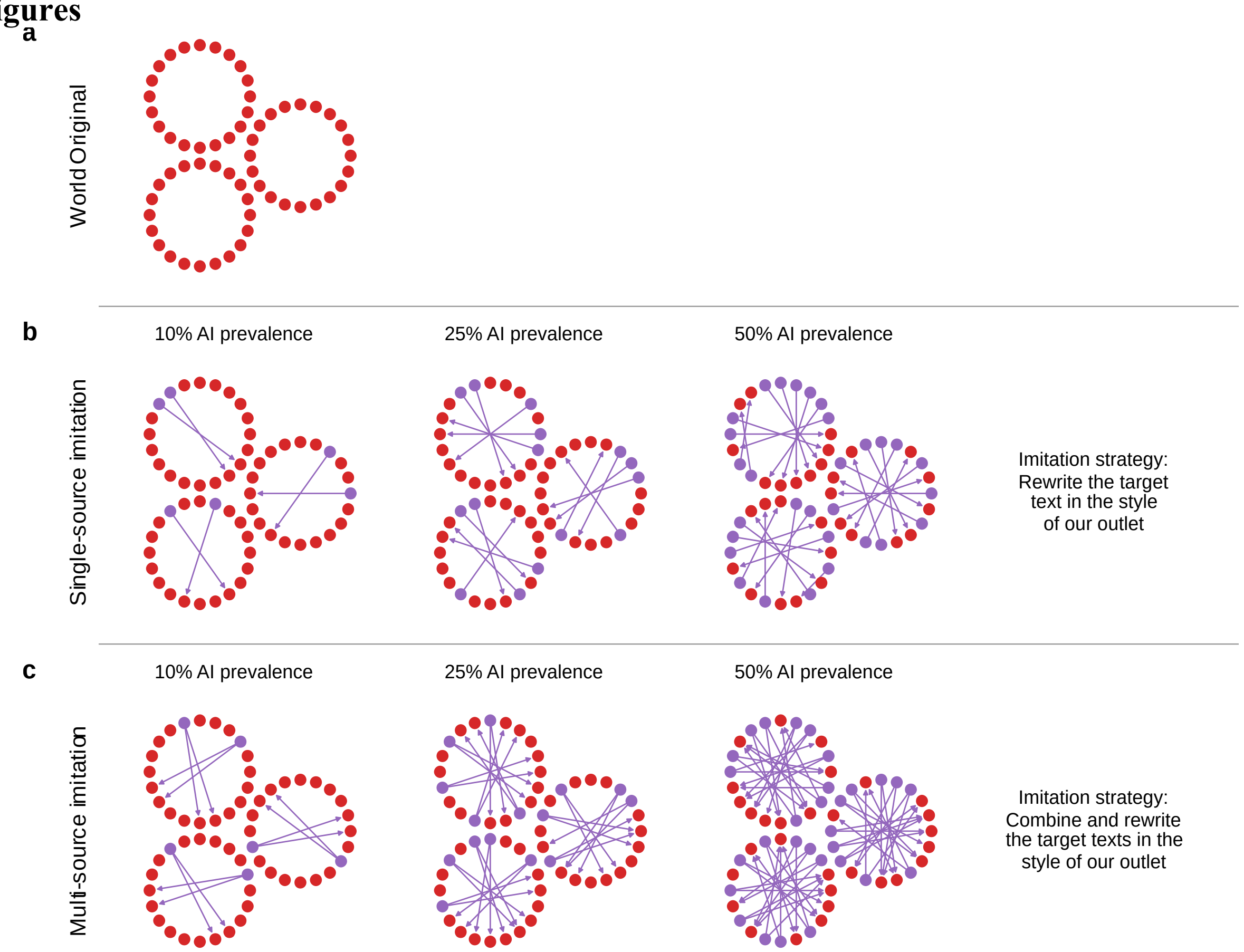

**Figure 1.** Conceptual illustration of the multi-world simulation design and AI imitation strategies. Within each imitation regime, columns show simulated worlds differing in AI prevalence (10%, 25%, 50%), while the underlying topic structure is held constant. Each row shows a different imitation regime: (a) World Original, containing only human-written articles; (b) Single-source imitation, where AI-generated articles replace human-written articles by imitating one randomly selected source within the same topic; and (c) Multi-source imitation, where AI-generated articles replace human-written articles by synthesizing two randomly selected sources from the same topic. Red nodes denote human-written articles and purple nodes denote AI-generated articles. Edges indicate the imitation target(s) randomly selected by the simulation procedure and are shown for illustrative purposes only. The figure depicts three hypothetical topics to visualize how imitation strategy and prevalence jointly define each simulated world. The 'rewrite in the style of our outlet' labels in (b) and (c) illustrate the benchmark task specification used for the results in Fig. 2, and the additional ablated and alternative task specifications examined in this study (Methods, 'Varying agents' task specifications' and Supplementary Text 2) are not depicted here.

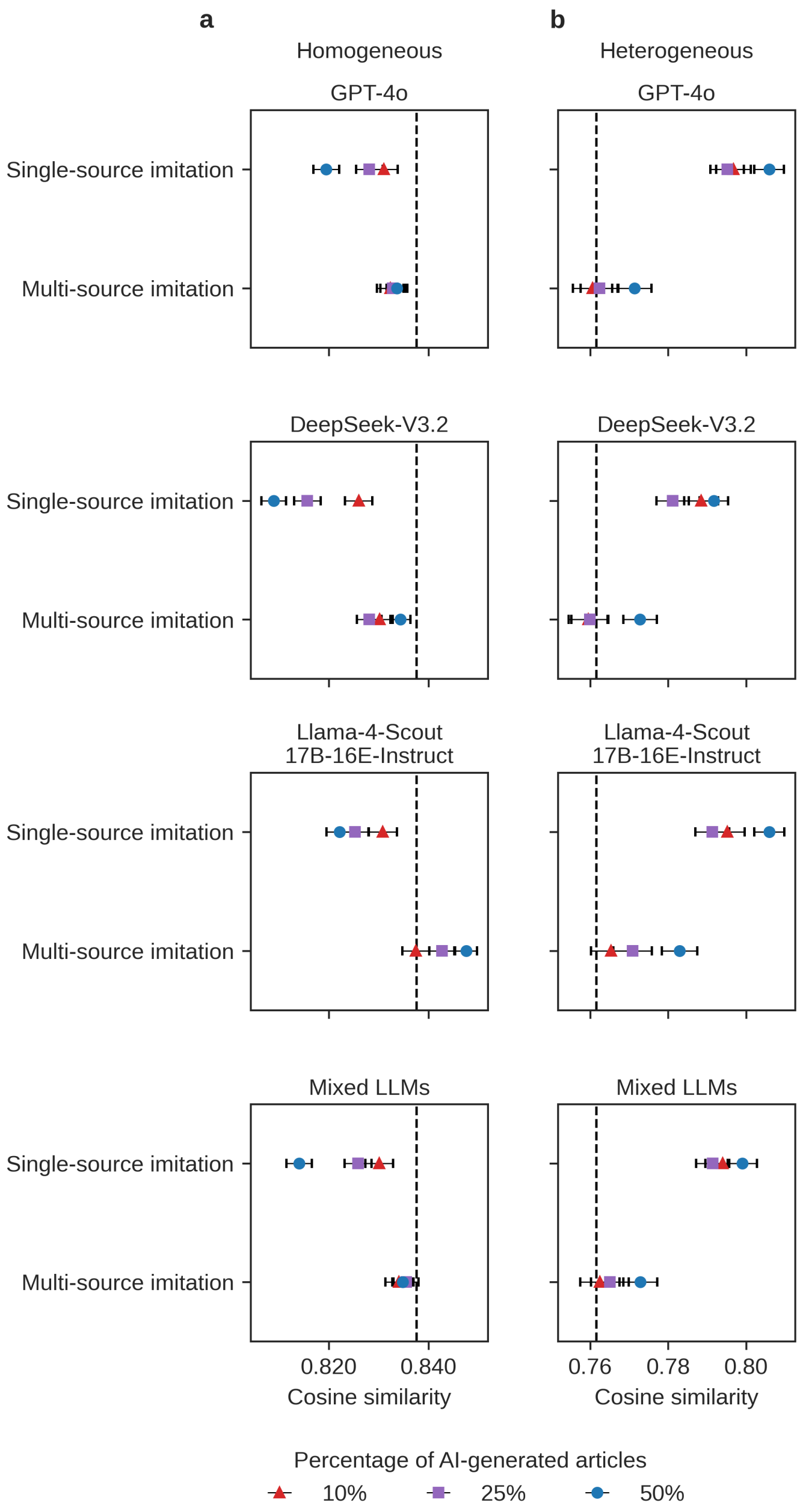

**Figure 2.** AI-driven imitation lowers within-topic similarity in the homogeneous environment and raises it in the heterogeneous environment under single-source imitation. Mean within-topic cosine similarity in simulated worlds defined by environment (columns), LLM (rows: GPT-4o, DeepSeek-V3.2, Llama-4-Scout-17B-16E-Instruct and a mixed-LLM condition combining the three), imitation strategy (single- vs multi-source) and AI prevalence (10%, 25%, 50%), under the benchmark task specification ($n$ = 48 worlds). In the homogeneous environment (a), single-source imitation lowers similarity (higher diversity) for every LLM and prevalence, while multi-source imitation changes it by at most 0.01 in either direction (slightly lower for GPT-4o, DeepSeek-V3.2 and the mixed-LLM condition, and higher for Llama at 25% and 50% prevalence). In the heterogeneous environment (b), single-source imitation raises similarity for every LLM and prevalence, and multi-source imitation raises it at 50% prevalence. Each point is the mean of per-topic average cosine similarity in one world. Error bars are 95% intervals across topics (9,267 topics in the homogeneous and 5,338 in the heterogeneous environment). Dashed lines mark the baseline similarity of the human worlds (homogeneous 0.838, heterogeneous 0.762). The homogeneous environment is the observed Danish news ecosystem in 2022, whereas the heterogeneous environment excludes articles sourced from the national news agency.

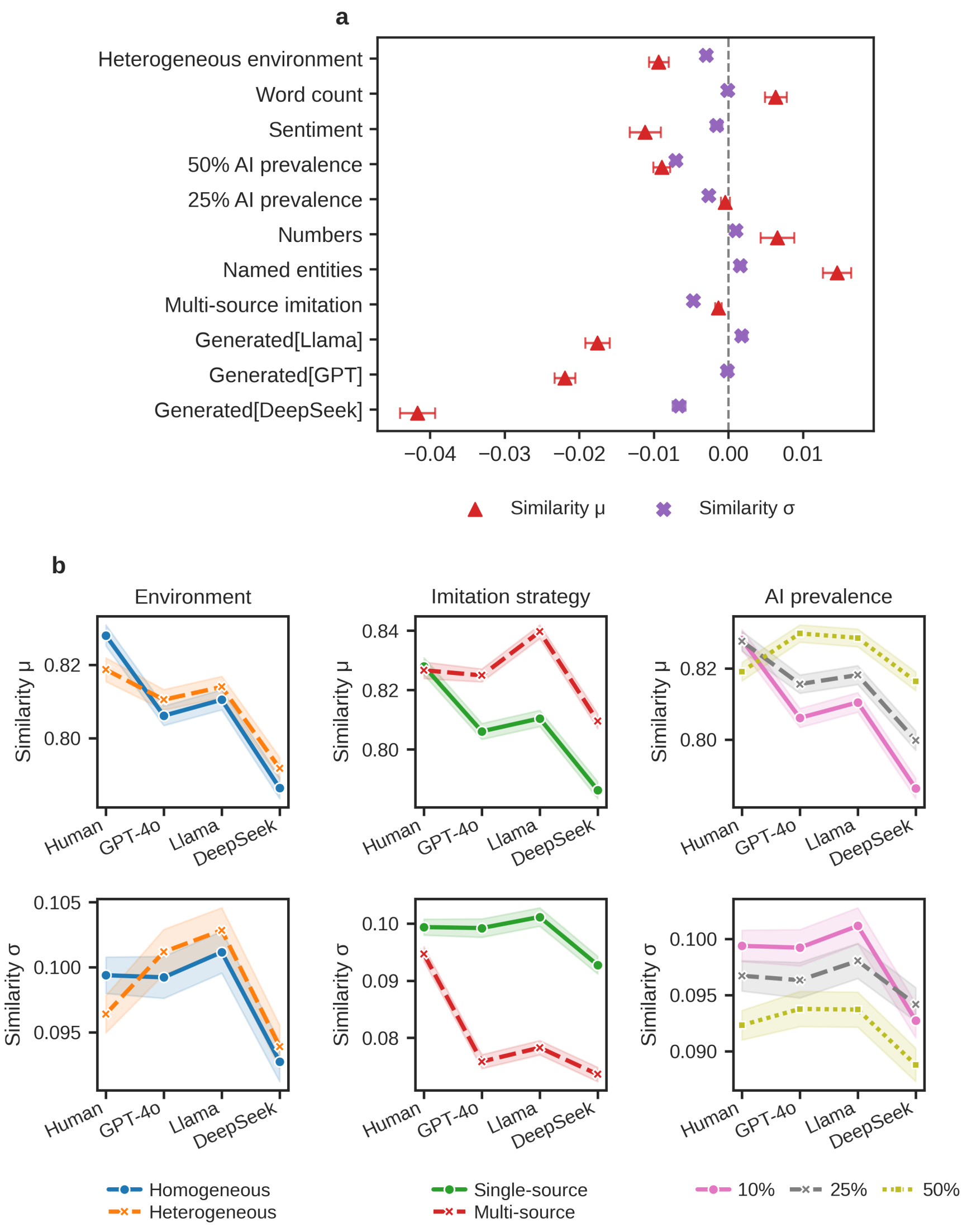

**Figure 3.** AI-generated articles are on average less similar to the other articles in their topic than human-written articles are. a, Coefficients from pooled linear mixed models of each article's mean within-topic cosine similarity ($\mu$, triangles) and the standard deviation of its similarities ($\sigma$, crosses), pooling all LLMs and conditions and adjusting for environment, imitation strategy, prevalence and article length, sentiment, named entities and numbers (random intercept for topic; normal-reference 95% intervals with standard errors clustered by topic, 9,267 clusters; $n$ = 1,455,960 for $\mu$ and 1,447,215 for $\sigma$; Extended Data Table 1). b, Predicted $\mu$ (upper row) and $\sigma$ (lower row) for human-written and AI-generated articles by environment, imitation strategy and prevalence, with 95% intervals. Because the adjusted characteristics may themselves change with AI rewriting, the coefficients are a conditional comparison, not the total effect of imitation (estimated without adjustment in Fig. 2; unadjusted estimates in Extended Data Table 1). Dispersion in similarity is lower for AI-generated than for human-written articles under multi-source imitation and does not differ consistently under single-source imitation. As AI prevalence increases toward 50%, the AI-human contrast narrows and, for GPT-4o and Llama, changes sign, because the remaining human-written articles are increasingly compared with AI-generated ones. In the homogeneous single-source condition shown in b, the change in within-topic similarity across the simulated world remains negative (Fig. 2).

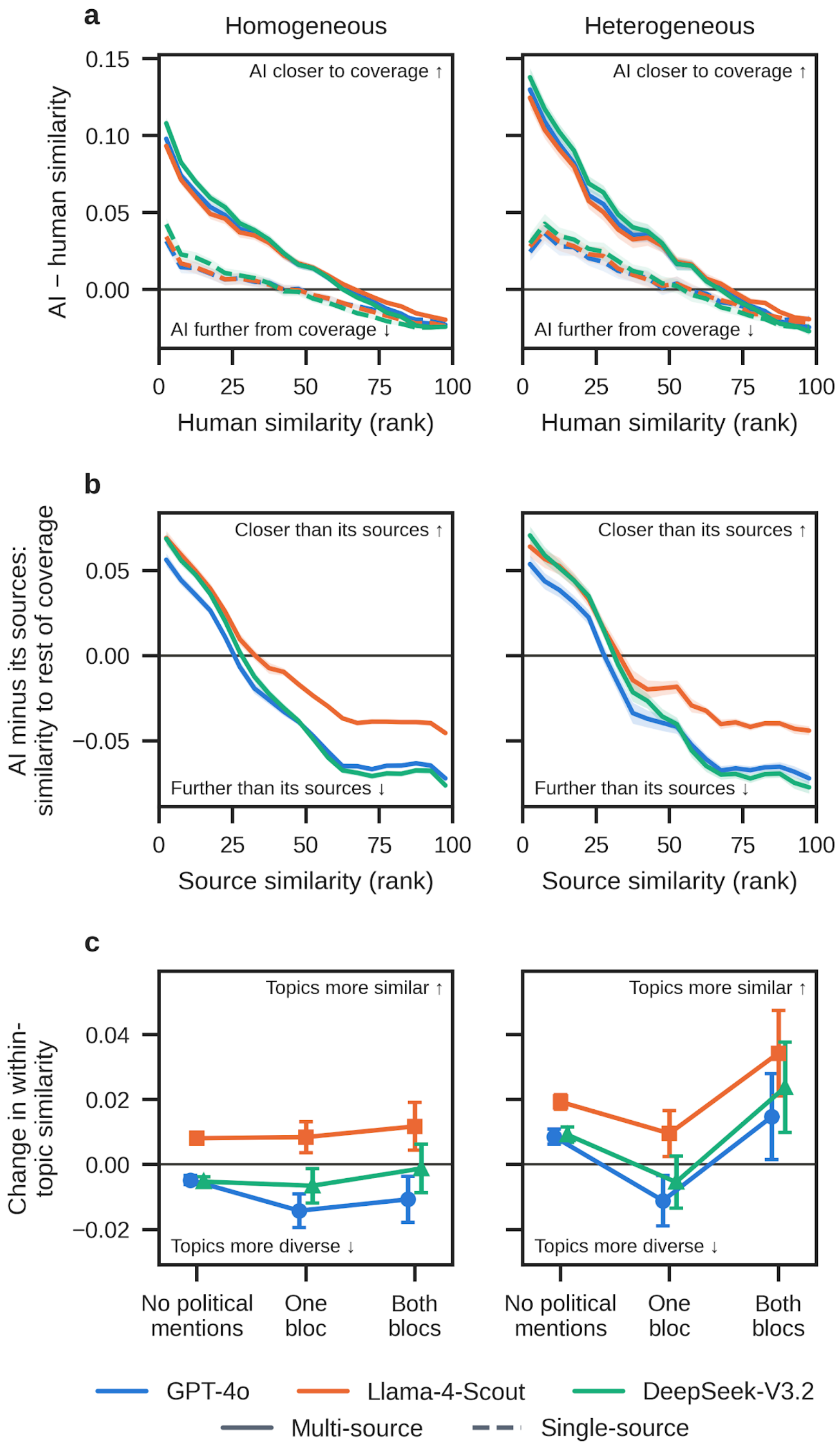

**Figure 4.** AI-generated articles stand out most where coverage is already alike, and multi-source imitation pulls coverage together where its sources differ. All panels: benchmark task specification, 50% AI prevalence; columns, environments; lines and points, LLMs. Shaded bands and error bars are 95% intervals across topics (9,267 topics in the homogeneous and 5,338 in the heterogeneous environment). The three rows show three different outcomes, and the labels inside each panel state what higher and lower values mean for that row; rows a and b compare articles within topics, and only c shows the change in a topic's overall similarity. a, AI-human article contrast: mean similarity of AI-generated articles to the rest of their topic's coverage minus that of the remaining human-written articles, with topics ranked within each environment by the similarity of their human-written articles before imitation (percentile rank, 20 equal-count bins). Solid lines indicate multi-source imitation, and dashed, single-source. Positive values mean AI articles are more similar to the coverage than human articles. The contrast falls from the lowest to the highest third of topics in all 36 combinations of LLM, environment, strategy and prevalence. b, Source-relative shift. For multi-source AI articles, similarity to the rest of the topic's coverage minus that of their two source articles, by percentile of the similarity between the two sources. Positive values mean the AI article sits closer to the coverage than its sources. c, Change in within-topic similarity from the all-human baseline under multi-source imitation, by topic type: no political mentions, political with parties or politicians from one bloc, and political with both blocs (from mentions of parties and politicians, see Methods). Under the two long-context embedding models (BGE-M3 and Qwen3-Embedding-4B; not shown), the contrast in a falls from the least to the most similar third in all 36 combinations under each model, the shift in b falls with source similarity in all 18 LLM × environment × model cells, and the both-bloc minus one-bloc difference in c in the heterogeneous environment has intervals excluding zero for all 9 LLM × model combinations.

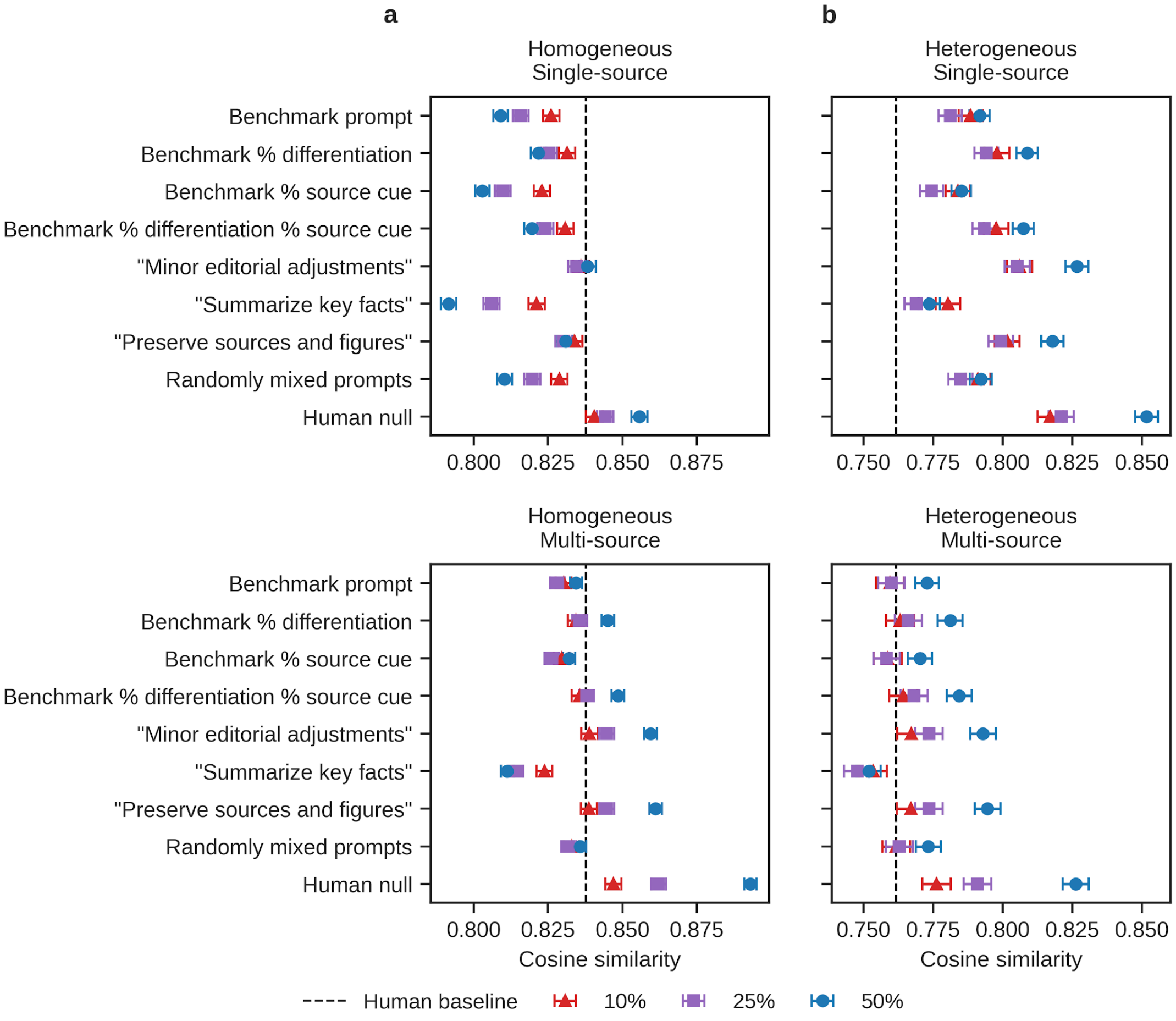

**Figure 5.** Task specification changes the size of the diversifying effect of single-source imitation, but does not change its direction within an environment. Mean within-topic cosine similarity in worlds generated by DeepSeek-V3.2 under seven task specifications, a mixed condition and a null model (rows, from the top): the benchmark task specification of Fig. 2; the benchmark without the differentiation instruction; without the outlet source cue; without both; a minor-editorial-adjustments instruction; a wire-agency-style summary; an instruction to preserve sources and figures; all of these randomly mixed within a world; and a human null model. Panels show environment (columns; a, homogeneous; b, heterogeneous) and imitation strategy (rows of panels); colors show AI prevalence ($n = 108$ worlds: 96 AI worlds and 12 null-model worlds). Under single-source imitation, every AI task specification lowers similarity in the homogeneous environment (except minimal edits at 50% prevalence, where the 95% interval includes zero) and raises it in the heterogeneous environment, with sizes that vary by specification. Under multi-source imitation, specifications diverge in direction in both environments. In the heterogeneous environment, every specification except the wire summary raises similarity at 50% prevalence. In the single-source human null model, the selected source article replaces the target (simple copying). In the multi-source human null model, the target is replaced by the centroid of the two source embeddings (simple combining). Both null models raise similarity above the baseline. Points and intervals as in Fig. 2. Dashed lines mark the human baseline.

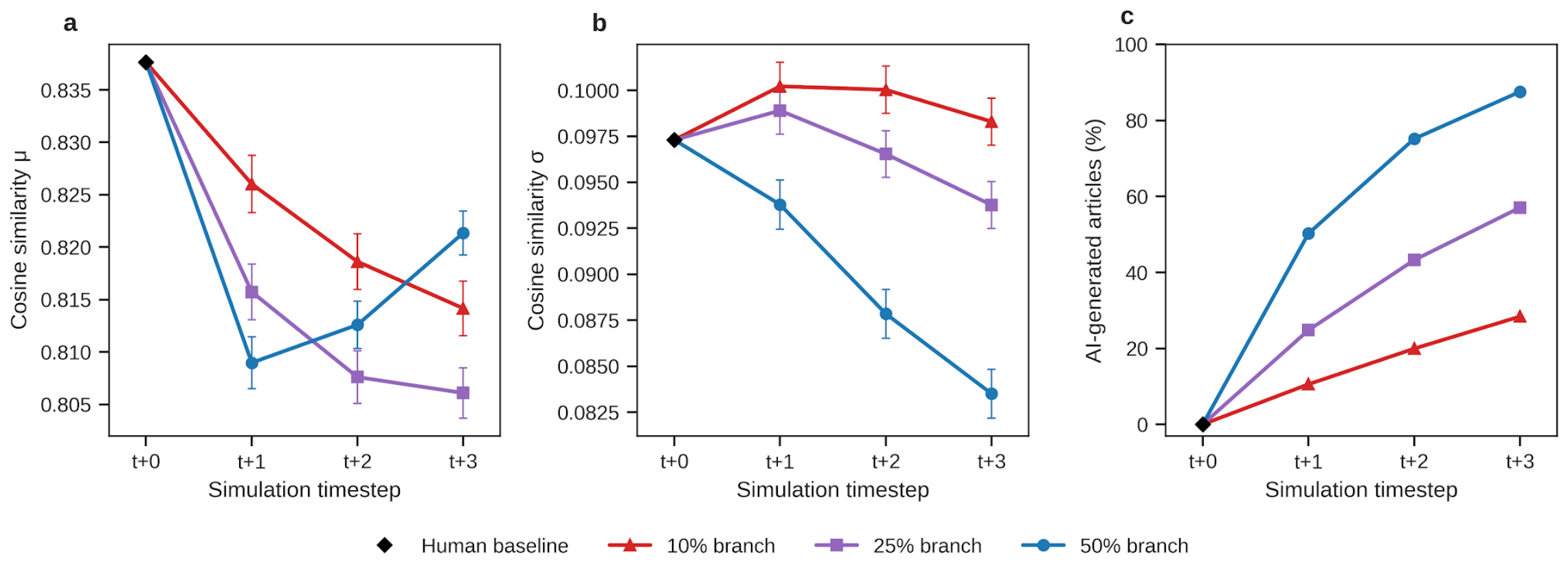

**Figure 6.** Repeated imitation keeps similarity below the human baseline. Longitudinal extension of the homogeneous single-source DeepSeek-V3.2 simulation under the benchmark task specification (Supplementary Text 13). At each time step, the indicated share of article slots in each independent branch is replaced. AI-generated articles from earlier rounds stay in the world and can become imitation targets, and branch labels are per-round replacement rates. a, Mean within-topic cosine similarity ($\mu$; lower values mean greater diversity). b, Mean within-topic dispersion of article-level similarity ($\sigma$). c, Realised share of AI-generated articles. The black diamond at t+0 is the human baseline. Error bars are 95% intervals across topics (9,267). In the 10% and 25% branches, mean similarity continues to decline through t+3. In the 50% branch, it partly recovers after the first round but remains below the human baseline. Dispersion falls with the replacement rate, but stays near or above the baseline in the 10% branch, falls below it after the first round in the 25% branch and falls steadily in the 50% branch.

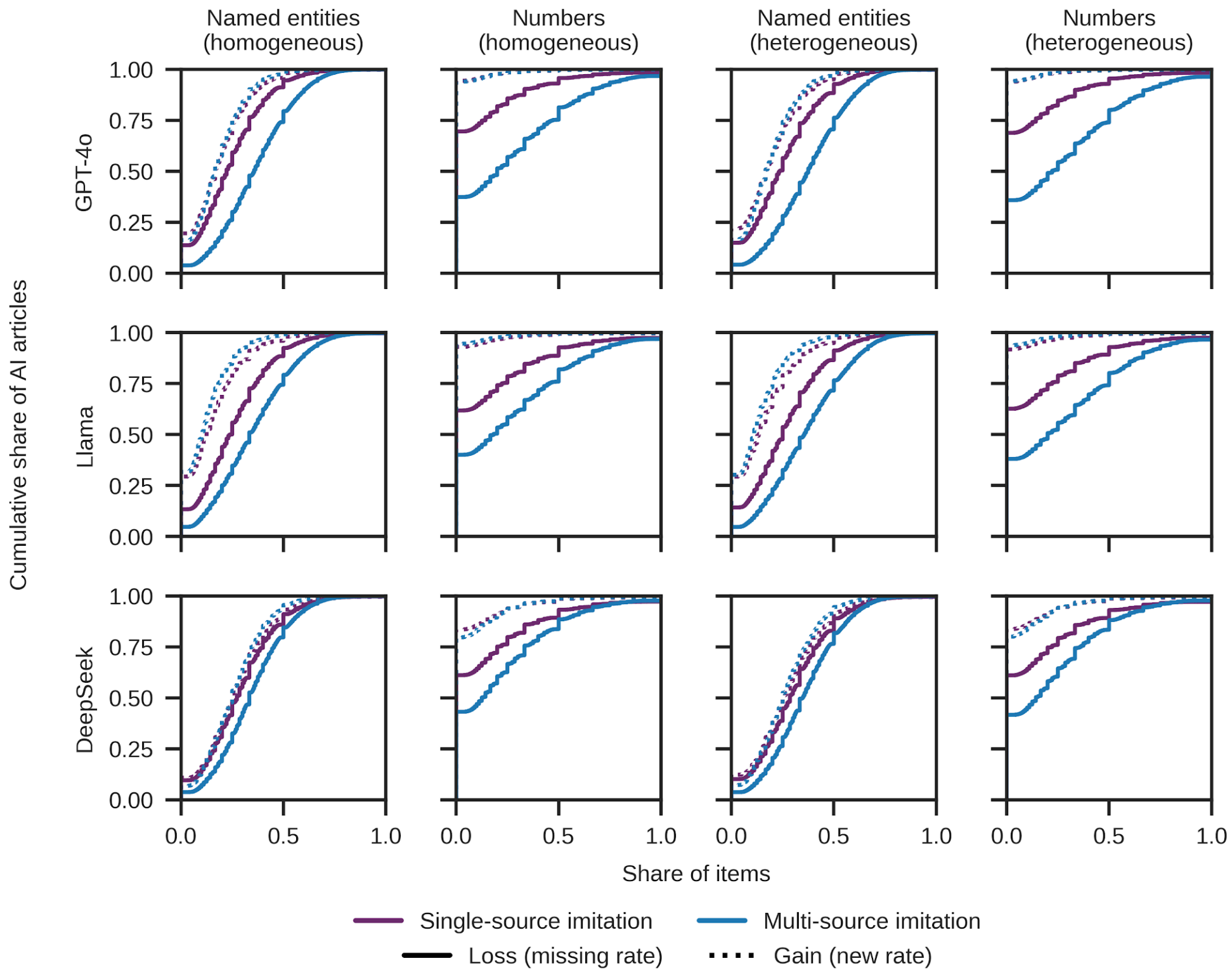

**Figure 7.** AI-driven imitation omits source named entities and numbers. Empirical cumulative distribution functions, across AI-generated articles (benchmark task specification, 50% prevalence; 462,219 article comparisons), of the loss rate (solid lines: share of the source article's unique named entities or numbers that are missing from the AI-generated article) and the gain rate (dotted lines: share of the AI-generated article's unique items that are new, i.e. absent from its source or sources). The two rates have different denominators, source items for the loss rate and output items for the gain rate, and are not netted, and a new item in an AI article is not necessarily a new fact. The x-axis gives the rate (share of items), and the y-axis the cumulative share of AI-generated articles. Rows are LLMs; columns are named entities and numbers in the homogeneous and heterogeneous environments. Purple indicates single-source imitation, blue is multi-source imitation. In a separate comparison, journalists adapting news agency articles omitted 11.8% of their named entities, against 25.4% for an LLM given the same agency articles (Extended Data Fig. 3).

## Extended Data figures and tables

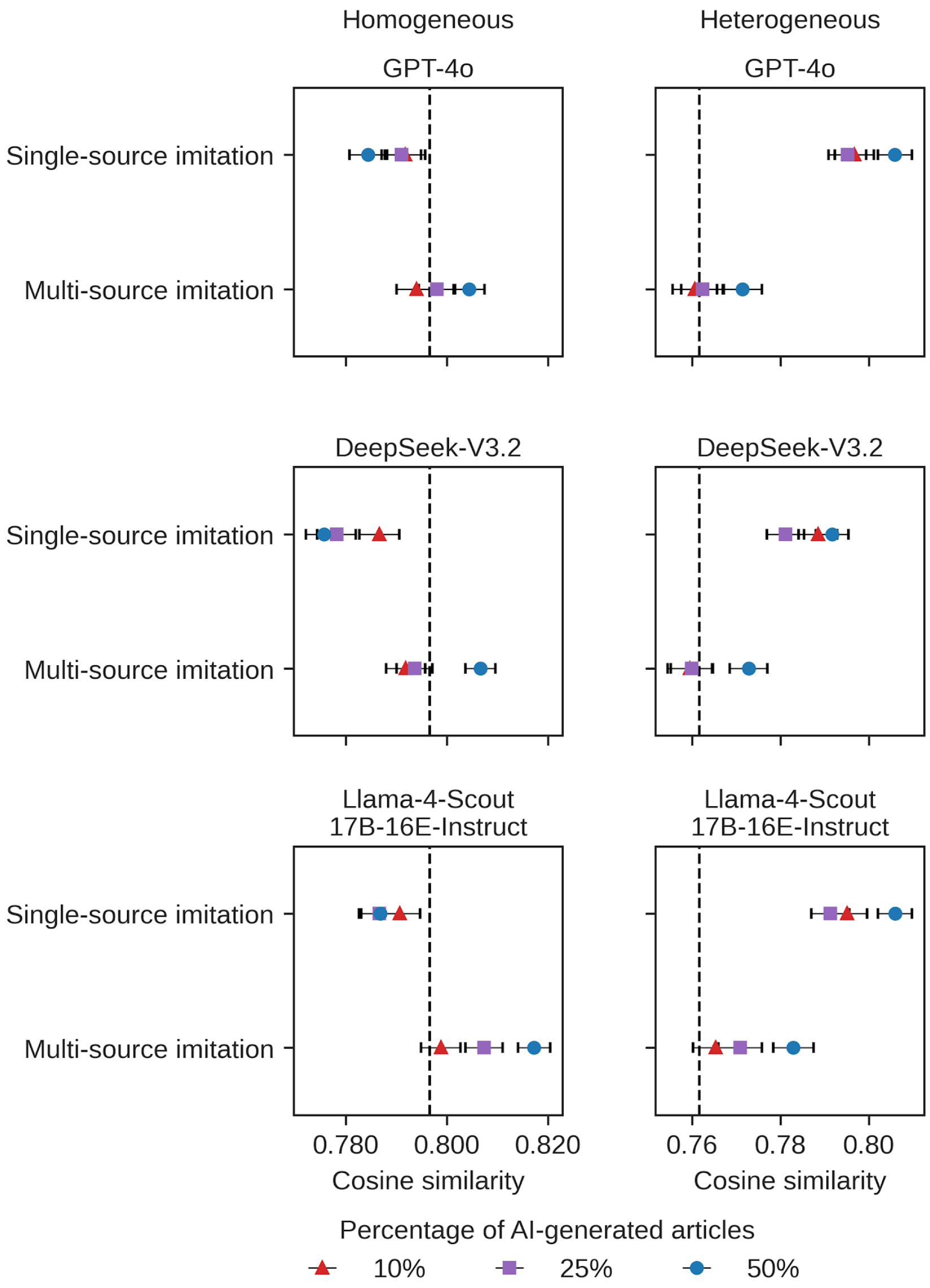

**Extended Data Fig. 1.** On topics covered in both environments, the environment pattern holds for single-source imitation, with a smaller effect in the homogeneous environment. $n = 36$ worlds (three LLMs). Error bars are 95% intervals across the 5,338 shared topics. As Fig. 2 (three LLMs), but restricted to topics present in both environments. Dashed lines are the all-human baselines on these topics (homogeneous 0.797; heterogeneous 0.762), which differ by 0.035, against 0.076 on all topics. At 50% prevalence under single-source imitation, the homogeneous reduction is −0.010 to −0.021 on shared topics, against −0.023 to −0.039 on topics covered mostly through news agency copy. Multi-source imitation changes similarity by at most 0.005 at 10% and 25% prevalence (Llama at 25%: about +0.01) and raises it at 50% in both environments (+0.008 to +0.021).

| | Mean similarity (μ) | | | | Dispersion (σ) | | | |
|---|---|---|---|---|---|---|---|---|
| Term | Adjusted | P | Unadjusted | P | Adjusted | P | Unadjusted | P |
| Intercept | 0.828 [0.825, 0.831] | <0.001 | 0.833 [0.830, 0.836] | <0.001 | 0.099 [0.098, 0.101] | <0.001 | 0.100 [0.099, 0.101] | <0.001 |
| Generated[GPT] | −0.022 [−0.023, −0.021] | <0.001 | −0.034 [−0.035, −0.033] | <0.001 | 0.000 [−0.001, 0.000] | 0.624 | −0.001 [−0.002, −0.001] | <0.001 |
| Generated[Llama] | −0.018 [−0.019, −0.016] | <0.001 | −0.026 [−0.027, −0.025] | <0.001 | 0.002 [0.001, 0.002] | <0.001 | 0.001 [0.000, 0.001] | 0.041 |
| Generated[DeepSeek] | −0.042 [−0.044, −0.039] | <0.001 | −0.048 [−0.049, −0.046] | <0.001 | −0.007 [−0.007, −0.006] | <0.001 | −0.008 [−0.009, −0.007] | <0.001 |
| Heterogeneous environment | −0.009 [−0.011, −0.008] | <0.001 | −0.010 [−0.011, −0.008] | <0.001 | −0.003 [−0.004, −0.002] | <0.001 | −0.003 [−0.004, −0.002] | <0.001 |
| Multi-source imitation | −0.001 [−0.002, −0.001] | <0.001 | −0.001 [−0.002, −0.001] | <0.001 | −0.005 [−0.005, −0.004] | <0.001 | −0.005 [−0.005, −0.004] | <0.001 |
| 25% prevalence | 0.000 [−0.001, 0.000] | 0.169 | 0.000 [−0.001, 0.000] | 0.190 | −0.003 [−0.003, −0.002] | <0.001 | −0.003 [−0.003, −0.002] | <0.001 |
| 50% prevalence | −0.009 [−0.010, −0.008] | <0.001 | −0.009 [−0.010, −0.008] | <0.001 | −0.007 [−0.008, −0.006] | <0.001 | −0.007 [−0.008, −0.006] | <0.001 |
| Generated[GPT] × heterogeneous | 0.014 [0.012, 0.015] | <0.001 | 0.014 [0.012, 0.015] | <0.001 | 0.005 [0.004, 0.006] | <0.001 | 0.005 [0.004, 0.006] | <0.001 |
| Generated[Llama] × heterogeneous | 0.013 [0.012, 0.014] | <0.001 | 0.013 [0.012, 0.015] | <0.001 | 0.005 [0.004, 0.005] | <0.001 | 0.005 [0.004, 0.005] | <0.001 |
| Generated[DeepSeek] × heterogeneous | 0.015 [0.013, 0.016] | <0.001 | 0.015 [0.013, 0.016] | <0.001 | 0.004 [0.004, 0.005] | <0.001 | 0.004 [0.004, 0.005] | <0.001 |
| Generated[GPT] × multi-source | 0.020 [0.019, 0.021] | <0.001 | 0.023 [0.022, 0.024] | <0.001 | −0.019 [−0.019, −0.018] | <0.001 | −0.019 [−0.019, −0.018] | <0.001 |
| Generated[Llama] × multi-source | 0.031 [0.030, 0.032] | <0.001 | 0.035 [0.034, 0.036] | <0.001 | −0.018 [−0.019, −0.018] | <0.001 | −0.018 [−0.018, −0.017] | <0.001 |
| Generated[DeepSeek] × multi-source | 0.025 [0.024, 0.026] | <0.001 | 0.037 [0.036, 0.037] | <0.001 | −0.014 [−0.015, −0.014] | <0.001 | −0.014 [−0.014, −0.013] | <0.001 |
| Generated[GPT] × 25% | 0.010 [0.009, 0.010] | <0.001 | 0.010 [0.009, 0.011] | <0.001 | 0.000 [−0.001, 0.000] | 0.224 | 0.000 [−0.001, 0.000] | 0.239 |
| Generated[Llama] × 25% | 0.008 [0.007, 0.009] | <0.001 | 0.006 [0.006, 0.007] | <0.001 | 0.000 [−0.001, 0.000] | 0.020 | −0.001 [−0.001, 0.000] | 0.004 |
| Generated[DeepSeek] × 25% | 0.014 [0.013, 0.015] | <0.001 | 0.010 [0.009, 0.011] | <0.001 | 0.004 [0.004, 0.005] | <0.001 | 0.004 [0.003, 0.004] | <0.001 |
| Generated[GPT] × 50% | 0.033 [0.032, 0.033] | <0.001 | 0.033 [0.032, 0.033] | <0.001 | 0.002 [0.001, 0.002] | <0.001 | 0.002 [0.001, 0.002] | <0.001 |
| Generated[Llama] × 50% | 0.027 [0.026, 0.028] | <0.001 | 0.026 [0.025, 0.027] | <0.001 | 0.000 [−0.001, 0.000] | 0.158 | 0.000 [−0.001, 0.000] | 0.100 |
| Generated[DeepSeek] × 50% | 0.039 [0.038, 0.040] | <0.001 | 0.037 [0.036, 0.038] | <0.001 | 0.003 [0.003, 0.004] | <0.001 | 0.003 [0.002, 0.004] | <0.001 |
| Word count (z) | 0.006 [0.005, 0.008] | <0.001 | – | | 0.000 [−0.001, 0.000] | 0.641 | – | |
| Named entities (z) | 0.015 [0.013, 0.016] | <0.001 | – | | 0.002 [0.001, 0.002] | <0.001 | – | |
| Sentiment (z) | −0.011 [−0.013, −0.009] | <0.001 | – | | −0.002 [−0.002, −0.001] | <0.001 | – | |
| Numbers (z) | 0.007 [0.004, 0.009] | <0.001 | – | | 0.001 [0.000, 0.002] | <0.001 | – | |
| Observations / topic clusters | 1,455,960 / 9,267 | | 1,455,960 / 9,267 | | 1,447,215 / 9,267 | | 1,447,215 / 9,267 | |
| $R^2$ marginal / conditional | 0.034 / 0.705 | | 0.011 / 0.695 | | 0.021 / 0.724 | | 0.020 / 0.722 | |

**Extended Data Table 1.** Pooled article-level models with and without adjustment for article characteristics. Coefficients from pooled linear mixed models of each article's mean within-topic cosine similarity ($\mu$) and the standard deviation of its within-topic similarities ($\sigma$), pooling all LLMs and conditions of the benchmark task

specification (main measure; random intercept for topic). The adjusted models include word count, named entities, sentiment and numbers (*z*-standardized), as in Fig. 3, while the unadjusted models leave them out (–). Because AI rewriting itself changes these characteristics, the adjusted coefficients compare AI-generated and human-written articles with similar characteristics, whereas the unadjusted coefficients give the overall difference between them. Neither is the change in within-topic similarity of the simulated world, which is shown in Fig. 2. Reference categories: human-written article, homogeneous environment, single-source imitation and 10% AI prevalence. Each cell gives the estimate with its 95% confidence interval below it, and the *P* value. Intervals and *P* values use CR2 standard errors clustered by topic (9,267 clusters) with a normal reference distribution.

| Environment | Imitation | LLM | Human–human | AI–own source | AI–other human | AI–AI | AI–AI, same source | Human–human, baseline world |
|---|---|---|---|---|---|---|---|---|
| Homogeneous | Single-source | DeepSeek-V3.2 | 0.769 | 0.885 | 0.721 | 0.773 | 0.915 | 0.774 |
| Homogeneous | Single-source | GPT-4o | 0.769 | 0.914 | 0.731 | 0.766 | 0.930 | 0.774 |
| Homogeneous | Single-source | Llama-4-Scout | 0.770 | 0.915 | 0.737 | 0.765 | 0.926 | 0.774 |
| Homogeneous | Multi-source | DeepSeek-V3.2 | 0.769 | 0.846 | 0.743 | 0.841 | 0.882 | 0.774 |
| Homogeneous | Multi-source | GPT-4o | 0.769 | 0.847 | 0.735 | 0.830 | 0.878 | 0.774 |
| Homogeneous | Multi-source | Llama-4-Scout | 0.769 | 0.874 | 0.754 | 0.826 | 0.879 | 0.774 |
| Heterogeneous | Single-source | DeepSeek-V3.2 | 0.737 | 0.878 | 0.689 | 0.748 | 0.910 | 0.741 |
| Heterogeneous | Single-source | GPT-4o | 0.736 | 0.909 | 0.695 | 0.737 | 0.926 | 0.741 |
| Heterogeneous | Single-source | Llama-4-Scout | 0.737 | 0.909 | 0.702 | 0.736 | 0.924 | 0.741 |
| Heterogeneous | Multi-source | DeepSeek-V3.2 | 0.733 | 0.831 | 0.709 | 0.819 | 0.871 | 0.741 |
| Heterogeneous | Multi-source | GPT-4o | 0.733 | 0.831 | 0.696 | 0.804 | 0.866 | 0.741 |
| Heterogeneous | Multi-source | Llama-4-Scout | 0.733 | 0.857 | 0.712 | 0.793 | 0.864 | 0.741 |

**Extended Data Table 2.** Similarity by type of article pair under the benchmark task specification. Mean cosine similarity within topics by pair type, at 50% prevalence, per LLM, environment and imitation strategy: human-human; AI-generated article and the article it rewrote (own source); AI-generated article and the other human-written articles; AI-AI (all); AI-AI pairs sharing a source; and human-human in the human baseline world. AI-generated articles are less similar to the human-written articles they did not rewrite than human-written articles are to each other in every condition, while remaining close to their own source. Human-human means in the imitation worlds can sit below the baseline value because the set of topics that keep more human pairs changes with replacement, i.e., numbers should be compared within the same world. Means are pooled over all article pairs within topics (each pair counts once), so their level differs from the topic-weighted means of Fig. 2. The corresponding values at 10% and 25% prevalence are reported in Supplementary Table 7.

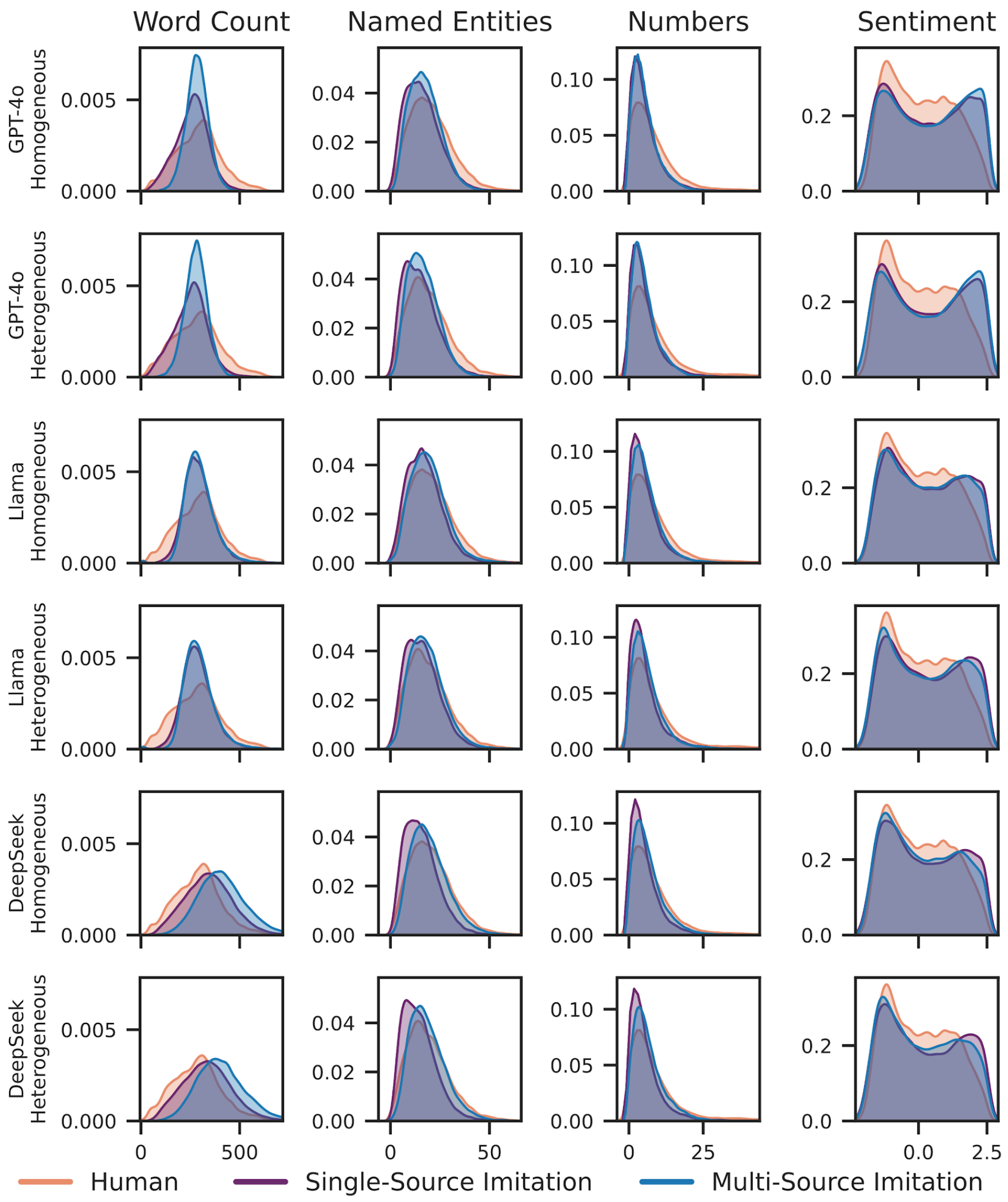

**Extended Data Fig. 2.** AI-generated articles differ systematically from human-written articles and generally exhibit more concentrated feature distributions and more affectively extreme sentiment distributions. Kernel density estimates (KDEs) show pooled distributions of word count, named entity count, number count, and sentiment for the human baseline ("Human", orange) versus AI-generated articles produced via single-source imitation (purple) and multi-source imitation (blue) across LLM conditions (Llama-4-Scout-17B-16E-Instruct, GPT-4o, DeepSeek-V3.2). 'Human' shows all human-written articles (104,761 homogeneous; 51,005 heterogeneous), and AI-generated articles are those in the 50% prevalence worlds, about half as many as human articles in each density. For each LLM, the top row shows homogeneous environments, and the bottom row shows heterogeneous environments. All plots use the benchmark task specification.

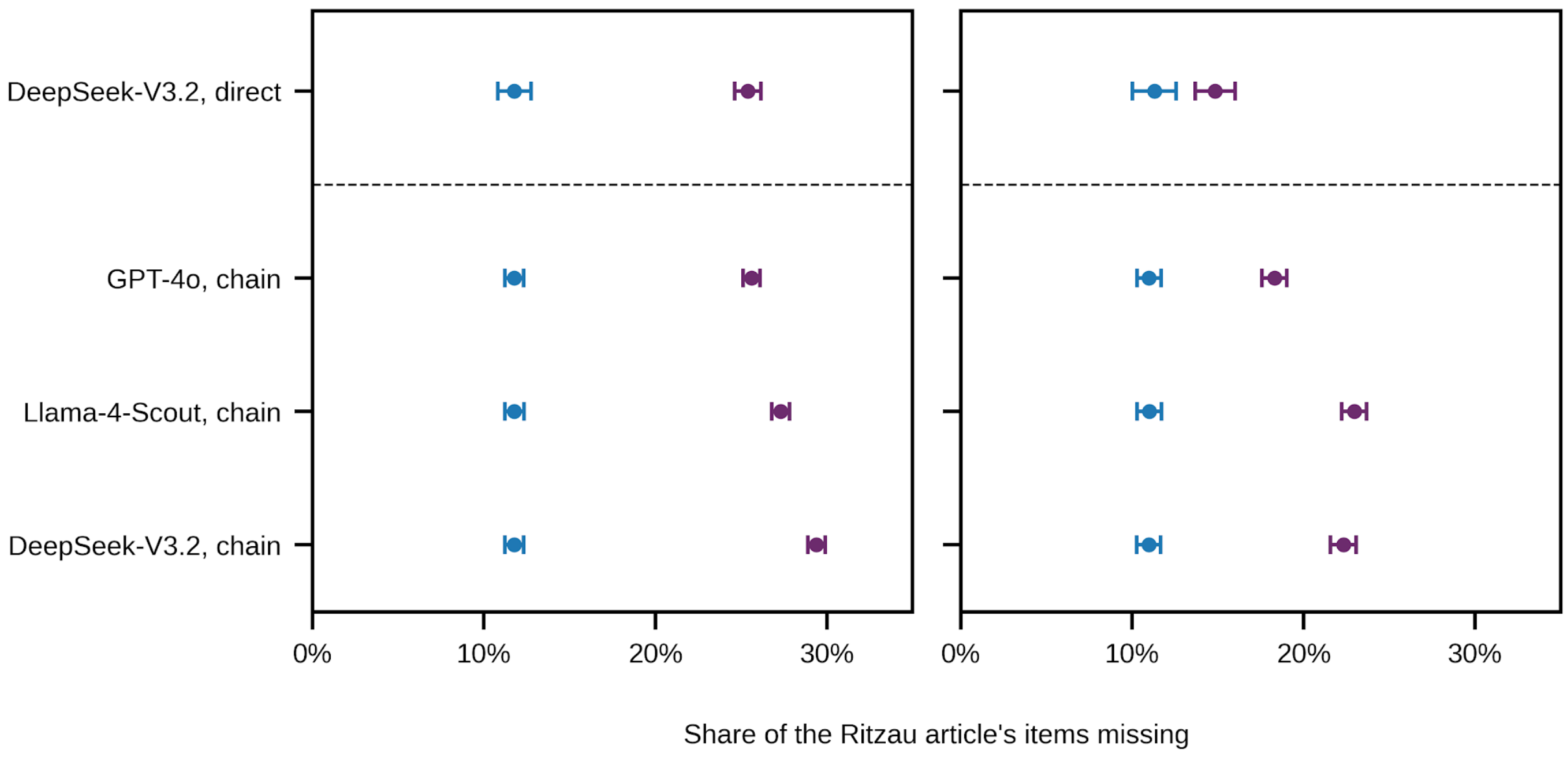

**Extended Data Fig. 3.** Given the same news agency text, AI rewriting omits more than twice as many of its named entities as journalists do. Share of the Ritzau article's named entities (a) and numbers (b) missing after the journalist's adaptation (blue) and after AI imitation (purple), each with its 95% interval (across articles for the direct test, clustered by Ritzau article for the chains). "Direct" (above the dashed line): DeepSeek-V3.2 rewrites the same Ritzau article for the same target outlet as the journalist (1,700 randomly sampled Ritzau articles, 1,593 containing at least one number). Entities: 11.8% vs 25.4%, difference +13.6 percentage points (95% interval 12.4 to 14.8 across articles). Numbers: 11.3% vs 14.8%, +3.6 points (2.0 to 5.1). "Chain" (below the line): each LLM rewrites the journalist's adaptation (Ritzau → journalist → AI; about 19,500 adaptations per LLM), and the share of the agency article's entities absent after the further AI rewriting step is compared with the share absent after the journalist's adaptation alone.

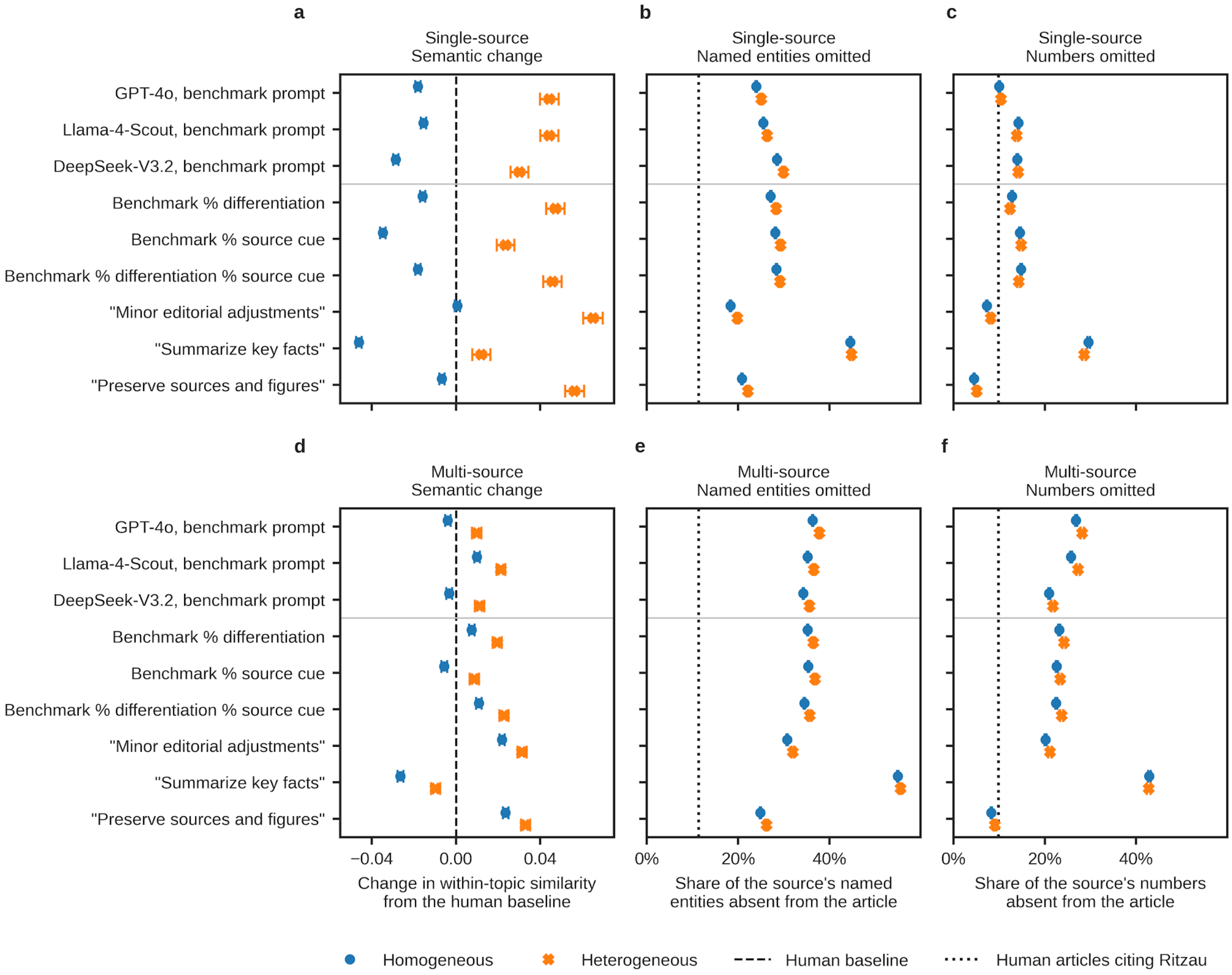

**Extended Data Fig. 4.** Semantic change and omission of source information, condition by condition. a–c, single-source imitation; d–f, multi-source imitation. a, d, Change in within-topic similarity from the all-human baseline (negative = more diverse; 95% intervals across topics; dashed line, human baseline, i.e. no change). b, e (named entities) and c, f (numbers), mean share of the source article's unique named entities and numbers absent from the AI-generated article (95% intervals across generated articles, not clustered by topic and mostly narrower than the markers; dotted lines, human articles citing the news agency Ritzau, measured against the Ritzau article: entities 11.4%, numbers 9.9%; 40,570 articles). Within each panel, rows are conditions at 50% prevalence: the three LLMs under the benchmark task specification (above the grey line; Fig. 2 worlds) and DeepSeek-V3.2 under the six further task specifications of Fig. 5 (below the line; same labels). Blue circles, homogeneous environment; orange crosses, heterogeneous environment. Within each environment and imitation strategy, a condition's change in similarity is negatively correlated with the share of its sources' named entities and numbers it omits (Spearman's $\rho = -0.83$ to $-0.88$ for entities, $P = 0.002$ to $0.005$, and $-0.67$ to $-0.82$ for numbers, $P = 0.007$ to $0.050$; 9 conditions each). Thus, the more a condition lowers similarity, the more source information it omits.

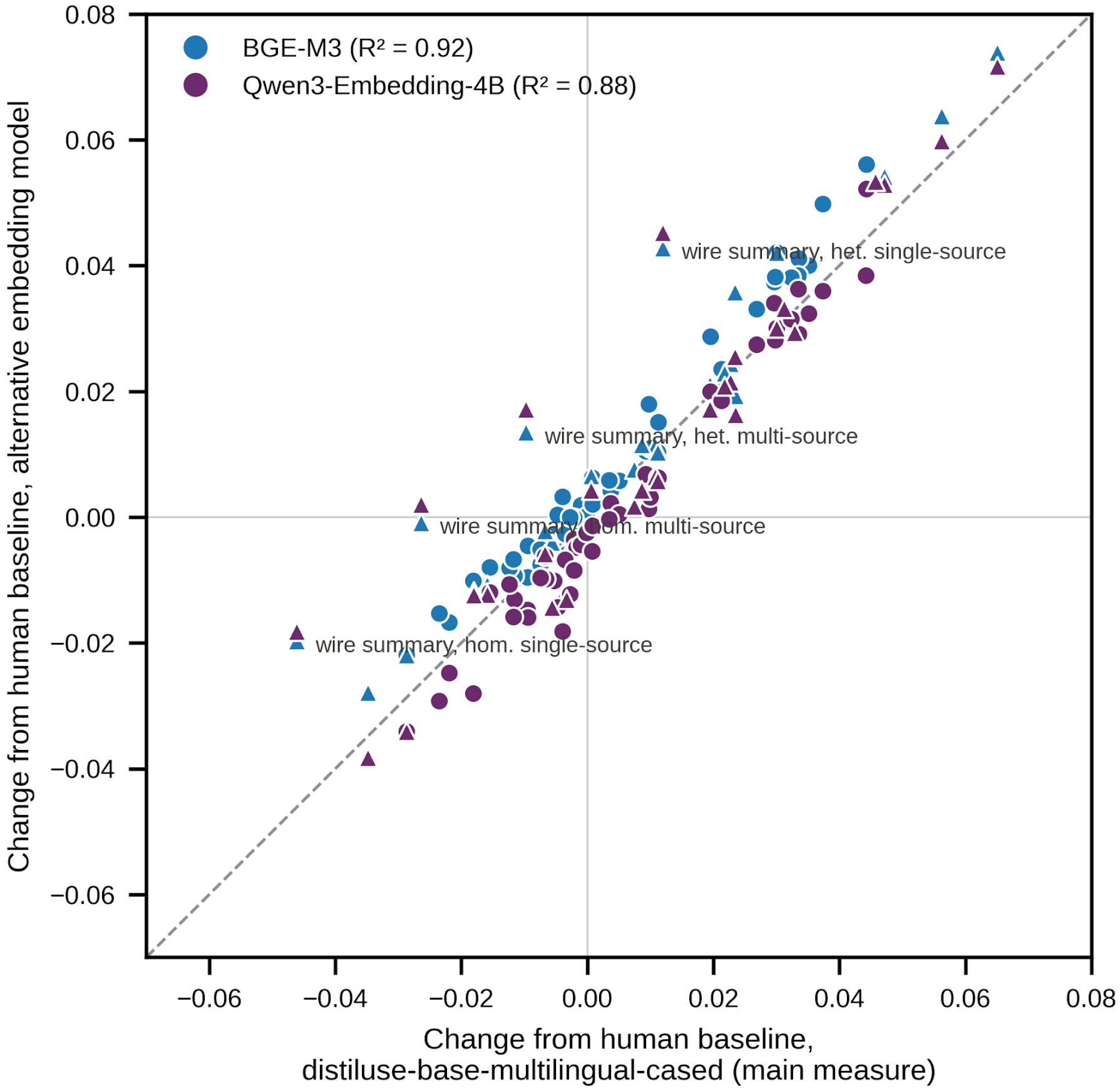

**Extended Data Fig. 5.** Condition-level results agree across three multilingual embedding models. Change in within-topic similarity from the all-human baseline for each simulated condition under the main measure (x) and under BGE-M3 (blue) or Qwen3-Embedding-4B (purple) (y). Circles, the 48 Fig. 2 conditions (three LLMs and the mixed-LLM condition); triangles, the Fig. 5 task-specification conditions at 50% prevalence (28); dashed, identity line. $R^2$ in the legend: share of variance in each model's condition-level changes explained by the main measure's (76 conditions; BGE-M3 0.92, Qwen3 0.88). Labelled points are the wire-summary conditions, where the models diverge most. Other disagreements (temperature 1.2, shared-topic GPT-4o and Llama under BGE-M3, near-zero multi-source cells) are listed in Supplementary Text 10.

## References

1. Rahwan, I. *et al.* Machine behaviour. *Nature* **568**, 477–486 (2019).
2. Akata, E. *et al.* Playing repeated games with large language models. *Nat. Hum. Behav.* **9**, 1380–1390 (2025).
3. Schroeder, D. T. *et al.* How malicious AI swarms can threaten democracy. *Science* **391**, 354–357 (2026).
4. Summerfield, C. *et al.* The impact of advanced AI systems on democracy. *Nat. Hum. Behav.* **9**, 2420–2430 (2025).
5. Sourati, Z. *et al.* The shrinking landscape of linguistic diversity in the age of large language models. *Nat. Hum. Behav.* **10**, 1925–1942 (2026).
6. DiMaggio, P. J. & Powell, W. W. The Iron Cage Revisited: Institutional Isomorphism and Collective Rationality in Organizational Fields. *Am. Sociol. Rev.* **48**, 147–160 (1983).
7. Posen, H. E., Lee, J. & Yi, S. The power of imperfect imitation. *Strateg. Manag. J.* **34**, 149–164 (2013).
8. Posen, H. E. & Martignoni, D. Revisiting the imitation assumption: Why imitation may increase, rather than decrease, performance heterogeneity. *Strateg. Manag. J.* **39**, 1350–1369 (2018).
9. Holland, J. H. *Adaptation in Natural and Artificial Systems: An Introductory Analysis with Applications to Biology, Control, and Artificial Intelligence*. (MIT Press, Cambridge, Mass., 1975).
10. Watts, D. J., Rothschild, D. M. & Mobius, M. Measuring the news and its impact on democracy. *Proc. Natl. Acad. Sci.* **118**, e1912443118 (2021).
11. Johansen, E. B. & Baumann, O. Platform Reliance: How Usage of Platform Content Shapes Product Similarity. SSRN Scholarly Paper at https://papers.ssrn.com/abstract=5178352 (2025).

12. Johansen, E. B. & Baumann, O. Platform Bypass and Societal Change: Informal Collaboration in the Danish News Ecosystem. SSRN Scholarly Paper at https://doi.org/10.2139/ssrn.5160253 (2025).

13. Russell, J. *et al.* AI use in American newspapers is widespread, uneven, and rarely disclosed. in *Proceedings of the 64th Annual Meeting of the Association for Computational Linguistics (Volume 1: Long Papers)* (eds Liakata, M., Moreira, V. P., Zhang, J. & Jurgens, D.) 14554–14580 (Association for Computational Linguistics, San Diego, California, United States, 2026). doi:10.18653/v1/2026.acl-long.663.

14. Russell, S. J. & Norvig, P. *Artificial Intelligence: A Modern Approach*. (Pearson, Boston, 2022).

15. Lu, J. G., Song, L. L. & Zhang, L. D. Cultural tendencies in generative AI. *Nat. Hum. Behav.* **9**, 2360–2369 (2025).

16. Brinkmann, L. *et al.* Machine culture. *Nat. Hum. Behav.* **7**, 1855–1868 (2023).

17. McQuail, D. *Media Performance: Mass Communication and the Public Interest*. (Sage Publications, London Thousand Oaks New Delhi, 1999).

18. Burton, J. W. *et al.* How large language models can reshape collective intelligence. *Nat. Hum. Behav.* **8**, 1643–1655 (2024).

19. Tsvetkova, M., Yasseri, T., Pescetelli, N. & Werner, T. A new sociology of humans and machines. *Nat. Hum. Behav.* **8**, 1864–1876 (2024).

20. Rahwan, I., Shariff, A. & Bonnefon, J.-F. The science fiction science method. *Nature* **644**, 51–58 (2025).

21. Kleinberg, J. & Raghavan, M. Algorithmic monoculture and social welfare. *Proc. Natl. Acad. Sci.* **118**, e2018340118 (2021).

22. Bommasani, R., Creel, K. A., Kumar, A., Jurafsky, D. & Liang, P. Picking on the Same Person: Does Algorithmic Monoculture lead to Outcome Homogenization? in *Advances in Neural Information Processing Systems 35 (NeurIPS 2022)* (2022).

23. Wenger, E. & Kenett, Y. N. Large language models are homogeneously creative. *PNAS Nexus* **5**, pgag042 (2026).

24. Sourati, Z., S. Ziabari, A. & Dehghani, M. The homogenizing effect of large language models on human expression and thought. *Trends Cogn. Sci.* **30**, 805–816 (2026).

25. Abdurahman, S. *et al.* Perils and opportunities in using large language models in psychological research. *PNAS Nexus* **3**, pgae245 (2024).

26. Boelaert, J., Coavoux, S., Ollion, É., Petev, I. & Präg, P. Machine Bias. How Do Generative Language Models Answer Opinion Polls? *Sociol. Methods Res.* **54**, 1156–1196 (2025).

27. Le Mens, G., Kovács, B., Hannan, M. T. & Pros, G. Uncovering the semantics of concepts using GPT-4. *Proc. Natl. Acad. Sci.* **120**, e2309350120 (2023).

28. Doshi, A. R. & Hauser, O. P. Generative AI enhances individual creativity but reduces the collective diversity of novel content. *Sci. Adv.* **10**, eadn5290 (2024).

29. Chen, Z. & Chan, J. Large Language Model in Creative Work: The Role of Collaboration Modality and User Expertise. *Manag. Sci.* **70**, 9101–9117 (2024).

30. Bisbee, J., Clinton, J. D., Dorff, C., Kenkel, B. & Larson, J. M. Synthetic Replacements for Human Survey Data? The Perils of Large Language Models. *Polit. Anal.* **32**, 401–416 (2024).

31. Mei, Q., Xie, Y., Yuan, W. & Jackson, M. O. A Turing test of whether AI chatbots are behaviorally similar to humans. *Proc. Natl. Acad. Sci.* **121**, e2313925121 (2024).

32. Kozlowski, A. C. & Evans, J. Simulating Subjects: The Promise and Peril of Artificial Intelligence Stand-Ins for Social Agents and Interactions. *Sociol. Methods Res.* **54**, 1017–1073 (2025).

33. Wang, A., Morgenstern, J. & Dickerson, J. P. Large language models that replace human participants can harmfully misportray and flatten identity groups. *Nat. Mach. Intell.* **7**, 400–411 (2025).

34. Meincke, L., Nave, G. & Terwiesch, C. ChatGPT decreases idea diversity in brainstorming. *Nat. Hum. Behav.* **9**, 1107–1109 (2025).
35. Wang, D., Huang, D., Shen, H. & Uzzi, B. A large-scale comparison of divergent creativity in humans and large language models. *Nat. Hum. Behav.* **10**, 531–540 (2026).
36. Xu, W., Jojic, N., Rao, S., Brockett, C. & Dolan, B. Echoes in AI: Quantifying lack of plot diversity in LLM outputs. *Proc. Natl. Acad. Sci.* **122**, e2504966122 (2025).
37. Argyle, L. P. *et al.* Out of One, Many: Using Language Models to Simulate Human Samples. *Polit. Anal.* **31**, 337–351 (2023).
38. Albert, D. & Billinger, S. Reproducing and Extending Experiments in Behavioral Strategy with Large Language Models. Preprint at https://doi.org/10.48550/arXiv.2410.06932 (2024).
39. Cao, X. & Kosinski, M. Large language models and humans converge in judging public figures' personalities. *PNAS Nexus* **3**, pgae418 (2024).
40. Meng, J. AI emerges as the frontier in behavioral science. *Proc. Natl. Acad. Sci.* **121**, e2401336121 (2024).
41. Farrell, H., Gopnik, A., Shalizi, C. & Evans, J. Large AI models are cultural and social technologies. *Science* **387**, 1153–1156 (2025).
42. Zhang, Y. *et al.* LLM as a Mastermind: A Survey of Strategic Reasoning with Large Language Models. Preprint at https://doi.org/10.48550/arXiv.2404.01230 (2024).
43. Hubert, K. F., Awa, K. N. & Zabelina, D. L. The current state of artificial intelligence generative language models is more creative than humans on divergent thinking tasks. *Sci. Rep.* **14**, 3440 (2024).
44. Meincke, L., Mollick, E. R. & Terwiesch, C. Prompting Diverse Ideas: Increasing AI Idea Variance. SSRN Scholarly Paper at https://doi.org/10.2139/ssrn.4708466 (2024).
45. Sourati, J. & Evans, J. A. Accelerating science with human-aware artificial intelligence. *Nat. Hum. Behav.* **7**, 1682–1696 (2023).

46. Zhou, E. & Lee, D. Generative artificial intelligence, human creativity, and art. *PNAS Nexus* **3**, pgae052 (2024).
47. Kumar, S., Flint, A., Aiello, L. M. & Baronchelli, A. Failure of contextual invariance in large language models. Preprint at https://doi.org/10.48550/arXiv.2603.23485 (2026).
48. Sclar, M., Choi, Y., Tsvetkov, Y. & Suhr, A. Quantifying Language Models' Sensitivity to Spurious Features in Prompt Design or: How I Learned to Start Worrying about Prompt Formatting. in *International Conference on Learning Representations* (2024).
49. Lazer, D. M. J. *et al.* The science of fake news. *Science* **359**, 1094–1096 (2018).
50. Flint, A., Aiello, L. M., Pastor-Satorras, R. & Baronchelli, A. Group size effects and collective misalignment in LLM multi-agent systems. *Proc. Natl. Acad. Sci.* **123**, e2531697123 (2026).
51. Yakura, H. *et al.* Empirical evidence of Large Language Model's influence on human spoken communication. Preprint at https://doi.org/10.48550/arXiv.2409.01754 (2024).
52. Xing, X. *et al.* On the caveats of AI autophagy. *Nat. Mach. Intell.* **7**, 172–180 (2025).
53. Shumailov, I. *et al.* AI models collapse when trained on recursively generated data. *Nature* **631**, 755–759 (2024).
54. Glickman, M. & Sharot, T. How human–AI feedback loops alter human perceptual, emotional and social judgements. *Nat. Hum. Behav.* **9**, 345–359 (2024).
55. Grootendorst, M. BERTopic: Neural topic modeling with a class-based TF-IDF procedure. Preprint at https://doi.org/10.48550/arXiv.2203.05794 (2022).
56. Nudo, J. *et al.* Generative exaggeration in LLM social agents: Consistency, bias, and toxicity. *Online Soc. Netw. Media* **51**, 100344 (2026).
57. Xu, F. *et al.* Large language models for computational social science. *Nat. Hum. Behav.* **10**, 2088–2104 (2026).
58. Chen, J. *et al.* M3-Embedding: Multi-Linguality, Multi-Functionality, Multi-Granularity Text Embeddings Through Self-Knowledge Distillation. Preprint at https://doi.org/10.48550/arXiv.2402.03216 (2025).

59. Zhang, Y. *et al.* Qwen3 Embedding: Advancing Text Embedding and Reranking Through Foundation Models. Preprint at https://doi.org/10.48550/arXiv.2506.05176 (2025).

60. Rosnes, D., Starke, A. D. & Trattner, C. Shaping the Future of Content-based News Recommenders: Insights from Evaluating Feature-Specific Similarity Metrics. in *Proceedings of the 32nd ACM Conference on User Modeling, Adaptation and Personalization* 201–211 (Association for Computing Machinery, New York, NY, USA, 2024). doi:10.1145/3627043.3659560.

61. Reimers, N. & Gurevych, I. Sentence-BERT: Sentence Embeddings using Siamese BERT-Networks. Preprint at https://doi.org/10.48550/arXiv.1908.10084 (2019).

62. Starke, A. D., Øverhaug, S. & Trattner, C. Predicting Feature-based Similarity in the News Domain Using Human Judgments. in *INRA'21: 9th International Workshop on News Recommendation and Analytics* (Amsterdam, Netherlands, 2021).

**Supplementary Information**

# AI-driven imitation can increase semantic diversity in news while omitting source information

## Supplementary Text

### Supplementary Text 1. Data processing

**Data cleaning**. The proprietary dataset was obtained from Infomedia, a Danish media surveillance company that archives nearly all digital articles published by mainstream news organizations, agencies, and magazines in Denmark (*n* = 362 outlets). We purchased access to Infomedia's Media Research API, obtained written consent from all Danish news organizations in the sample (with only one media house declining consent), and retrieved all articles from the database for these organizations between January 1, 2022, and December 31, 2022. To prepare the dataset, we deleted within-publisher duplicates, keeping only the latest version, and removed HTML artifacts using BeautifulSoup4.

**Topic modeling**. We applied the topic modeling approach outlined in Johansen & Baumann[1], clustering articles into daily topics using BERTopic[2] with pre-trained multilingual embeddings (distiluse-base-multilingual-cased). Articles from each day were processed without a predefined number of topics, allowing BERTopic to infer optimal clusters automatically. The goal was to capture "news event"-level topics, such as "Borussia Dortmund loses narrowly to Bayern Munich in a thrilling Bundesliga match" and "Politicians condemn terrorist attack in European city," to support accurate and context-specific similarity estimation. As a check that topics group articles reporting the same event, we compared the named entities, extracted from the full article text, of randomly sampled pairs of articles from different outlets (one pair per topic, five draws). Pairs from the same topic share at least two named entities in 81% of cases in the homogeneous environment and 68% in the heterogeneous environment, compared with 3% and 2% for pairs from different topics published on the same day.

Our workflow involved the following steps:

- **Embedding**: Obtaining article embeddings using a sentence-transformer (distiluse-base-multilingual-cased).
- **Dimensionality reduction**: Using UMAP (n_components=5, n_neighbors=15, min_dist=0.0) to reduce dimensionality.
- **Clustering**: Applying HDBSCAN (min_samples=5) to cluster the reduced embeddings.
- **Keyword refinement**: Implementing a CountVectorizer with c-TF-IDF, after removing Danish stopwords, to refine topic keywords.

To refine the dataset, we removed "junk" topics (articles that could not be assigned a cluster of a minimum of 5 articles), articles shorter than 15 words, and articles in the upper 90th percentile for word length to filter out potential corrupted data.

The final dataset contains 104,761 articles, clustered into 9,267 topics. Supplementary Fig. 1 displays the word count distribution for the dataset.

**Similarity measure.** To evaluate similarity among articles within the same topic, we computed cosine similarity scores on article embeddings derived from article text (merging its headline, byline, and body). The same sentence-transformer embedding model (distiluse-base-multilingual-cased) used for topic modeling was applied for consistency across analyses. Because the model reads at most 128 tokens and most of the distinct human-written texts are longer (median 523 tokens), we split each article into consecutive 126-token windows, embedded each window and represented the article by the length-weighted mean of its window embeddings, normalized to unit length (the main measure; Supplementary Text 10).

Within each topic cluster – defined by shared date and inferred topic – we calculated pairwise cosine similarity scores between all articles authored by different media outlets or AI agents, specific to a given world. For each article, this yielded a list of cosine similarity scores to all other original and AI-generated articles in the same topic and world (excluding self-comparisons). Because article composition varies across worlds (e.g., due to different prevalence levels of AI-generated articles and imitation strategies), each article's similarity list is world-specific.

To quantify within-topic similarity at the article level, we computed the average of each article's pairwise similarity scores (*AvgSim*). To quantify within-topic variability, we also calculated the standard deviation (*StdSim*) of these scores.

Finally, to capture system-level diversity, we first compute the mean pairwise cosine similarity within each date-specific topic in a given world, then take the unweighted mean of these topic-level values. Each topic therefore contributes equally to the world-level estimate, regardless of its number of articles. Lower values indicate greater semantic diversity in coverage of the same news events.

We check for our similarity measure's sensitivity to embedding choice and find a strong correlation in cosine similarity scores (Pearson's $r = 0.95$, $P < 0.0001$) between the main measure and embeddings obtained using OpenAI's API (see Supplementary Text 10 for more detail). All embedding-based results were also re-estimated with two long-context embedding models, BGE-M3 and Qwen3-Embedding-4B (Supplementary Text 10).

## Supplementary Text 2: Prompt templates

All AI-generated articles in the main analyses were produced using prompt templates designed to operationalize imitation under fixed rules. The templates specified (i) the article(s) to be imitated, (ii) the imitating outlet, (iii) the imitation strategy (single-source vs. multi-source), and (iv) the prompting strategy (zero-shot vs. chain-of-thought).

### *Benchmark prompts*

In prompts for the benchmark task specification, we use GPT-4o (gpt-4o-2024-08-06) accessed via OpenAI's API and the DeepSeek-V3.2 and Llama-4-Scout-17B-16E-Instruct models via the DeepInfra API. Across models, prompts were kept identical in structure and content, with substitutions only for the model identifier and API endpoint. Temperature settings were matched across models to ensure comparable levels of output randomness. CoT-prompting was used as a sensitivity check for GPT-4o only.

Below we document the exact templates used in the benchmark task specifications, which serve as the reference specification for all LLM generators used in the study.

original_article_id = unique hex code for article being substituted by AI-generated article
target_source_id = unique hex code for article being imitated
original_source = website domain of imitating firm
target_source = website domain of imitated firm
target_text = headline, byline, and bodytext of imitated article
gen_article = headline, byline, and bodytext of AI-generated article
temperature = set temperature in code (default = 0.7) (OpenAI's temperature parameter controls the degree of randomness in its outputs. Lower values make responses more deterministic and focused, while higher values increase variability and "creativity".)

**Zero-shot single-source imitation:** request_json = { "custom_id": f"task-{group_date}-{topic}-{original_article_id}", "method": "POST", "url": "/v1/chat/completions", "body": { "model": "gpt-4o-2024-08-06", "messages": [ {"role": "system", "content": '''You are a journalist that specializes in adapting content to align with your newspaper's unique style.'''}, { "role": "user", "content": f'''You will write a verbose, restructured, and expansive version of this article: "{target_text}" based on the articleid: "{target_source_id}". Write the article in Danish and in the usual style of my news website: {original_source}. Make sure that the new article length matches the word count of the original! You will output a JSON object containing the following information: {{ "articleid": "{original_article_id}", "original_source": "{original_source}", "target_source": "{target_source}", "target_source_id": "{target_source_id}", "gen_article": "", // Replace with your new version of the original article "temperature": {temperature} }} ''' } ], 'temperature': temperature } }

**Zero-shot multi-source imitation:** request_json = { "custom_id": f"task-{group_date}-{topic}-{original_article_id}", "method": "POST", "url": "/v1/chat/completions", "body": { "model": "gpt-4o-2024-08-06", "messages": [ {"role": "system", "content": '''You are a journalist that specializes in adapting content to align with your newspaper's unique style.'''}, { "role": "user", "content": f'''You will write a verbose, restructured, and expansive article that uses rewritten elements of these articles: "{target_texts[0]}" and "{target_texts[1]}" based on their article IDs: "{target_source_ids[0]}" and "{target_source_ids[1]}". Write the article in Danish and in the usual style of my news website: {original_source}. Make sure that the new article length matches the word count of the original article! You will output a JSON object containing the following information: {{ "articleid": "{original_article_id}", "original_source": "{original_source}", "target_sources": {target_sources}, "target_source_ids": {target_source_ids}, "gen_article": "", // Replace with your new version of the original article "temperature": {temperature} }} ''' } ], 'temperature': temperature } }

**Chain-of-thought single-source imitation:** request_json = { "custom_id": f"task-{group_date}-{topic}-{original_article_id}", "method": "POST", "url": "/v1/chat/completions", "body": { "model": "gpt-4o-2024-08-06", "messages": [ {"role": "system", "content": '''You are a journalist that specializes in adapting content to align with your newspaper's unique style.'''}, { "role": "user", "content": f'''First, use internal chain of thought reasoning to come up with a detailed plan for rewriting the following article in Danish to maintain its facts and length, but differentiate it in style and tone to match the typical writing language of my newspaper {original_source}. Do not output your plan and reasoning, keep your thoughts internal. Article Details: - Original Article ID: {original_article_id} - Original Source: {original_source} - Target Source: {target_source} - Target Source ID: {target_source_id} Article: "{target_text}" Next, you will rewrite the article based on that plan and output only a JSON object containing the following information: {{ "articleid": "{original_article_id}", "original_source": "{original_source}", "target_source": "{target_source}", "target_source_id": "{target_source_id}", "gen_article": "", // Replace with your new version of the original article "temperature": {temperature} }} ''' } ], 'temperature': temperature } }

**Chain-of-thought multi-source imitation:** { "role": "user", "content": f'''First, use internal chain of thought reasoning to come up with a detailed plan for rewriting the two following articles into a new Danish news article that maintains their facts and length, but is differentiated in style and tone to match the typical writing language of my newspaper {original_source}. Do not output your plan and reasoning, keep your thoughts internal. Article Details: - Original Article ID: {original_article_id} - Original Source: {original_source} - Target Article1 ID: "{target_source_ids[0]}" - Target Article2 ID: "{target_source_ids[1]}" - Target Source Article1: "{target_sources[0]}" - Target Source Article2: "{target_sources[1]}" Article1: "{target_texts[0]}" Article2: "{target_texts[1]}" Next, you will rewrite the article based on that plan and output only a JSON object containing the following information: {{ "articleid": "{original_article_id}", "original_source": "{original_source}", "target_sources": {target_sources}, "target_source_ids": {target_source_ids}, "gen_article": "", // Replace with your new version of the original article "temperature": {temperature} }} ''' } ], 'temperature': temperature } }

### *Ablation prompts*

To test whether the inclusion of specific elements in the benchmark task specification was driving the results reported in the manuscript, we conducted an ablation study of the benchmark task specification prompts, using DeepSeek-V3.2 via the DeepInfra API and zero-shot prompting techniques. In each ablation, we removed language related to (i) differentiation aims from the original article, (ii) cues about which source mobilized the AI agent, or (iii) both these aspects. We repeat the ablation study across information environment (heterogeneous or homogeneous) and imitation strategy (single- or multi-source).

Below we document the exact templates used in each ablation prompt.

### No explicit differentiation language

Zero-shot single-source imitation: request_json = { "custom_id": f"task-{group_date}-{topic}-{original_article_id}", "method": "POST", "url": "/v1/chat/completions", "body": { "model": "deepseek-ai/DeepSeek-V3.2", "messages": [ { "role": "system", "content": """You are a journalist writing an article for publication on a Danish news website.""" }, { "role": "user", "content": f"""You will write a new version of this article. Write the article in Danish for my news website: {original_source}. Make sure that the new article length matches the word count of the original! Article to use: "{target_text}" Base the output on the articleid: "{target_source_id}". You will output a JSON object containing the following information, replacing only gen_article with your new version of the original article: {{ "articleid": "{original_article_id}", "original_source":

"{original_source}", "target_source": "{target_source}", "target_source_id": "{target_source_id}", "gen_article": "", "temperature": {temperature} }} """ } ], "temperature": temperature } }

Zero-shot multi-source imitation: request_json = { "custom_id": f"task-{group_date}-{topic}-{original_article_id}", "method": "POST", "url": "/v1/chat/completions", "body": { "model": "deepseek-ai/DeepSeek-V3.2", "messages": [ { "role": "system", "content": """You are a journalist writing an article for publication on a Danish news website.""" }, { "role": "user", "content": f"""You will write a new article based on these articles. Write the article in Danish for my news website: {original_source}. Make sure that the new article length matches the word count of the original! Articles to use: "{target_texts[0]}" and "{target_texts[1]}" Base the output on their article IDs: "{target_source_ids[0]}" and "{target_source_ids[1]}". Return only a valid JSON object containing the following information, replacing only gen_article with your new Danish generated text: {{ "articleid": "{original_article_id}", "original_source": "{original_source}", "target_sources": {target_sources}, "target_source_ids": {target_source_ids}, "gen_article": "", "temperature": {temperature} }} """ } ], "temperature": temperature } }

**No imitating-outlet/source cue**

Zero-shot single-source imitation: request_json = { "custom_id": f"task-{group_date}-{topic}-{original_article_id}", "method": "POST", "url": "/v1/chat/completions", "body": { "model": "deepseek-ai/DeepSeek-V3.2", "messages": [ { "role": "system", "content": """You are a journalist that specializes in adapting content to align with your newspaper's unique style.""" }, { "role": "user", "content": f"""You will write a verbose, restructured, and expansive version of this article. Write the article in Danish. Make sure that the new article length matches the word count of the original! Article to use: "{target_text}" Base the output on the articleid: "{target_source_id}". You will output a JSON object containing the following information, replacing only gen_article with your new version of the original article: {{ "articleid": "{original_article_id}", "original_source": "{original_source}", "target_source": "{target_source}", "target_source_id": "{target_source_id}", "gen_article": "", "temperature": {temperature} }} """ } ], "temperature": temperature } }

Zero-shot multi-source imitation: request_json = { "custom_id": f"task-{group_date}-{topic}-{original_article_id}", "method": "POST", "url": "/v1/chat/completions", "body": { "model": "deepseek-ai/DeepSeek-V3.2", "messages": [ { "role": "system", "content": """You are a journalist that specializes in adapting content to align with your newspaper's unique style.""" }, { "role": "user", "content": f"""You will write a verbose, restructured, and expansive article that uses rewritten elements of these articles. Articles to use: "{target_texts[0]}" and "{target_texts[1]}" Base the output on their article IDs: "{target_source_ids[0]}" and "{target_source_ids[1]}". Write the article in Danish. Make sure that the new article length matches the word count of the original article! Return only a valid JSON object containing the following information, replacing only gen_article with your new Danish generated text: {{ "articleid": "{original_article_id}", "original_source": "{original_source}", "target_sources": {target_sources}, "target_source_ids": {target_source_ids}, "gen_article": "", "temperature": {temperature} }} """ } ], "temperature": temperature } }

**No explicit differentiation language and no imitating-outlet/source cue**

Zero-shot single-source imitation: request_json = { "custom_id": f"task-{group_date}-{topic}-{original_article_id}", "method": "POST", "url": "/v1/chat/completions", "body": { "model": "deepseek-ai/DeepSeek-V3.2", "messages": [ { "role": "system", "content": """You are a journalist writing an article for publication on a Danish news website.""" }, { "role": "user", "content": f"""You will write a new version of this article. Write the article in Danish. Make sure that the new article length matches the word count of the original! Article to use: "{target_text}" Base the output on the articleid: "{target_source_id}". You will output a JSON object containing the following information, replacing only gen_article with your new version of the original article: {{ "articleid": "{original_article_id}", "original_source": "{original_source}", "target_source": "{target_source}", "target_source_id": "{target_source_id}", "gen_article": "", "temperature": {temperature} }} """ } ], "temperature": temperature } }

Zero-shot multi-source imitation: request_json = { "custom_id": f"task-{group_date}-{topic}-{original_article_id}", "method": "POST", "url": "/v1/chat/completions", "body": { "model": "deepseek-ai/DeepSeek-V3.2", "messages": [ { "role": "system", "content": """You are a journalist writing an article for publication on a Danish news website.""" }, { "role": "user", "content": f"""You will write a new article that uses elements of these articles. Articles to use: "{target_texts[0]}" and "{target_texts[1]}" Base the output on their article IDs: "{target_source_ids[0]}" and "{target_source_ids[1]}". Write the article in Danish. Make sure that the new article length matches the word count of the original article! Return only a valid JSON object containing the following information, replacing only gen_article with your new Danish generated text: {{ "articleid": "{original_article_id}", "original_source": "{original_source}", "target_sources": {target_sources}, "target_source_ids": {target_source_ids}, "gen_article": "", "temperature": {temperature} }} """ } ], "temperature": temperature } }

***Alternative prompts***

While the benchmark task specifications represent likely imitation strategies by news outlets, alternative task specifications, and thus prompts, would likely occur in realized environments. Thus, we tested alternative prompts to understand whether different task specifications result in varying diversity outcomes, using DeepSeek-V3.2 via the DeepInfra API and zero-shot prompting techniques. The task specified in the alternative prompts requested either (i) do minor editorial adjustments to the original article as a cost-minimizing outlet, (ii) summarize key facts as a short wire news story, or (iii) preserve sources and facts as a high-quality outlet during rewriting. We repeat the use of alternative prompts across information environment (heterogeneous or homogeneous) and imitation strategy (single- or multi-source).

Below we document the exact templates used in each alternative prompt.

**Cost-minimizing outlet: minor editorial adjustments**

Zero-shot single-source imitation: request_json = { "custom_id": f"task-{group_date}-{topic}-{original_article_id}", "method": "POST", "url": "/v1/chat/completions", "body": { "model": "deepseek-ai/DeepSeek-V3.2", "messages": [ { "role": "system", "content": f"""You are a Danish newsroom editor at a cost-conscious local news outlet called "{original_source}". Your job is to do the minimal editorial work needed to make an article from another news outlet ready to publish on our own website, while preserving its substance and avoiding unnecessary rewriting.""" }, { "role": "user", "content": f"""Reproduce the article with only minor editorial adjust-ments. Do the minimum work needed to make it publishable for our own news outlet. Article to use: "{target_text}" Base the output on the articleid: "{target_source_id}". Write the article in Danish as publishable newsroom copy. Return only a valid JSON object. Do not include markdown, comments, explanatory text, or fields beyond the JSON object. The JSON object must use exactly this structure and metadata, replacing only gen_article with your Danish generated text: {{ "articleid": "{original_article_id}", "original_source": "{original_source}", "target_source": "{target_source}", "target_source_id": "{target_source_id}", "gen_article": "", "temperature": {temperature} }} """ } ], "temperature": temperature } }

Zero-shot multi-source imitation: request_json = { "custom_id": f"task-{group_date}-{topic}-{original_article_id}", "method": "POST", "url": "/v1/chat/completions", "body": { "model": "deepseek-ai/DeepSeek-V3.2", "messages": [ { "role": "system", "content": f"""You are a Danish newsroom editor at a cost-conscious local news outlet called "{original_source}". Your job is to do the minimal editorial work needed to make an article from another news outlet ready to publish on our own website, while preserving its substance and avoiding unnecessary rewriting.""" }, { "role": "user", "content": f"""Reproduce the articles with only minor editorial adjustments. Do the minimum work needed to make them publishable for our own news outlet. Articles to use: "{target_texts[0]}" and "{target_texts[1]}" Base the output on their article IDs: "{target_source_ids[0]}" and "{target_source_ids[1]}". Write the article in Danish as publishable newsroom copy. Return only a valid JSON object. Do not include markdown, comments, explanatory text, or fields beyond the JSON object. The JSON object must use exactly this structure and metadata, replacing only gen_article with your Danish generated text: {{ "articleid": "{original_article_id}", "original_source": "{original_source}", "target_sources": {target_sources}, "target_source_ids": {target_source_ids}, "gen_article": "", "temperature": {temperature} }} """ } ], "temperature": temperature } }

**Wire-service aggregator: approximately 200-word summary**

Zero-shot single-source imitation: request_json = { "custom_id": f"task-{group_date}-{topic}-{original_article_id}", "method": "POST", "url": "/v1/chat/completions", "body": { "model": "deepseek-ai/DeepSeek-V3.2", "messages": [ { "role": "system", "content": """You are a Danish wire-service editor tasked with quickly and efficiently summarizing the key facts of news articles clearly and neutrally for redistribution to other news publishers.""" }, { "role": "user", "content": f"""Summarize the key facts in about 200 words as neutral Danish wire-service copy suitable for pushing to other news publishers. Article to use: "{target_text}" Base the output on the articleid: "{target_source_id}". Write the summary in Danish as clear, neutral wire-service copy. Keep the output concise and factual. Return only a valid JSON object. Do not include markdown, comments, explanatory text, or fields beyond the JSON object. The JSON object must use exactly this structure and metadata, replacing only gen_article with your Danish generated text: {{ "articleid": "{original_article_id}", "original_source": "{original_source}", "target_source": "{target_source}", "target_source_id": "{target_source_id}", "gen_article": "", "temperature": {temperature} }} """ } ], "temperature": temperature } }

Zero-shot multi-source imitation: request_json = { "custom_id": f"task-{group_date}-{topic}-{original_article_id}", "method": "POST", "url": "/v1/chat/completions", "body": { "model": "deepseek-ai/DeepSeek-V3.2", "messages": [ { "role": "system", "content": """You are a Danish wire-service editor tasked with quickly and efficiently summarizing the key facts of news articles clearly and neutrally for redistribution to other news publishers.""" }, { "role": "user", "content": f"""Summarize the key facts from these articles in about 200 words as neutral Danish wire-service copy suitable for pushing to other news publishers. Articles to use: "{target_texts[0]}" and "{target_texts[1]}" Base the output on their article IDs: "{target_source_ids[0]}" and "{target_source_ids[1]}". Write the summary in Danish as clear, neutral wire-service copy. Keep the output concise and factual. Return only a valid JSON object. Do not include markdown, comments, explanatory text, or fields beyond the JSON object. The JSON object must use exactly this structure and metadata, replacing only gen_article with your Danish generated text: {{ "articleid": "{original_article_id}", "original_source": "{original_source}", "target_sources": {target_sources}, "target_source_ids": {target_source_ids}, "gen_article": "", "temperature": {temperature} }} """ } ], "temperature": temperature } }

**Quality-focused desk: preservation of sources and figures**

Zero-shot single-source imitation: request_json = { "custom_id": f"task-{group_date}-{topic}-{original_article_id}", "method": "POST", "url": "/v1/chat/completions", "body": { "model": "deepseek-ai/DeepSeek-V3.2", "messages": [ { "role": "system", "content": f"""You are a Danish editor working at a reputable high-quality news desk called "{original_source}". When reporting on other news stories, your desk is careful to preserve all named sources, figures, chronology, and factual claims.""" }, { "role": "user", "content": f"""Rewrite the article for publication while preserving all named sources, figures, chronology, and factual claims from the original. Article to use: "{target_text}" Base the output on the articleid: "{target_source_id}". Write the article in Danish as publishable newsroom copy. Return only a valid JSON object. Do not include markdown, comments, explanatory text, or fields beyond the JSON object. The JSON object must use exactly this structure and metadata, replacing only gen_article with your Danish generated text: {{ "articleid": "{original_article_id}", "original_source": "{original_source}", "target_source": "{target_source}", "target_source_id": "{target_source_id}", "gen_article": "", "temperature": {temperature} }} """ } ], "temperature": temperature } }

Zero-shot multi-source imitation: request_json = { "custom_id": f"task-{group_date}-{topic}-{original_article_id}", "method": "POST", "url": "/v1/chat/completions", "body": { "model": "deepseek-ai/DeepSeek-V3.2", "messages": [ { "role": "system", "content": f"""You are a Danish editor working at a reputable high-quality news desk called "{original_source}". When reporting on other news stories, your desk is careful to preserve all named sources, figures, chronology, and factual claims.""" }, { "role": "user", "content": f"""Rewrite the articles for publication while preserving all named sources, figures, chronology, and factual claims from the originals. Articles to use: "{target_texts[0]}" and "{target_texts[1]}" Base the output on their article IDs: "{target_source_ids[0]}" and "{target_source_ids[1]}". Write the article in Danish as publishable newsroom copy. Return only a valid JSON object. Do not include markdown, comments, explanatory text, or fields beyond the JSON object. The JSON object must use exactly this structure and metadata, replacing only gen_article with your Danish generated text: {{ "articleid": "{original_article_id}", "original_source": "{original_source}", "target_sources": {target_sources}, "target_source_ids": {target_source_ids}, "gen_article": "", "temperature": {temperature} }} """ } ], "temperature": temperature } }

## Supplementary Text 3: Pooled mixed-effects models on article-level cosine similarity means and variance

To assess how AI-generated imitation affects article similarity, we calculated each article's average pairwise cosine similarity (*AvgSim*) with all other articles in the same date-topic combination within each world. We also computed the sample standard deviation of these similarity scores *(StdSim)*, using denominator (n-1). *StdSim* was undefined when an article had fewer than two comparison articles. We use *AvgSim* and *StdSim* as dependent variables and include the following independent variables and interaction terms in our regression models.

### Independent Variables

- *Generated*: A four-level categorical variable indicating the generator of each article: human (reference category), GPT (for GPT-4o), Llama (for Llama-4-Scout-17B-16E-Instruct), and DeepSeek (for DeepSeek-V3.2). This variable allows us to estimate whether and how AI agents using different LLMs systematically differ from human-written articles in similarity outcomes. All LLMs use the benchmark task specification and a zero-shot approach.

- *Environment*: A categorical variable distinguishing between two information environments: homogeneous (news agency articles included in the information pool) and heterogeneous (news agency articles excluded). This distinction allows us to examine how the preexisting semantic diversity in the environment influences similarity patterns.

- *Imitation*: A categorical variable indicating the imitation strategy used: single-source imitation (one reference article) and multi-source imitation (two reference articles). This helps analyze how different imitation strategies affect similarity.

- *Percentage*: A categorical variable representing the proportion of AI-generated articles in each date-topic combination (10%, 25%, or 50%). This variable tests whether an increasing share of AI-generated articles influences similarity.

### Control Variables

The adjusted models also include the below four article-level features. Because AI rewriting changes these features, the adjusted coefficients compare articles with similar features and are not the total effect of imitation. Unadjusted models are reported alongside (Extended Data Table 1).

- *Word Count*: The total number of words in the article, which may influence cosine similarity (see Supplementary Text 5 for details on extraction of word counts).

- *Named Entities*: The number of named entities (people, organizations, locations) in an article (see Supplementary Text 5 for details on named entity recognition).

- *Sentiment*: A sentiment polarity measure ranging from −2 (highly negative) to 2 (highly positive), capturing emotional tone in the articles (see Supplementary Text 5 for details on sentiment estimation).

- *Number Count*: The number of numerical expressions (e.g., dates, figures), which may indicate fact-heavy content (see Supplementary Text 5 for details on extracting number count).

**Interaction terms**

- *Generated* × *Environment*: Tests whether the impact of AI-generated content on similarity differs between homogeneous and heterogeneous environments.

- *Generated* × *Imitation*: Examines whether single-source and multi-source imitation strategies influence similarity differently.

- *Generated* × *Percentage*: Assesses whether the proportion of AI-generated articles within a date-topic combination moderates similarity patterns.

We standardized our control variables (*Word Count, Named Entities, Sentiment, Number Count*) using pooled standardization. To account for the hierarchical data structure, we created a grouping variable *DateTopic* (a concatenation of publication date and topic), which served as a random effect in our model.

We then fitted pooled linear mixed-effects models with the following specifications:

$$AvgSim_{ij} = \beta_0 + \beta_M^T M_{ij} + \beta_W W_{ij} + \beta_I I_{ij} + \beta_P^T P_{ij} + \beta_{MW}^T MW_{ij} + \beta_{MI}^T MI_{ij} + \beta_{MP}^T MP_{ij} + \theta^T X_{ij} + u_j + \epsilon_{ij}$$

and

$$StdSim_{ij} = \beta_0 + \beta_M^T M_{ij} + \beta_W W_{ij} + \beta_I I_{ij} + \beta_P^T P_{ij} + \beta_{MW}^T MW_{ij} + \beta_{MI}^T MI_{ij} + \beta_{MP}^T MP_{ij} + \theta^T X_{ij} + u_j + \epsilon_{ij}$$

In these equations, $i$ indexes an article-world observation and $j$ a date-specific news topic (DateTopic). $M$ contains three indicators for GPT-4o, Llama, and DeepSeek, with human-written articles as the reference. $W$ indicates the heterogeneous environment (reference: homogeneous), $I$ indicates multi-source imitation reference: single-source, and $P$ contains indicators for 25% and 50% AI prevalence (reference: 10%). $MW$, $MI$, and $MP$ contain the Generated[Model] × Environment, Generated[Model] × Imitation, and Generated[Model] × Percentage interaction indicators, respectively. $X$ contains the four standardized article controls: word count, named-entity count, sentiment, and number count. The superscript $T$ denotes a vector transpose, so each coefficient vector is multiplied by its corresponding indicator or control vector. $u_j$ is the DateTopic random intercept, and $\epsilon_{ij}$ is the residual error.

Mixed-effects models assume normally distributed residuals and random effects. However, article-level *AvgSim* and *StdSim* observations were not fully independent because articles within a *DateTopic* shared underlying pairwise comparisons and because the same underlying articles recurred across simulated conditions. We include a *DateTopic* random intercept to model systematic between-topic heterogeneity and calculated cluster-robust standard errors clustered by *DateTopic* (CR2 cluster-robust covariance estimator) to accommodate arbitrary residual correlation and heteroskedasticity within each DateTopic. Inference uses Z-statistics referenced to the standard normal distribution (95% CI = estimate ± 1.96×SE), and was therefore based on the 9,267 *DateTopic* clusters rather than treating the millions of article-world observations as independent residual units.

Given the large sample size and the exploratory nature of the analyses, no formal correction for multiple comparisons was applied. These models estimate the interaction effects between the 4-level indicator *Generated* and categorical predictors while controlling for standardized features.

Extended Data Table 1 includes the results of our analysis.

**Robustness check for pooled mixed-effects models**

As a robustness analysis, we changed the unit of analysis to the news-topic-by-simulated-world level. Within every LLM-specific *DateTopic*, we calculated the difference between the mean outcome among generated articles and the mean outcome among remaining human articles. Each AI-minus-human difference constituted one regression observation. This prevents topics containing many articles from contributing disproportionate numbers of correlated observations. Because each *DateTopic* occurs in multiple simulated worlds, standard errors remained clustered by *DateTopic.* We estimate the following equations:

$$\Delta AvgSim_{jk} = \alpha_M^T M_{jk} + \gamma_{MW}^T MW_{jk} + \gamma_{MI}^T MI_{jk} + \gamma_{MP}^T MP_{jk} + u_j + \epsilon_{jk}$$

and

$$\Delta StdSim_{jk} = \alpha_M^T M_{jk} + \gamma_{MW}^T MW_{jk} + \gamma_{MI}^T MI_{jk} + \gamma_{MP}^T MP_{jk} + u_j + \epsilon_{jk}$$

Here $j$ indexes a DateTopic and $k$ a model-specific simulated world within that DateTopic. Each outcome is the mean among generated articles minus the mean among remaining human articles in that same world. $M$ contains indicators for GPT-4o, Llama, and DeepSeek, $MW$, $MI$, and $MP$ contain their respective interactions with environment, imitation strategy, and AI-prevalence level. The model has no fixed intercept, and the three coefficients in $\alpha_M$ directly estimate the model-specific generated-minus-human differences in the homogeneous, single-source, 10% reference condition. The superscript $T$ denotes a vector transpose, so each coefficient vector is multiplied by its corresponding indicator vector. $u_j$ is a DateTopic random intercept and $\epsilon_{jk}$ is the residual error.

The robustness analysis reproduced the main conclusions. In the homogeneous, single-source, 10% reference condition, generated-minus-human AvgSim differences were −0.0323 for GPT-4o (95% CI [−0.0333, −0.0312], $z = -61.43$), −0.0216 for Llama (95% CI [−0.0226, −0.0206], $z = -41.23$), and −0.0412 for DeepSeek (95% CI [−0.0423, −0.0401], $z = -75.73$; all $P < .001$). Cor-

responding StdSim differences were −0.0031 for GPT-4o (95% CI [−0.0037, −0.0025], $z = -10.36$, $P < .001$), −0.0008 for Llama (95% CI [−0.0014, −0.0002], $z = -2.66$, $P = .008$), and −0.0114 for DeepSeek (95% CI [−0.0119, −0.0108], $z = -38.26$, $P < .001$). Effects remained conditional on environment, imitation strategy, and AI prevalence.

Supplementary Table 5 shows the full results.

**Distribution and effect-size interpretation**

The cosine outcomes were bounded and not exactly normally distributed. Normality of the dependent variable is not required for the clustered large-sample coefficient inference, which is based on 9,267 independent DateTopic clusters. However, across the underlying pairwise cosine scores, skewness was −0.863 and excess kurtosis was −0.094. Article-level *AvgSim* was left-skewed (skewness −1.120), whereas article-level *StdSim* was right-skewed (skewness 1.026). We therefore report raw cosine-scale coefficients as the primary effect sizes, and residual-standard-deviation scaling is only reported as descriptive context.

**Pair-type decomposition**

The article-level contrast compares AI-generated and human-written articles within a simulated world, whereas Fig. 2 reports the change in the whole environment relative to the human baseline. To connect the two, we split each topic's change in mean pairwise similarity from its baseline into contributions from five types of article pairs: human-human, an AI-generated article and its own source, an AI-generated article and the other human-written articles, AI-AI pairs sharing a source, and other AI-AI pairs. Each contribution is the share of the topic's pairs of that type multiplied by the difference between their mean similarity and the topic's baseline mean. Averaged over topics, the contributions sum to the change in Fig. 2 (Supplementary Table 8, mean similarities by pair type are in Extended Data Table 2 and Supplementary Table 7). In the homogeneous environment under single-source imitation, pairs of AI-generated articles and other human-written articles account for roughly the whole reduction (for GPT-4o at 50% prevalence, −0.021 of a total change of −0.018), partly offset by the closeness of AI-generated articles to their own sources (+0.007). In the heterogeneous environment, the same pairs also lower similarity (−0.016 to −0.023 at 50% prevalence), but the closeness to the own source contributes more (+0.049 to +0.057), and therefore similarity rises. Under multi-source imitation, AI-AI pairs sharing a source add +0.007 to +0.010 at 50% prevalence, which is why multi-source imitation lowers similarity less than single-source imitation, or raises it.

**Supplementary Text 4: Within-world mixed-effects models on article-level cosine similarity**

To assess how AI-generated imitation affects article similarity within each world, we first computed each article's average pairwise cosine similarity (*AvgSim*) relative to all other articles within the same date-topic combination in that world. We then fitted a separate mixed-effects model within each world, using a generated-versus-human indicator and a DateTopic random intercept.

In these models, the outcome variable is *AvgSim* and the key predictor is *Generated[GPT]*, a binary indicator denoting whether an article was AI-generated by GPT-4o (1) or was written by a human (0). To account for unobserved heterogeneity in daily topic clusters, we included a random intercept for each date-topic combination. Because article-level outcomes within the same news topic share pairwise inputs, we calculated cluster-robust standard errors clustered by *DateTopic*. We also distinguish between zero-shot and chain-of-thought prompting in this analysis.

For each world, the model is specified as:

$$AvgSim_{ij} = \beta_0 + \beta_1 G_{ij} + u_j + \epsilon_{ij}$$

Here $i$ indexes an article in a particular simulated world and $j$ its DateTopic. $G_{ij}$ equals 1 when the article was generated by GPT-4o and 0 when it was wCritten by a human. The model is fitted separately within each environment × imitation strategy × prompting strategy × AI-prevalence world. $u_j$ is a DateTopic random intercept and $\epsilon_{ij}$ is the residual error. Standard errors are clustered by DateTopic.

To examine the within-world impact of AI-generated articles, we plotted the coefficient for *Generated[GPT]* = 1 in each world (Supplementary Fig. 2).

**Supplementary Text 5: Feature extraction and source information retention**

We extracted four article characteristics to examine textual changes accompanying AI-driven imitation: word count, numerical references, named entities, and sentiment.

- Word count was calculated by tokenizing the article text using simple whitespace splitting, counting only non-empty tokens as words.
- Numbers were identified using a regular expression to capture numeric tokens, which were transformed to yield the total count of numerical values in the text.
- Named entities were extracted using the Danish transformer model from spaCy (da_core_news_trf) to identify persons, organizations, and locations. We then counted the total named entities per article.
- Sentiment scores were computed using a multilingual transformer-based sentiment analysis model (agentlans/multilingual-e5-small-aligned-sentiment), which returns a score from −2 (highly negative) to 2 (highly positive).

Under the benchmark task specification, Extended Data Fig. 2 shows the feature distribution for AI-generated articles compared to human articles.

**Source information retention.**

To assess the degree to which AI-generated articles retained the named entities and numerical references present in the human-written article(s) they imitated, we compared, for each generated article, the set of named entities and the set of numbers it contained against the corresponding sets extracted from its source article(s).

- Omission and addition rates were computed over the set of distinct named entities (or numbers) appearing in an article, not the number of times each occurred. Repeated mentions of the same entity or number within an article were therefore counted once, and comparisons were based on set membership rather than token-level counts.
- Named-entity mentions (PER, ORG, and LOC spans returned by da_core_news_trf) were normalized by lowercasing and collapsing internal whitespace to a single space before comparison; two mentions were treated as the same entity only if their normalized strings were identical, with no fuzzy matching, alias resolution, or cross-reference to external knowledge bases. Numerical references were identified directly from the raw article text as digit sequences (regular expression \b\d+\b) and compared as exact digit strings.
- Under single-source imitation, the source reference for an article was the entity/number set extracted from the one human-written article the AI agent was instructed to imitate. Under multi-source imitation, the source reference was the union of the entity/number sets extracted from all human-written source articles supplied to the AI agent for that generation task.
- For each article comparison and each feature (entities, numbers), the omission (loss) rate was defined as the share of source items absent from the generated article, with the source set as denominator. The addition (gain) rate was defined as the share of generated-article items not present in the source, with the generated article's own set as denominator. Loss and gain rates therefore describe retention relative to the source and novelty relative to the generated article, respectively.
- When a source article's set was empty (e.g., an article with no detected named entities), the omission rate for that comparison was defined as 0, and when a generated article's set was empty, the addition rate was defined as 0.
- Aggregation across articles. Loss and gain rates were computed separately for named entities and for numbers at the article level and then aggregated within each model × environment × imitation-strategy condition, both as empirical cumulative distribution functions (Fig. 7 and Supplementary Fig. 7) and as arithmetic means across articles (reported in the Results).

**Retention of shared and source-specific references under multi-source imitation.**
For each multi-source AI-generated article in the benchmark worlds, we classified the named entities in its two source articles as shared, when they appear in both sources, or source-specific, when they appear in only one, using the same extraction and matching as above, and computed the share of each that appears in the generated article. Across LLMs and environments, generated articles retained 78–83% of shared entities and 42–47% of source-specific entities. For comparison, single-source rewrites retained 70–76% of the entities in their source.

We applied the same comparison to political blocs, using the classification of parties and politicians described in Methods (also see "Political mentions", Supplementary Text 5, below). A party was also counted as retained when the generated article contained the full name of a person who carried that party's tag in a source. We distinguish three situations in which the source material names both blocs. In the first, each of the two sources names politicians from only one bloc, and the blocs differ. In the second, the two sources together name both blocs, and at least one of them names both. In the third, a single source naming both blocs is rewritten under single-source imitation. The generated article retained both blocs in 83% of cases in the first situation (197 of 238 articles, pooled across LLMs and environments; 95% CI [77, 87]), in 86% in the second ($n$ = 6,373) and in 93% in the third ($n$ = 4,951). For comparison, when both sources named the same single bloc, the generated article retained that bloc in 98% of cases ($n$ = 8,678). When one bloc was omitted, the bloc that was retained differed between the three situations. In the first situation, the blue bloc was retained in 24 of the 36 such cases and the red bloc in 12, and the bloc named in the source listed first in the prompt was retained in 25. In the second and third situations, which had many more cases, the red bloc was retained more often (469 against 239 cases, and 187 against 119). Because the direction differs between situations, and the first situation includes few cases, we cannot reliably establish whether either bloc is omitted more often. The more frequent retention of the red bloc in the second and third situations may reflect the prominence of the governing Social Democrats in 2022 coverage. References that appear in both sources do not establish that the sources agree about the facts or their interpretation, and retaining a politician's name does not imply retaining that politician's position.

**Political mentions**

An article mentions a Danish party when it contains a party-letter tag after a name (e.g. "(S)"), the full name of a member of the Folketing in 2019–2026 (from Wikidata), or a written-out party name. Parties are grouped into the conventional Danish political blocs: red (S, SF, EL, RV, ALT, FG) and blue (V, K, DF, LA, NB, KD, DD); Moderaterne (M) counts as political but belongs to neither bloc. A topic has no political mentions when none of its articles mentions a party or politician, and is political when at least one-third of its articles do. Political topics are one-bloc when at least 80% of the articles naming a bloc name the same one, and both-bloc otherwise. Topics with some mentions below one-third (370 homogeneous, 212 heterogeneous) and topics naming only Moderaterne (8 and 6) are not analysed. In counts: no mentions 7,994 / 4,667; one bloc 587 / 294; both blocs 308 / 159 (homogeneous / heterogeneous). With 90% and 100% thresholds the heterogeneous both-minus-one difference is +0.025 to +0.030 (all intervals above zero). In retention analyses a bloc counts as kept when the generated article names any party of that bloc, by tag, party name or the full name of a politician tagged in a source.

**Source distance on political topics**

For multi-source imitation at 50% prevalence, we compared the source-relative shift (see "Topic- and source-level analyses", Methods) between political topics involving both blocs and those involving one bloc. In the heterogeneous environment, the two sources combined on both-bloc topics are less similar to each other than on one-bloc topics (mean cosine distance 0.237 against 0.189 on the main measure), and the shift is larger on both-bloc topics by 0.012 for GPT-4o, 0.008 for Llama and 0.020 for DeepSeek. We then removed the part of the shift predicted by source distance, using a linear fit of the shift on source distance within each LLM and environment. After this adjustment, the differences are 0.003, 0.001 and 0.009. Under BGE-M3 and Qwen3-Embedding-4B, they fall from 0.009–0.014 to −0.001–0.003. In the homogeneous environment, source distance hardly differs between the two topic types (0.176 against 0.174), and the shift differs by less than 0.002. These comparisons are descriptive, as topics involving both blocs may differ from one-bloc topics in respects other than source distance.

**Supplementary Text 6: Pooled mixed-effects models on article features**

To assess how AI-generated imitation affects article-level features, we fitted separate mixed-effects models for word count, number count, named entity count, and sentiment scores, using the same variable specifications as outlined in Supplementary Text 3.

"Generated" uses the human-written, unreplaced articles in each simulated world as the reference category. For these feature models, each human-written article contributes one observation, while generated articles contribute observations under their corresponding simulated conditions, which is also the reason for DateTopic-clustered inference (Supplementary Text 3). Groups entering Supplementary Tables 3–4 require at least two articles with complete feature values in that DateTopic-condition cell.

For each feature, the model was specified as:

$$F_{ij}^{(f)} = \beta_0 + \beta_M^T M_{ij} + \beta_W W_{ij} + \beta_I I_{ij} + \beta_P^T P_{ij} + u_j + \epsilon_{ij}$$

Here $i$ indexes an article observation, $j$ its DateTopic, and $F_{ij}^{(f)}$ the standardized value of feature $f$. Separate additive models are fitted for word count, named-entity count, number count, and sentiment. $M$ contains indicators for GPT-4o, Llama, and DeepSeek (reference: human), $W$ indicates the heterogeneous environment (reference: homogeneous), $I$ indicates multi-source imitation (reference: single-source), and $P$ contains indicators for 25% and 50% AI prevalence (reference: 10%). The superscript $T$ denotes a vector transpose. $u_j$ is the DateTopic random intercept and $\epsilon_{ij}$ is the residual error. Standard errors are clustered by DateTopic.

Supplementary Table 1 includes the results of our analysis.

**Supplementary Text 7: Mixed-effects models on within-topic feature variance**

To assess within-topic feature variance, we use the within-topic feature dispersion as a dependent variable. We used the article sample and human-reference construction described in Supplementary Text 6. We grouped observations by DateTopic, information environment, imitation strategy, AI prevalence and generator. For each group containing at least two articles with complete feature values, we calculated the sample standard deviation of word count, named-entity count, numerical-reference count and sentiment. Each SD outcome was standardized across eligible groups. We fitted separate additive linear mixed-effects models with generator, environment, imitation strategy and prevalence as fixed effects and a DateTopic random intercept. Inference used DateTopic-clustered standard errors (Supplementary Table 2).

The models were specified as:

$$S_{jc}^{(f)} = \beta_0 + \beta_M^T M_{jc} + \beta_W W_{jc} + \beta_I I_{jc} + \beta_P^T P_{jc} + u_j + \epsilon_{jc}$$

Here $j$ indexes a DateTopic and $c$ a cell defined by its environment, imitation strategy, AI-prevalence level, and generator. $S_{jc}^{(f)}$ is the standardized sample standard deviation of feature $f$ among articles in that cell. Separate additive models are fitted for the standard deviations of word count, named-entity count, number count, and sentiment. $M$ contains indicators for GPT-4o, Llama, and DeepSeek (reference: human), $W$ indicates the heterogeneous environment (reference: homogeneous), $I$ indicates multi-source imitation (reference: single-source), and $P$ contains indicators for 25% and 50% AI prevalence (reference: 10%). The superscript $T$ denotes a vector transpose. $u_j$ is the DateTopic random intercept and $\epsilon_{jc}$ is the residual error. Standard errors are clustered by DateTopic.

Across GPT-4o, Llama, and DeepSeek conditions, within-topic dispersion is lower for word count (GPT-4o: $\beta = -1.147$, 95% CI [−1.170, −1.124], $z = -97.02$, $P < .001$; Llama: $\beta = -0.944$, 95% CI [−0.969, −0.920], $z = -75.48$, $P < .001$; DeepSeek: $\beta = -0.336$, 95% CI [−0.360, −0.313], $z = -27.62$, $P < .001$), named entities (GPT-4o: $\beta = -1.443$, 95% CI [−1.467, −1.420], $z = -118.30$, $P < .001$; Llama: $\beta = -1.108$, 95% CI [−1.133, −1.084], $z = -87.52$, $P < .001$; DeepSeek: $\beta = -1.084$, 95% CI [−1.108, −1.060], $z = -88.08$, $P < .001$, numerical references (GPT-4o: $\beta = -0.784$, 95% CI [−0.806, −0.762], $z = -68.76$, $P < .001$; Llama: $\beta = -0.796$, 95% CI [−0.819, −0.772], $z = -67.13$, $P < .001$; DeepSeek: $\beta = -0.822$, 95% CI [−0.847, −0.797], $z = -63.43$, $P < .001$), and sentiment (GPT-4o: $\beta = -0.333$, 95% CI [−0.352, −0.315], $z = -34.87$, $P < .001$; Llama: $\beta = -0.285$, 95% CI [−0.304, −0.265], $z = -28.72$, $P < .001$; DeepSeek: $\beta = -0.259$, 95% CI [−0.278, −0.240], $z = -26.28$, $P < .001$) relative to the human-reference condition (Supplementary Table 2).

**Supplementary Text 8: Mixed-effects models on topic-level feature means**

To assess within-topic feature means, we used the article sample and human-reference construction described in Supplementary Text 6. We grouped observations by DateTopic, information environment, imitation strategy, AI prevalence and generator. For each group containing at least two articles with complete feature values, we calculated the mean of each feature and standardized these means across eligible groups. We fitted separate additive linear mixed-effects models with generator, environment, imitation strategy and prevalence as fixed effects and a DateTopic random intercept. Inference used DateTopic- clustered standard errors (Supplementary Table 3).

The models were specified as:

$$A_{jc}^{(f)} = \beta_0 + \beta_M^T M_{jc} + \beta_W W_{jc} + \beta_I I_{jc} + \beta_P^T P_{jc} + u_j + \epsilon_{jc}$$

Here $j$ indexes a DateTopic and $c$ a cell defined by its environment, imitation strategy, AI-prevalence level, and generator. $A_{jc}^{(f)}$ is the standardized mean of feature $f$ among articles in that cell. Separate additive models are fitted for mean word count, named-entity count, number count, and sentiment. $M$ contains indicators for GPT-4o, Llama, and DeepSeek (reference: human), $W$ indicates the heterogeneous environment (reference: homogeneous), $I$ indicates multi-source imitation (reference: single-source), and $P$ contains indicators for 25% and 50% AI prevalence (reference: 10%). The superscript $T$ denotes a vector transpose. $u_j$ is the DateTopic random intercept and $\epsilon_{jc}$ is the residual error. Standard errors are clustered by DateTopic.

Supplementary Table 3 shows the results of the analysis.

### Supplementary Text 9: Sensitivity check: LLM temperature setting

In our main analysis, we generated text with a default temperature setting of 0.7, which balances randomness in the model's outputs while maintaining coherence. However, temperature adjustments can influence text generation and, consequently, our results.

To assess the sensitivity of our results to temperature variations, we conducted a robustness check by comparing text generated at temperature 0.7 with text generated at temperatures 0.2, 0.5, and 1.2, using GPT-4o. This comparison focuses on a single world: single-source imitation with zero-shot prompting, allowing us to isolate the effect of temperature variation without introducing confounding factors from other world conditions.

Supplementary Fig. 3 displays that the overall impact of AI-driven imitation remains consistent across percentage levels between the temperature settings, although the diversifying effect is weakest at temperature 0.2 and strongest at 1.2. We also observe strong alignment in the feature distributions of AI-generated articles across the three lowest temperature settings. However, articles generated using temperature 1.2 tend to include fewer numbers and named entities, likely due to increased output randomness (Supplementary Fig. 4).

### Supplementary Text 10: Full-text embeddings and sensitivity to the embedding model

Because cosine similarity depends on how articles are represented, we examined how our estimates depend on the embedding. The main measure uses the sentence-transformer distiluse-base-multilingual-cased (revision d20bcb6), the model used for topic modeling. The model reads at most 128 tokens, and most articles are longer: 83,886 distinct human-written texts in the homogeneous environment exceed this limit (median 523 tokens; heterogeneous environment: 43,248 texts, median 510 tokens). We therefore tokenized each article's concatenated headline, byline and body once, split the tokens into consecutive, non-overlapping windows of 126 tokens (the limit without the model's two special tokens), encoded each window, normalized each window vector to unit length, and represented the article by the mean of its window vectors, weighted by the number of tokens in each window and normalized to unit length. Articles of 126 tokens or fewer receive the model's ordinary embedding. The pooling rule was fixed before any results were computed, and no alternative was tried. Topic assignments, environment membership and the simulated worlds were held fixed, so all representations are compared on the same articles and topics.

As robustness checks, we re-estimated all embedding-based results with two long-context models that encode each article in a single pass: BGE-M3 (dense output with CLS pooling; input limit 8,192 tokens, above the longest text of 4,140 tokens) and Qwen3-Embedding-4B (last-token pooling with one fixed instruction; configured limit 8,192 tokens). Article-level mean similarities under the main measure correlate at $r = 0.94$ with both models (1,878,034 article-world observations), per-topic changes from the baseline in the Fig. 2 worlds at $r = 0.95$ (BGE-M3) and $r = 0.94$ (Qwen3-Embedding-4B), and the condition-level changes across the 76 conditions of Figs 2 and 5 give $R^2 = 0.92$ and 0.88 (Extended Data Fig. 5). The main results are the same under all three representations. They differ mainly in the size of some effects and, in a few conditions, in sign.

The differences in results from choice of embedding model are as follows. (i) Wire-service summary at 50% prevalence: similarity falls on the main measure but changes little or rises under the long-context models (heterogeneous multi-source: −0.010 against +0.014 under BGE-M3 and +0.017 under Qwen3-Embedding-4B; homogeneous multi-source: −0.026 against −0.001 and +0.002). The four wire-service conditions are the only ones more than 0.02 from equality in Extended Data Fig. 5. (ii) Temperature 1.2 at 50% prevalence: the reduction is −0.104 on the main measure against −0.054 and −0.042. (iii) Topics shared by both environments: homogeneous single-source imitation at 50% by GPT-4o and Llama changes similarity by −0.002 (95% CI [−0.004, −0.0003]) and −0.001 ([−0.002, 0.001]) under BGE-M3, so only the Llama interval includes zero. (iv) Of the 48 Fig. 2 conditions, the sign of the change differs in five under BGE-M3 and three under Qwen3-Embedding-4B. All are multi-source conditions with changes below 0.005 on the main measure. (v) The GPT-4o contrast in the pooled article model (Extended Data Table 1) is −0.022 on the main measure, −0.014 under BGE-M3 and −0.039 under Qwen3-Embedding-4B. (vi) In the longitudinal simulation (Fig. 6), dispersion of similarity falls below the human baseline from t+2 in the 25% branch and at every round in the 50% branch on the main measure and under BGE-M3, whereas under Qwen3-Embedding-4B it stays above the baseline in the 10% and 25% branches and falls below it in the 50% branch only at t+3. The ordering by replacement rate and the decline over rounds hold under all three. (vii) Minimal editorial adjustment at 50% prevalence (homogeneous environment, single-source imitation): similarity changes by less than 0.005 on the main measure (+0.001, 95% CI [−0.001, 0.002]) but rises slightly under BGE-M3 (+0.007, 95% CI [0.005, 0.008]) and Qwen3-Embedding-4B (+0.004, 95% CI [0.003, 0.006]). Correlations between representations are supporting evidence, and the list above shows the cases where conclusions depend on the representation.

We compared the main measure with OpenAI's text-embedding-3-large in one simulated world (GPT-4o, homogeneous environment, single-source imitation, zero-shot prompting, 50% AI prevalence). Both representations cover the full article text. Within each topic, we averaged the cosine similarities of all article pairs under each representation, giving 9,267 paired topic means. These were strongly correlated (Pearson's $r = 0.954$, 95% CI [0.952, 0.956], $t(9{,}265) = 307.32$, $P < .001$; Spearman's $\rho = 0.934$; Supplementary Fig. 5a). Regressing the main measure on the text-embedding-3-large values gave an intercept of 0.02, a slope of 0.96 and $R^2 = 0.91$. text-embedding-3-large gave slightly higher similarity on average (0.833 against 0.820). The difference is largest for topics of intermediate similarity (about 0.02 to 0.03 between 0.78 and 0.92) and absent among the most similar topics. The distributions of the topic means are similar (Supplementary Fig. 5b). text-embedding-3-large places more topics above 0.9 (46% against 36%), and the main measure more between 0.75 and 0.9 (38% against 30%). This comparison covers one world, whereas the comparison with BGE-M3 and Qwen3-Embedding-4B above covers all conditions. Thus, it shows agreement in the configuration examined and does not establish that the representations are interchangeable. We use the main measure throughout because it applies the encoder used for topic modeling to the full text of every article.

### Supplementary Text 11. Sanity check: design and exact permutation testing

**Survey administration.** The survey was fielded online by Epinion between 6 and 16 June 2025. The final dataset includes 1,201 completed responses from members of Epinion's online panel (ages 18–92), yielding a completion rate of 55.6%. Quotas and post-stratification were used to approximate the adult Danish population by gender, age, and region.

**Survey design and stimuli.** Respondents were randomly assigned to one of five stimulus pairs. Each stimulus pair comprised two short snippets drawn from news articles from the same topic in a simulated world: one snippet from a human-written article and one snippet from an AI-generated imitation (GPT-4o). The five pairs were not drawn at random from this world's full set of article pairs, but were selected to span a range of embedding similarity (cosine similarity 0.38 to 0.88). Cosine similarity for each pair was calculated from the concatenated headline, byline, and body using the embedding procedure described in Supplementary Text 10, although the displayed survey snippets were separately truncated to their first 133 words. Participants were blind to the authorship behind the snippets throughout the tasks.

**Measures.** Respondents completed two tasks. First, they rated perceived similarity between the two snippets on a scale from 0 ("not at all similar") to 10 ("very similar"). Ratings were rescaled to [0,1] via division by 10. Second, respondents indicated which snippet they believed to be AI-generated by choosing among four options: "Article A" (ground truth: AI article), "Article B" (ground truth: human article), "Both," or "Neither". Only "Article A" is considered a correct choice.

**Aggregation.** Because embedding-based cosine similarity is constant within each stimulus pair, analyses were conducted at the pair level ($n = 5$) by computing the mean perceived similarity rating for each stimulus pair by respondents. Confidence intervals shown in Supplementary Fig. 9a were computed from respondent-level ratings within each pair.

**Exact permutation test.** We report two-sided exact p-values as the proportion of the 120 permuted test statistics with absolute values greater than or equal to the absolute value of the observed statistic. No additional continuity correction is applied, since the full permutation distribution is enumerated rather than sampled. This yields an exact p-value of 0.0083 (1 of 120 permutations at least as extreme as observed) for the Pearson correlation. Because the ranking of |correlation| and |slope| is identical across permutations of a fixed pairing, the pair-level OLS slope (observed slope = 0.330, scaled rating units per cosine unit) yields the same exact p-value, 0.0083.

**Leave-one-pair-out sensitivity diagnostic.** Due to only five stimulus pairs in this analysis, we examined whether the conclusion depended on any single pair. We recomputed the Pearson correlation five times, each time omitting one stimulus pair. The resulting correlations ranged from $r = 0.869$ to $r = 0.972$ and were positive in all five cases. This diagnostic indicates that the positive association across the available stimuli is not attributable solely to any single pair. However, this does not address the limited number of stimuli or establish generalizability beyond them.

**Interpretation.** Given that the association is estimated across only five stimulus pairs, we treat this analysis as a limited directional sanity check. It assesses whether embedding-based similarity and mean perceived textual similarity covary across these stimuli, but does not establish generalizability beyond these pairs or identify whether factual, stylistic, framing, or other textual differences account for either measure.

**Supplementary Text 12: Human null conditions**

To assess whether changes in semantic diversity could arise mechanically from the simulation's replacement procedure rather than from LLM generation, we constructed two procedural null models. Both retained the replacement masks, prevalence levels, and source-target assignments used in the corresponding DeepSeek-V3.2 simulations. In the single-source null, each article selected for replacement was substituted with the complete human-written article assigned as its imitation target. This preserves the replacement structure and target selection of the AI simulation while introducing no generative transformation. In the multi-source null, each replaced article was represented by the L2-normalized mean of the normalized embeddings of its two assigned human target articles. This embedding-space centroid represents a mechanical combination of the two source positions without generating synthetic text. Both null models were evaluated at 10%, 25%, and 50% replacement in homogeneous and heterogeneous environments using the same within-DateTopic cosine-similarity procedure as the corresponding AI-generated worlds. The null models therefore provide procedural benchmarks for the convergence that can arise from copying or averaging selected human sources independently of LLM generation. These benchmarks compare generative rewriting with direct copying and mechanical source averaging; they do not define or test a separate imitation-error process. For results, we refer to Fig. 5.

**Supplementary Text 13: Longitudinal simulation of repeated AI-driven imitation**

To examine whether the diversity effects observed after a single round of AI-driven imitation persist when generated articles remain in the simulated world, we conducted a longitudinal extension of the homogeneous-world simulation. The purpose of this analysis was to model repeated replacement and imitation, allowing AI-generated articles to become imitation targets in subsequent rounds. The analysis therefore represents simulated recursive replacement over discrete timesteps rather than observed change over calendar time.

The longitudinal simulation uses DeepSeek-V3.2, a homogeneous information environment, single-source imitation, zero-shot prompting, temperature 0.7, and the differentiation-oriented benchmark prompt used in the main analysis. We simulate three separate replacement branches corresponding to 10%, 25%, and 50% replacement per round. Each branch contains the same 104,761 article slots distributed across 9,267 topics as the original homogeneous information environment. The three branches are different experimental conditions (replacement rates), and are run as single simulated trajectories, i.e., not repeated simulations of the same condition.

At (t+0), all article slots contain the original human-written articles. At (t+1), the simulation corresponds to the original one-round DeepSeek-V3.2 worlds reported in Fig. 2. At (t+2), a new fraction of article slots within each branch is selected for replacement using articles available in the corresponding (t+1) world as potential imitation targets. The procedure is repeated at (t+3), using the corresponding (t+2) world as potential imitation targets. The three branches evolve independently. For example, the 25% (t+3) world descends exclusively from the 25% (t+2) world. Target and slot selection at each round used a deterministic, seed-based procedure, so the reported trajectory is a single reproducible realization of each branch rather than an average over independent statistical replications of the same design. The focal article was excluded from its own candidate target pool at every round.

Previously generated articles remain in the simulated world unless their slots are subsequently replaced, causing the share of AI-generated content to accumulate over time. Realized AI prevalence was 10.6%, 19.9%, and 28.4% across (t+1), (t+2), and (t+3) in the 10% branch. Corresponding prevalence was 24.8%, 43.2%, and 57.1% in the 25% branch, and 50.2%, 75.2%, and 87.5% in the 50% branch. By the third round in the 50% replacement branch, approximately half of all article slots contained second- or third-generation imitations, in which generated content had itself been rewritten (Supplementary Text 13).

Semantic similarity was estimated using the same cosine-similarity procedure as in the main analyses (Supplementary Text 10).

**Replacement and recursive target selection**

Within each topic, replacement slots and imitation targets were determined from the immediately preceding world in the same branch. Because generated articles remain part of later worlds, imitation targets could be either human-written or AI-generated. An article generated from a human target is defined as generation depth 1. An article generated by imitating a depth-1 AI article is depth 2, and an article generated from a depth-2 target is depth 3.

By (t+3), the 10% branch contained 27.9% depth-1 and 0.5% depth-2 articles, with no depth-3 articles. The 25% branch contained 41.7% depth-1, 15.2% depth-2, and 0.15% depth-3 articles. The 50% branch contained 37.1% depth-1, 37.7% depth-2, and 12.7% depth-3 articles.

**Limitations**

This longitudinal extension represents one simulated trajectory under a homogeneous environment, single-source imitation, DeepSeek-V3.2, and the benchmark task specification. The observed persistence of diversification should therefore not be interpreted as a general trajectory of repeated AI imitation across models, task specifications, or information environments.

### Supplementary Text 14: Descriptive comparison of homogeneous and heterogeneous environments

The heterogeneous environment was constructed by excluding articles identified as relying on content from the Danish national news agency, Ritzau. Because this exclusion could affect the composition of the article universe as well as baseline semantic similarity, and thus the simulations, we compared the human articles underlying the homogeneous and heterogeneous environments.

The exclusion reduced the information environment from 104,761 to 51,005 articles and the number of topics containing at least two articles from 9,267 to 5,338. All 5,338 usable topics in the heterogeneous environment were also represented in the homogeneous environment. We therefore conducted the principal balance comparison on these shared topics to hold topical and temporal support constant (see Supplementary Table 4A).

Within the shared topics, baseline within-topic cosine similarity was lower in the heterogeneous environment than in the homogeneous environment (0.762 versus 0.797). The heterogeneous environment also contained fewer articles per topic (mean = 9.36, SD = 6.07) than the homogeneous environment (mean = 11.98, SD = 6.19), as well as fewer outlets per topic (mean = 7.01,

SD = 3.36, versus mean = 8.60, SD = 2.81). These differences reflect the removal of news-agency-dependent content.

However, measured article characteristics were closely balanced. The mean article length was 285.43 words (SD = 117.42) in the homogeneous environment and 284.04 words (SD = 118.69) in the heterogeneous environment. Differences in named entities, unique named entities, numerical expressions, and sentiment were similarly small. Across these measures, all standardized mean differences were smaller than 0.04 in absolute magnitude (see Supplementary Table 4B).

The overall mixture of outlets was also highly similar: the total-variation distance between the marginal outlet distributions was 0.034, meaning that approximately 3.4% of article share would have to be redistributed among outlets to make the distributions identical. The exclusion had a clearer effect on the relative representation of the shared topics (total-variation distance = 0.187). Thus, although both environments contained the same 5,338 topics, some topics lost proportionally more articles than others. At the more granular topic-by-outlet level, the total-variation distance was 0.203, indicating that particular outlet-topic combinations were also affected unevenly (see Supplementary Table 4C).

As an additional robustness check, we replicated the analysis reported in Fig. 2 using only the 5,338 topics represented in both environments (Extended Data Fig. 1). The principal single-source result was unchanged. On the common topics, single-source imitation reduced similarity in the homogeneous environment and increased it in the heterogeneous environment. Multi-source imitation changed similarity by at most 0.005 at 10% and 25% prevalence (Llama at 25%: about +0.01) and increased it at 50% in both environments (+0.008 to +0.021). Thus, on the main measure, the direction of the single-source effect in each environment does not depend on the topics represented only in the homogeneous environment, whereas the size of the reduction in the homogeneous environment does (for differences under BGE-M3, see Supplementary Text 10).

The construction therefore did not alter baseline semantic similarity in complete isolation, but also reduced coverage and within-topic density and changed the relative representation of some topics. Nevertheless, within common topic support, lower baseline similarity was not accompanied by substantively important differences in the measured article characteristics or the overall mixture of outlets. Moreover, the principal context-dependent effect of AI imitation remained when both environments were restricted to identical topic support.

**Supplementary Text 15. Generator composition, replacement sets, and system-level dispersion**

To examine whether the system-level findings depend on using a single underlying LLM, we extended the simulations to include mixed-LLM worlds combining articles generated by GPT-4o, DeepSeek-V3.2, and Llama-4-Scout-17B-16E-Instruct. These worlds used the benchmark task specification and covered both information environments, both imitation strategies, and the three AI prevalence levels. Each article chosen for replacement was assigned to one of the three models. Assignment used a stable hash of a fixed seed, the environment, imitation strategy, AI-prevalence level, and base article identifier, mapping each replacement article slot to one of the three LLMs with approximately equal (1/3) probability. Fig. 2 reports the resulting system-level mean semantic similarities for both single-LLM and mixed-LLM worlds.

We additionally examined whether mixed-LLM worlds exhibited greater or lower dispersion than single-LLM worlds. Within each *DateTopic*, we calculated the standard deviation of unique unordered pairwise cosine similarities and the standard deviations of article word counts, named-entity counts, numerical-reference counts, and sentiment scores. These calculations included all articles present in each world, including retained human articles. We used population standard deviations (ddof = 0), consistent with the existing world-level similarity calculations. *DateTopics* required at least two articles and complete values for the relevant outcome.

We compared the corrected mixed- and single-LLM worlds separately for each information environment, imitation strategy, and AI prevalence level. For each comparison, we calculated mixed-minus-single differences in standard deviation within shared eligible *DateTopics* and averaged these differences with equal weight per *DateTopic*. Accordingly, the outcome measures average dispersion within shared news topics rather than dispersion pooled across unrelated topics. Each comparison included 9,267 *DateTopics* in the homogeneous environment or 5,338 in the heterogeneous environment. We obtained 95% paired t confidence intervals using the across-*DateTopic* standard error of the mean difference. Positive differences indicate greater dispersion in the mixed world, whereas negative differences indicate lower dispersion.

Mixing LLMs did not uniformly increase system-level dispersion (Supplementary Table 6). Similarity dispersion was not consistently higher in mixed worlds. On the main measure, relative to GPT-4o it was higher in 5 conditions, lower in 2 and unresolved in 5, Relative to Llama higher in 6, lower in 2 and unresolved in 4, and relative to DeepSeek higher in 2, lower in 7 and unresolved in 3. The per-model counts differ under the two long-context embedding models, while the finding that mixing does not uniformly increase dispersion holds under all three.Word-count dispersion was higher in mixed worlds than in GPT-4o and Llama worlds in all 12 point estimates. It was lower than in DeepSeek worlds in all 12 conditions. Named-entity dispersion was higher than in GPT-4o and Llama worlds in all point estimates, with intervals above zero in 10 conditions for each comparator, and lower than in DeepSeek worlds with intervals below zero in nine conditions. Numerical-reference and sentiment dispersion showed less consistent patterns across models and conditions.

Thus, these findings suggest that combining LLMs does not necessarily broaden the range of textual characteristics or similarity relationships within news topics. Mixed-LLM worlds can be more dispersed than those produced by one model and less dispersed than those produced by another. Generator composition therefore matters alongside the characteristics of the constituent models, rather than implying a simple relationship between the number of generators and system-level dispersion.

**Replacement sets across LLMs**

At 50% prevalence, all three LLMs replaced the same articles. At 10% and 25%, the articles to be replaced were drawn separately for DeepSeek-V3.2 (both imitation strategies) and for Llama-4-Scout under single-source imitation. These worlds therefore differ from the corresponding GPT-4o worlds both in the LLM and in which articles were replaced (overlap between replace-ment sets, measured as Jaccard similarity, about 0.12 at 10% and 0.33 at 25%). To separate the two, we rebuilt each 10% and 25% world with a common replacement set, using either the set drawn for GPT-4o or the set drawn for DeepSeek-V3.2. Because an LLM's rewrite of a given article is identical across prevalence levels, each replaced article was filled with that LLM's existing rewrite from its 50% world, and no new text was generated.

With common replacement sets, no single-source condition changes sign under any of the three embedding models. In the homogeneous environment, similarity remains below baseline and the reduction grows with prevalence. In the heterogeneous environment, it remains above baseline. Under both sets, DeepSeek-V3.2 shows the largest reduction in the homogeneous environment and the smallest increase in the heterogeneous environment. For any LLM and condition, the change from baseline moves by at most 0.005. Under multi-source imitation at 10% and 25% prevalence, where changes are close to zero, their sign can depend on which articles were replaced.

**Supplementary Figures and Tables**

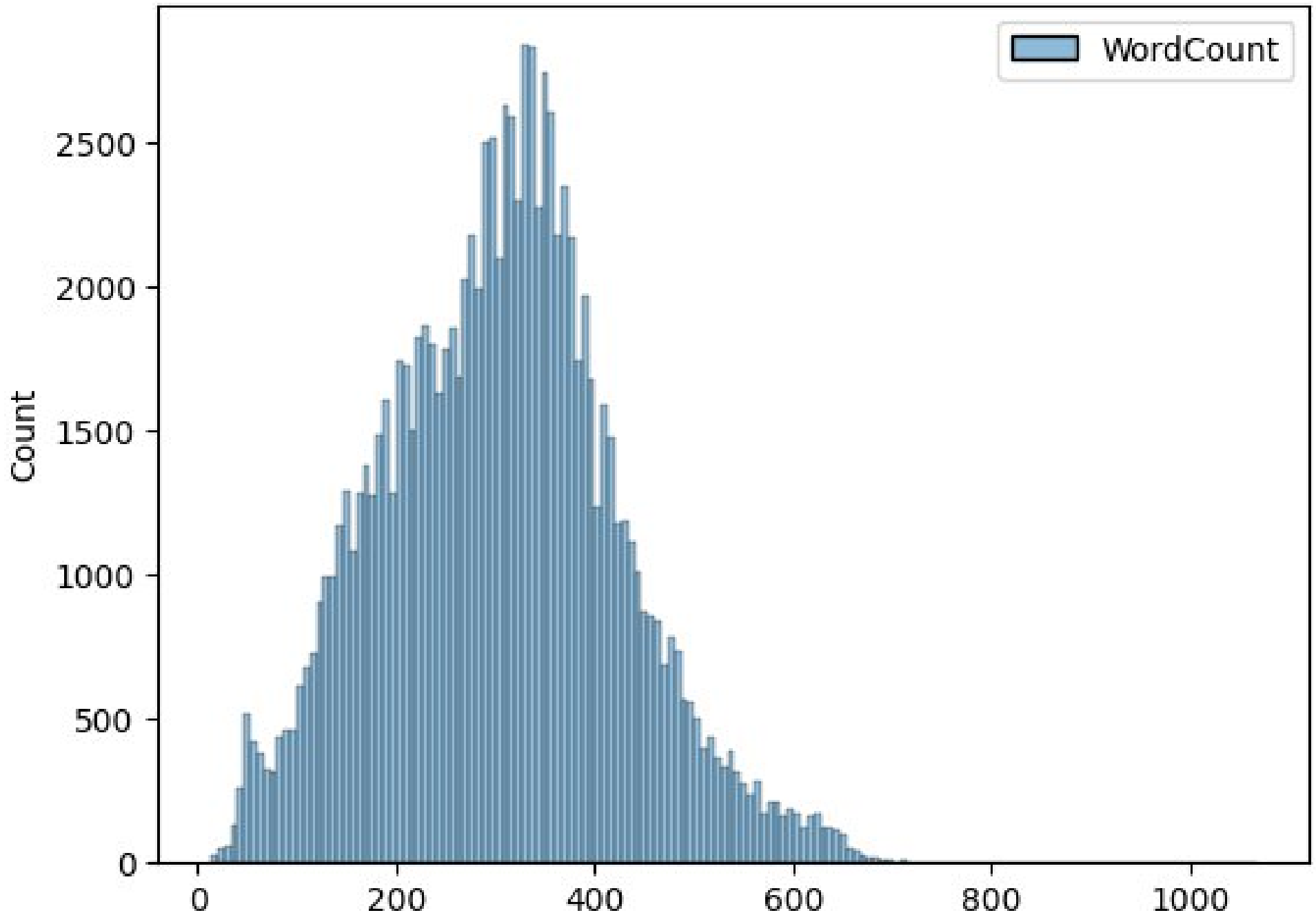

**Supplementary Figure 1.** Histogram of word counts in all original news articles ($n$ = 104,761).

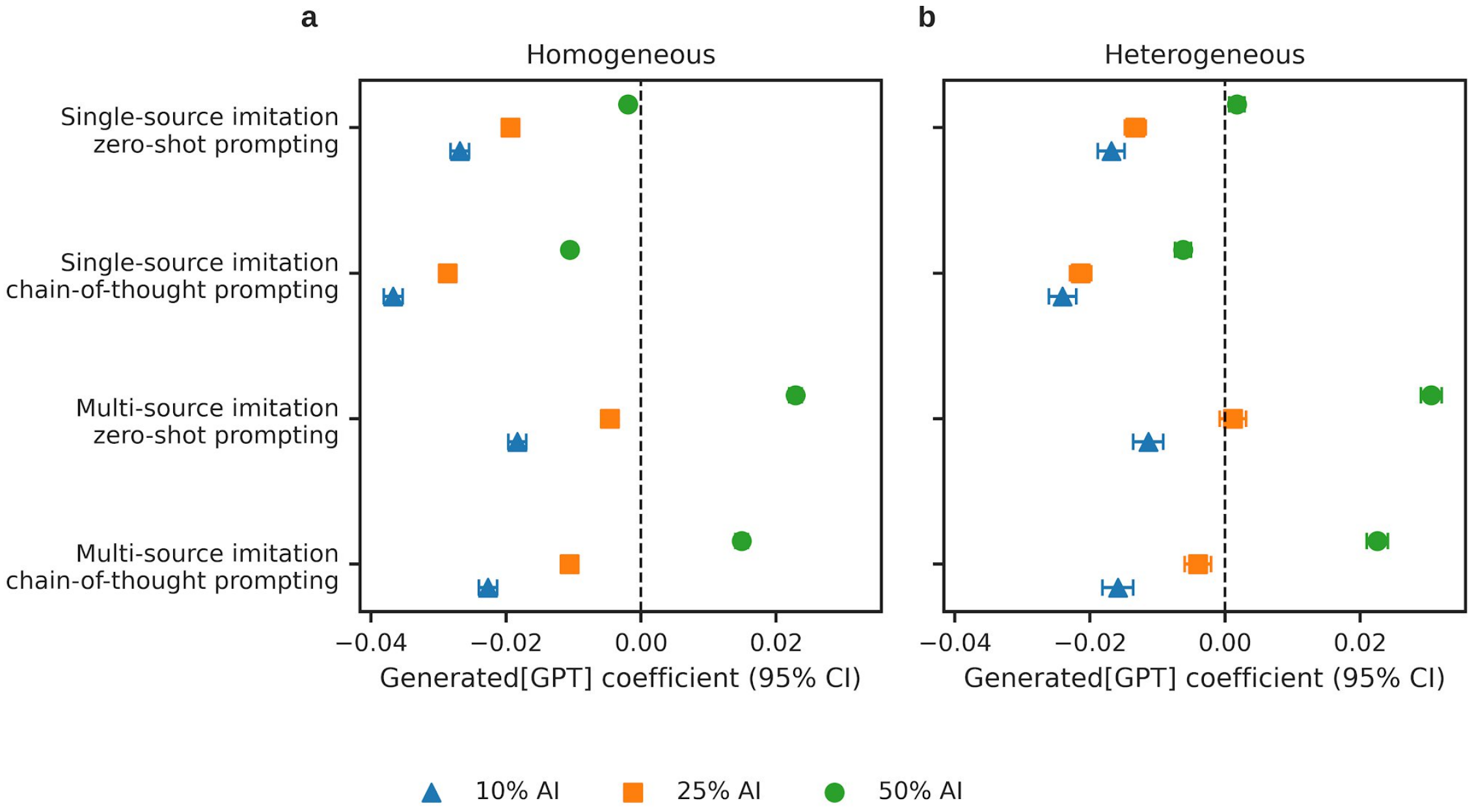

**Supplementary Figure 2.** Coefficient plot displaying the estimated coefficient for *Generated[GPT]* = 1 on *AvgSim*, capturing how AI-generated articles influence within-world, within-topic cosine similarity across (a) homogeneous and (b) heterogeneous environments, as well as variations in imitation strategies (single-source vs. multi-source), prompting strategies (zero-shot vs. chain-of-thought, CoT), and AI prevalence (10%, 25%, 50%). All worlds used GPT-4o and the benchmark task specification. The negative effect weakens as prevalence rises: similarity reduction is most pronounced at 10% prevalence, and at 50% the coefficient turns positive under multi-source imitation in both environments. Error bars are 95% confidence intervals calculated from cluster-robust covariance estimates clustered by DateTopic.

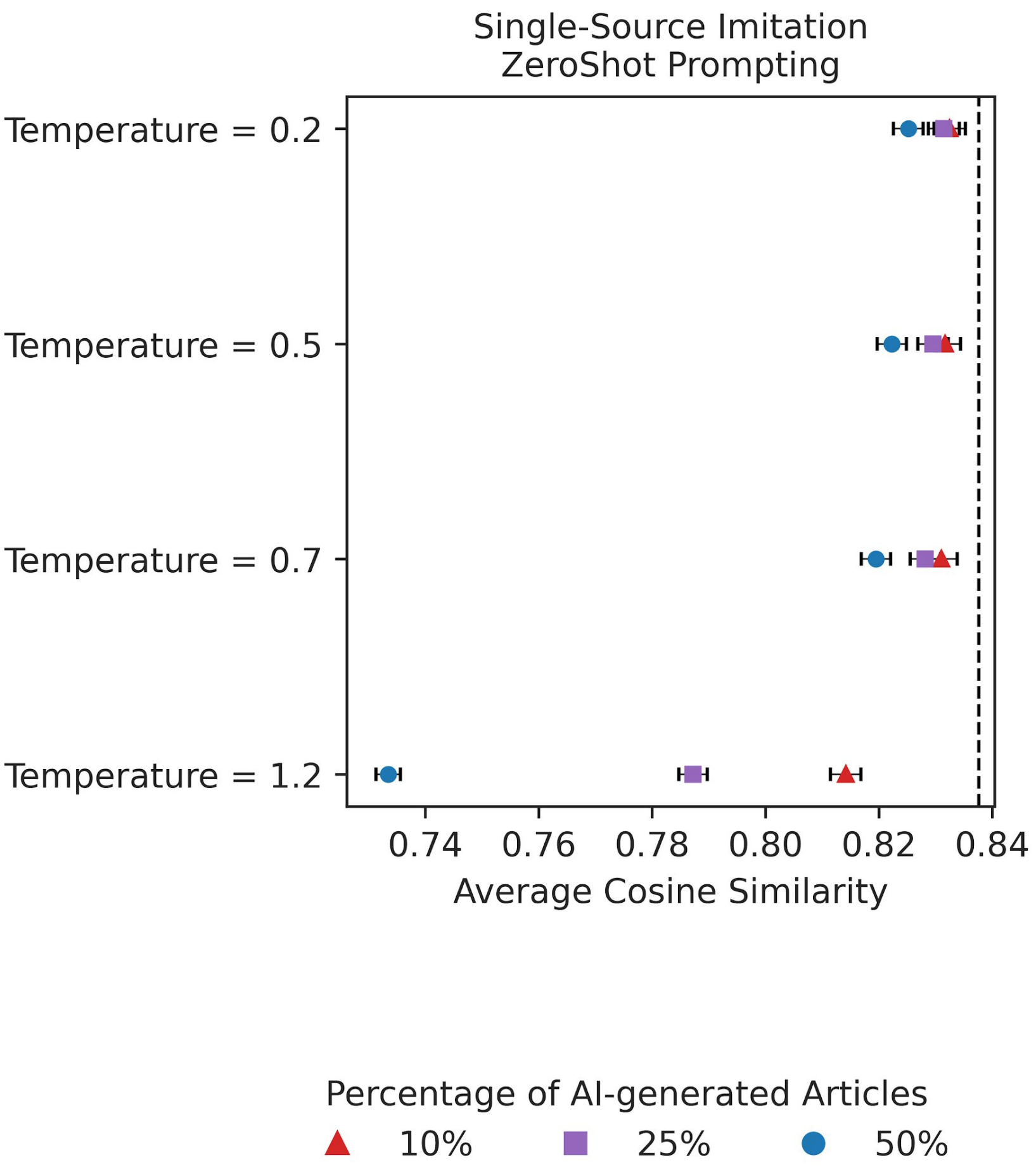

**Supplementary Figure 3.** Mean within-topic cosine similarity scores across temperature conditions (GPT-4o) in zero-shot Single worlds ($n$ = 12). The vertical dashed line indicates the mean similarity in World Original. Means with 95% confidence intervals (error bars) are plotted for each simulation. The plot shows that increasing the proportion of AI-generated articles leads to lower similarity in homogeneous worlds, regardless of temperature settings, although the effect is strongest for articles generated at the highest temperature settings.

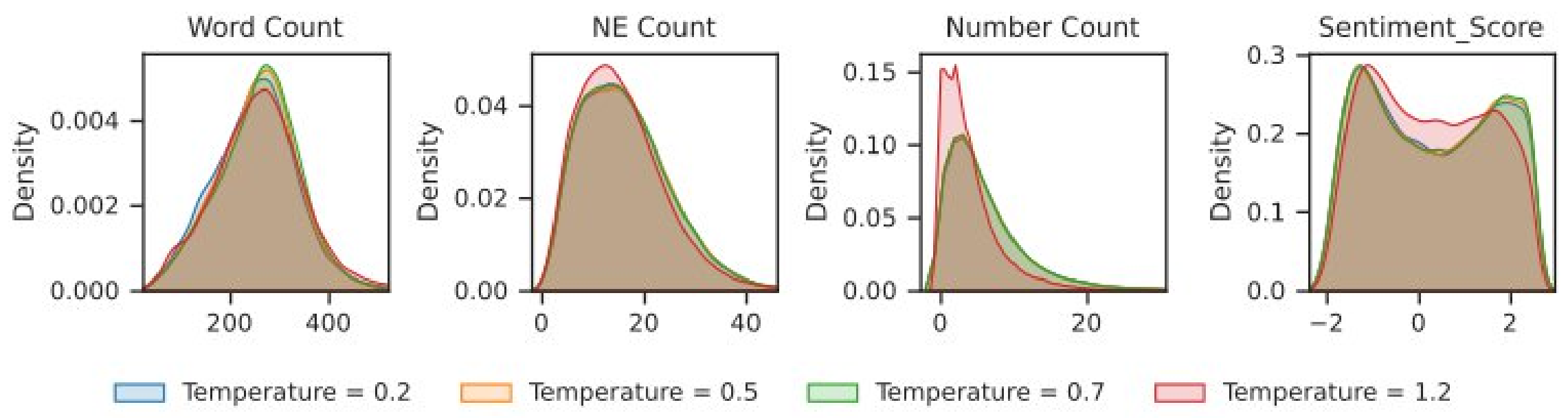

**Supplementary Figure 4.** Kernel density estimates (KDEs) of feature distributions for AI-generated news articles at temperature 0.2, (blue), 0.5 (orange) 0.7 (green), and 1.2 (red), generated using single-source imitation and zero-shot prompting (GPT-4o). Each plot represents a specific feature: Word Count, Named Entities, Number Count, and Sentiment. KDEs are truncated at the 1st and 99th percentiles for improved visualization.

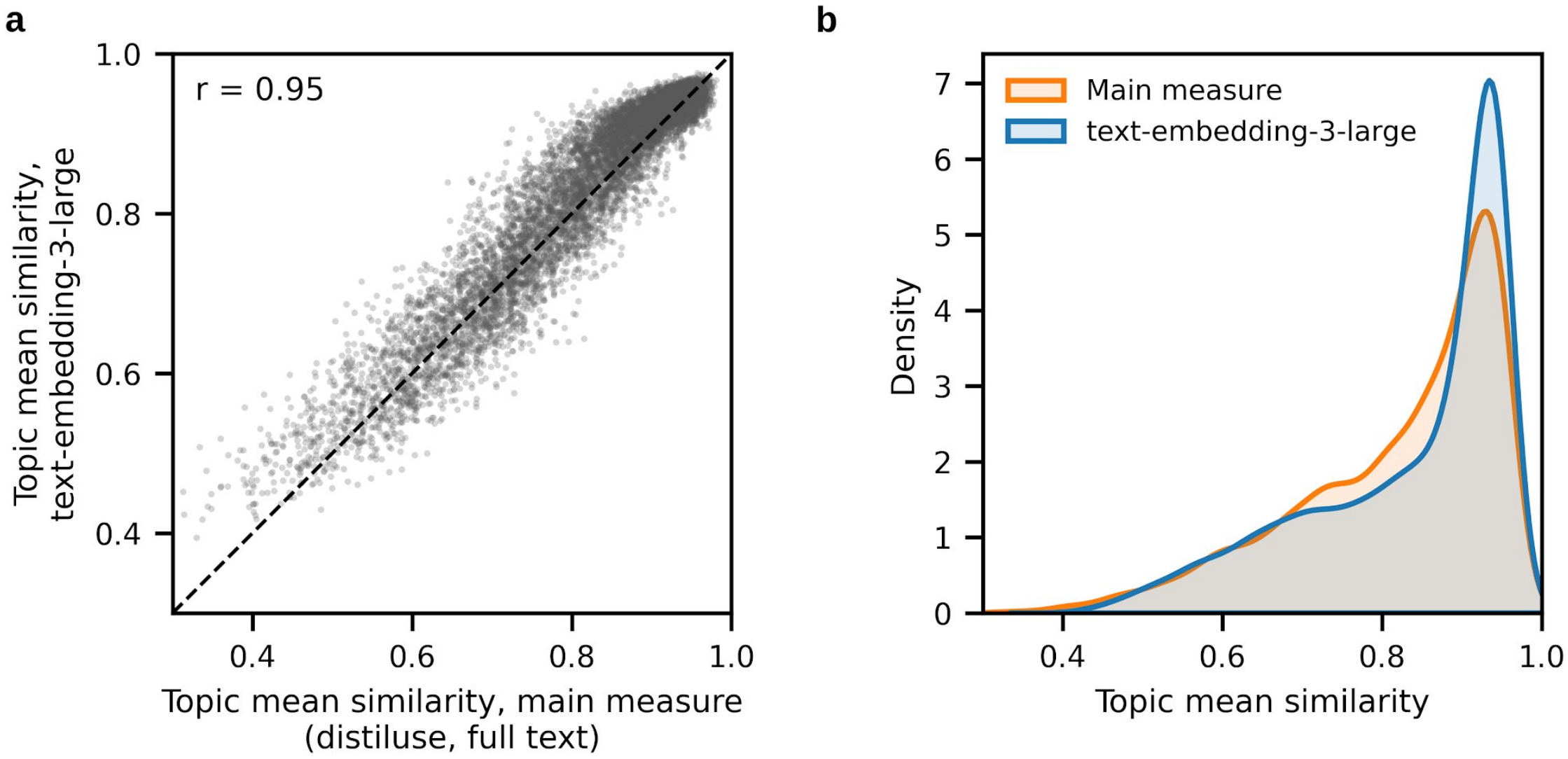

**Supplementary Figure 5.** Topic-level similarity under the main measure and under text-embedding-3-large. One simulated world (GPT-4o, homogeneous environment, single-source imitation, zero-shot prompting, 50% AI prevalence). For each date-specific topic (*n* = 9,267), the mean cosine similarity of its article pairs under the main measure (distiluse-base-multilingual-cased, full text in 126-token chunks) and under OpenAI's text-embedding-3-large. (a) Each point is one topic; dashed line, equal values. Pearson's *r* = 0.954 (95% CI [0.952, 0.956]). (b) Kernel density estimates of the topic means under the main measure (orange) and text-embedding-3-large (blue).

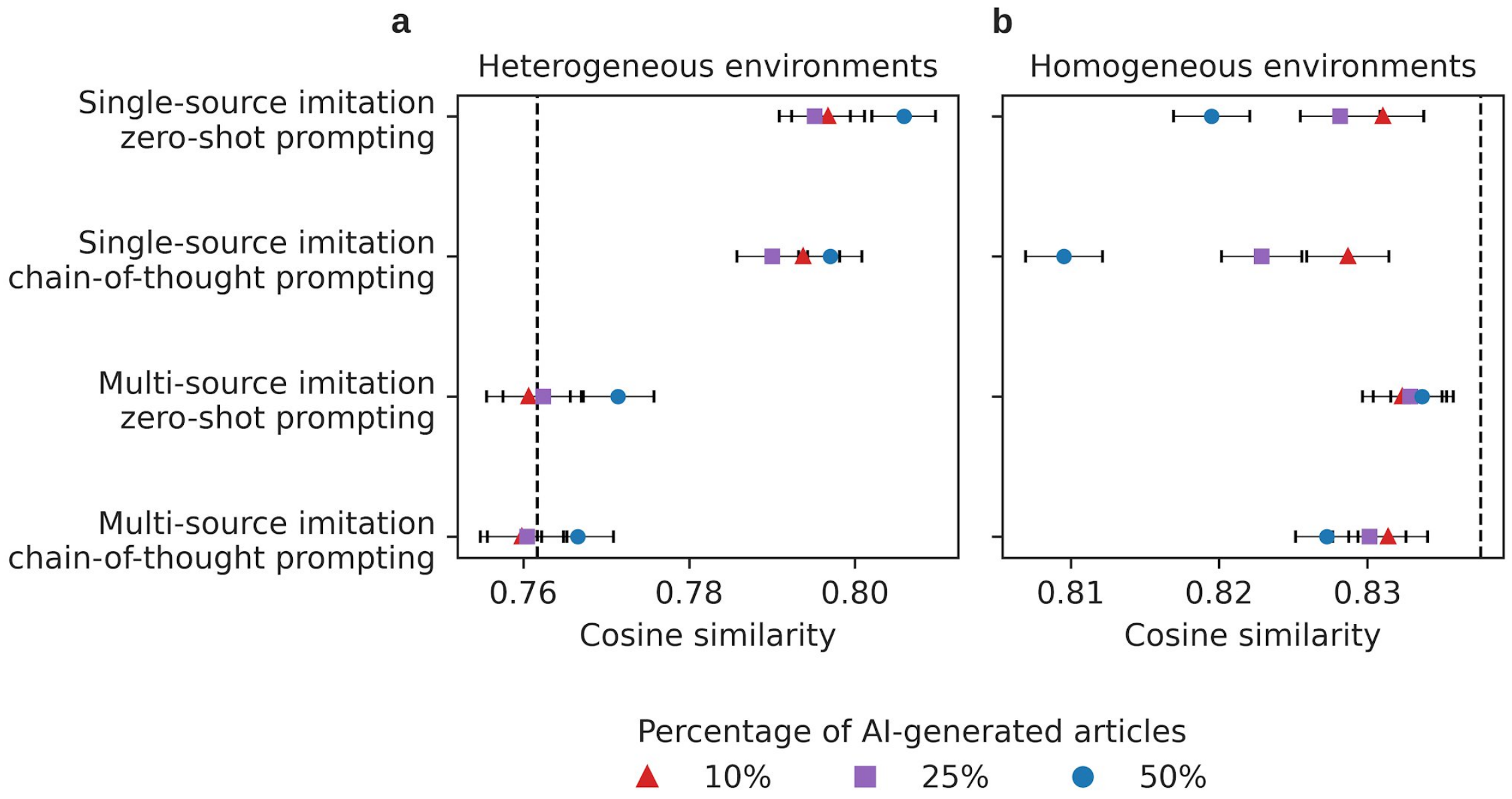

**Supplementary Figure 6.** Sensitivity check of prompting strategy effects on semantic diversity. Mean within-topic cosine similarity across simulated worlds under zero-shot and chain-of-thought prompting, separately for heterogeneous (a) and homogeneous (b) information environments (all based on GPT-4o). Results are shown for single-source and multi-source imitation at three AI prevalence levels (10%, 25%, 50%). Vertical dashed lines indicate baseline similarity in the corresponding original worlds. Across prompting strategies, the qualitative pattern remains unchanged: imitating AI agents increase semantic diversity in homogeneous environments (lower similarity relative to baseline in all 12 conditions), while in heterogeneous environments single-source imitation increases similarity and multi-source imitation leaves it close to baseline at 10% and 25% prevalence and raises it at 50%. Error bars represent 95% confidence intervals around mean within-topic similarity estimates.

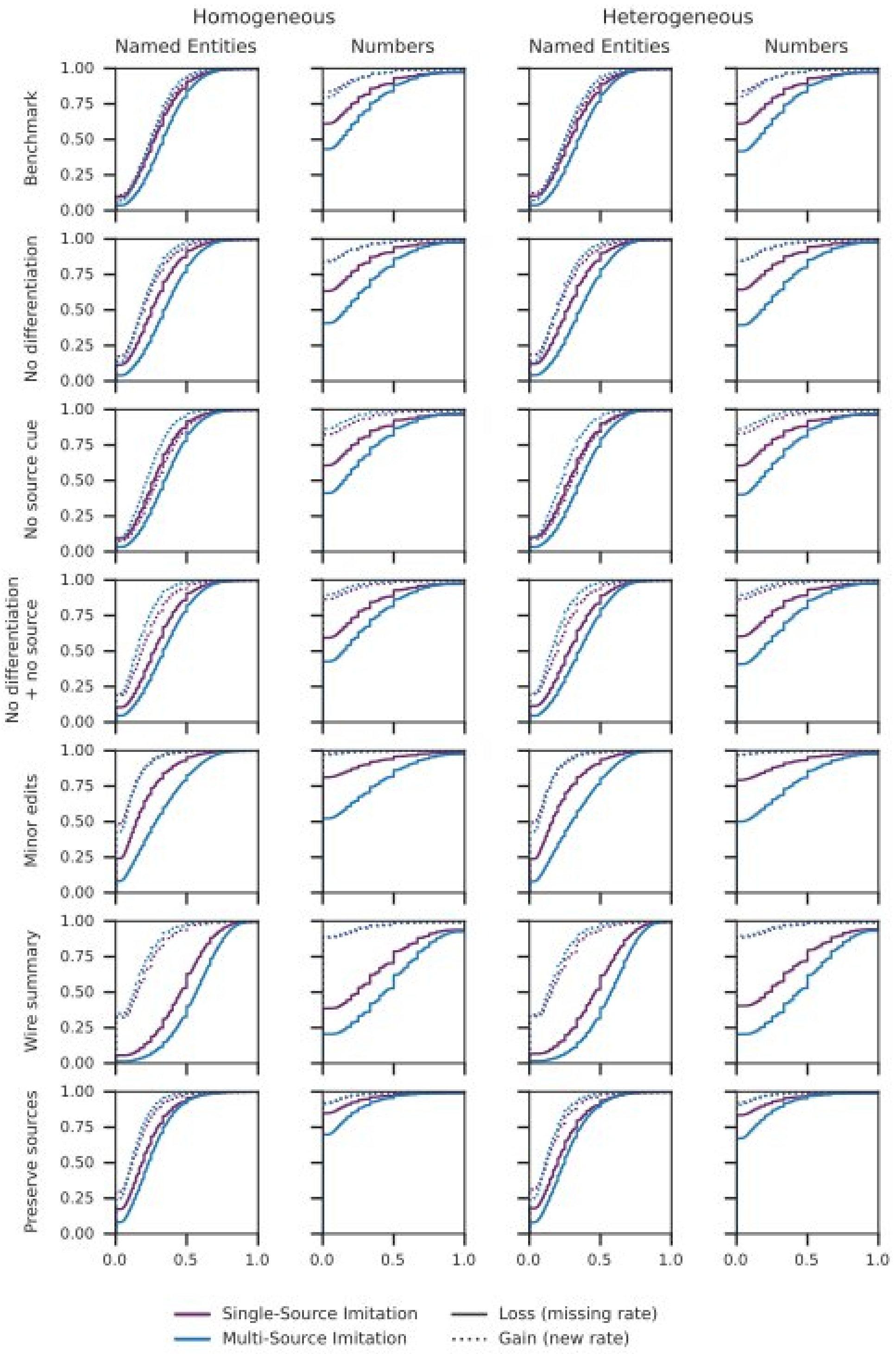

**Supplementary Figure 7.** Information loss and gain across task specifications. Empirical cumulative distribution functions (ECDFs) of information loss (solid lines) and information gain (dotted lines) in DeepSeek-V3.2-generated news articles across seven task specifications, shown separately for homogeneous and heterogeneous information environments and for named entities

and numerical references. Rows correspond to the benchmark task specification, three prompt ablations removing differentiation and/or outlet-source cues, and three alternative newsroom-oriented specifications emphasizing minimal editorial adjustment, wire-service summarization, or factual preservation. Information loss is defined as the share of named entities or numerical references present in the human-written source article(s) but absent from the AI-generated article. Information gain is defined as the share present in the AI-generated article but absent from its human-written source article(s). Purple lines indicate single-source imitation and blue lines multi-source imitation. Across task specifications, environments, and imitation strategies, information-loss rates are generally higher than gain rates (the two have different denominators, so this is not a net information balance), indicating that AI-generated articles more frequently omit factual elements than introduce new ones. The magnitude of this imbalance varies with task specification, with particularly high loss under wire-service summarization and lower loss under minimal-edit and factual-preservation specifications.

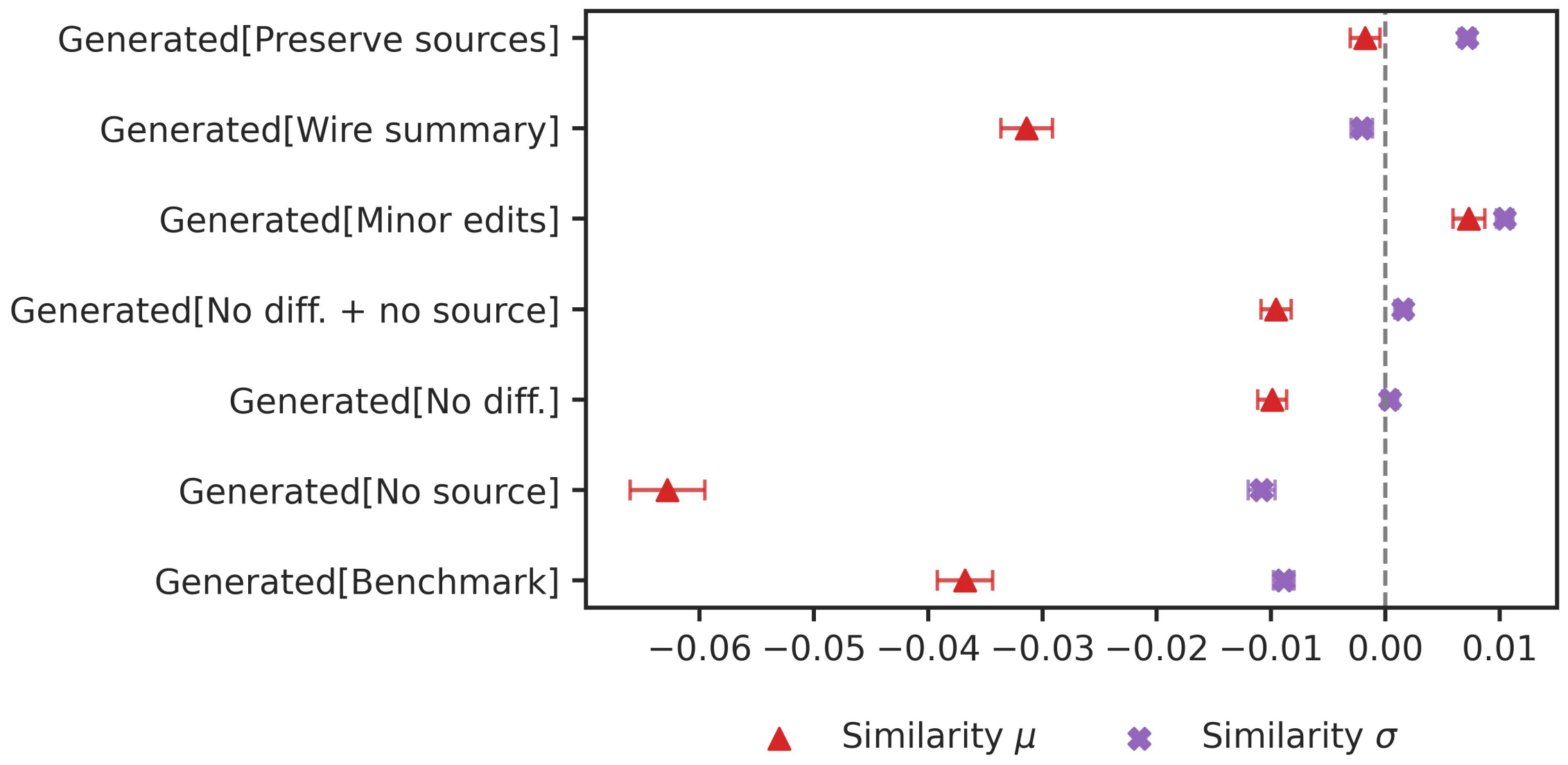

**Supplementary Figure 8.** Pooled mixed-effects regression coefficients predicting each article's mean within-topic pairwise cosine similarity (μ, triangles) and the standard deviation of those pairwise similarities (σ, crosses) under different task specifications (DeepSeek-V3.2). Error bars are 95% confidence intervals calculated using topic-clustered covariance estimates. The model pools all imitating AI agent setups and experimental conditions while controlling for environment, imitation strategy, AI prevalence, and article-level covariates. Negative coefficients indicate lower mean similarity (greater semantic diversity) or lower dispersion in similarity, respectively.

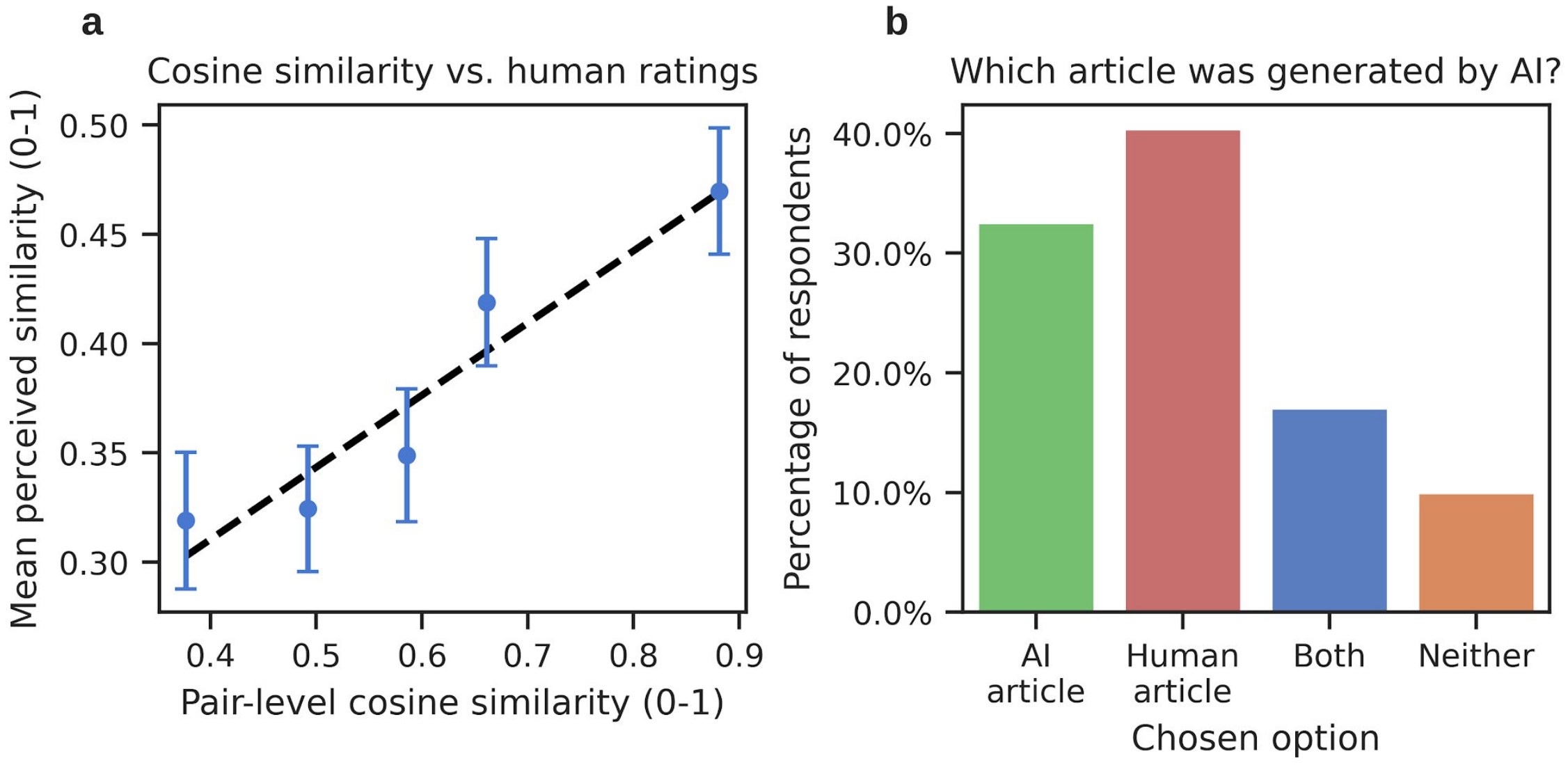

**Supplementary Figure 9**. Limited sanity check of embedding-based similarity with human judgments. (a) Pair-level relationship between embedding cosine similarity and mean perceived similarity ratings from an approximately representative survey of Danish respondents ($n$ = 1,201). Each point corresponds to one stimulus pair ($n$ = 5), with error bars indicating 95% confidence intervals across respondents. The dashed line shows a linear fit for descriptive purposes. Higher cosine similarity is associated with higher perceived similarity between article snippets. (b) Distribution of responses to a four-option identification task asking which snippet was generated by AI. Only one option ("AI article") was correct, yet respondents selected incorrect options more often than the correct one, most frequently labeling the human-written snippet as AI. This pattern indicates that AI-generated snippets are not trivially distinguishable from human-written text in this setting.

**Supplementary Table 1.** Outcomes are article-level textual features, each Z-standardized across the analysis sample. Predictors are categorical: human articles, homogeneous environments, single-source imitation and 10% AI prevalence are the reference categories. DateTopic-clustered standard errors are reported in parentheses.

| | **Word Count** | | **Named Entities** | | **Number Count** | | **Sentiment** | |
|---|---|---|---|---|---|---|---|---|
| | coef. | P | coef. | P | coef. | P | coef. | P |
| Generated[GPT] | −.30576 (.00576) | <.001 | −.49431 (.00489) | <.001 | −.43211 (.00857) | <.001 | .18471 (.00386) | <.001 |
| Generated[Llama] | −.06045 (.00684) | <.001 | −.33394 (.00489) | <.001 | −.45923 (.01105) | <.001 | .14537 (.00324) | <.001 |
| Generated[DeepSeek] | .75038 (.00686) | <.001 | −.40612 (.00516) | <.001 | −.43817 (.01121) | <.001 | .07320 (.00336) | <.001 |
| Environment[Heterogeneous] | −.01134 (.00312) | <.001 | −.01066 (.00396) | .007 | −.00846 (.00294) | .004 | .00593 (.00242) | .014 |
| Imitation[Multi] | .33407 (.00437) | <.001 | .23643 (.00375) | <.001 | .08573 (.00341) | <.001 | .01178 (.00233) | <.001 |
| Percentage[25] | −.09730 (.00184) | <.001 | −.03889 (.00161) | <.001 | −.01611 (.00152) | <.001 | .00117 (.00121) | .332 |
| Percentage[50] | −.05260 (.00222) | <.001 | −.02362 (.00190) | <.001 | −.00870 (.00183) | <.001 | .00086 (.00142) | .547 |
| Intercept | −.04041 (.00820) | <.001 | .38217 (.00988) | <.001 | .34462 (.01405) | <.001 | −.12495 (.00833) | <.001 |
| N Observations | 1,093,344 | | 1,093,344 | | 1,093,344 | | 1,093,344 | |
| N Groups | 9267DateTopic | | 9267DateTopic | | 9267DateTopic | | 9267DateTopic | |
| Marginal $R^2$ / Conditional $R^2$ | .228 / .490 | | .004 / .720 | | .003 / .612 | | .002 / .818 | |
| $\sigma^2$ | .55 | | .36 | | .39 | | .19 | |
| $\tau_{00}$ | .26DateTopic | | .62DateTopic | | .54DateTopic | | .81DateTopic | |
| ICC | .34 | | .63 | | .59 | | .81 | |

**Supplementary Table 2.** Outcomes are the within-cell standard deviation of each textual feature, computed across articles within each DateTopic × environment × imitation strategy × AI prevalence × model cell, and then Z-standardized across cells. Predictors are categorical, with human articles, homogeneous environments, single-source imitation and 10% AI prevalence as reference categories. DateTopic-clustered standard errors are reported in parentheses.

| | **StdWord Count** | | **StdNamed Entities** | | **StdNumber Count** | | **StdSentiment** | |
|---|---|---|---|---|---|---|---|---|
| | coef. | P | coef. | P | coef. | P | coef. | P |
| Generated[GPT] | −1.14721 (.01182) | <.001 | −1.44344 (.01220) | <.001 | −.78404 (.01140) | <.001 | −.33330 (.00956) | <.001 |
| Generated[Llama] | −.94443 (.01251) | <.001 | −1.10846 (.01267) | <.001 | −.79558 (.01185) | <.001 | −.28475 (.00991) | <.001 |
| Generated[DeepSeek] | −.33647 (.01218) | <.001 | −1.08414 (.01231) | <.001 | −.82197 (.01296) | <.001 | −.25900 (.00986) | <.001 |
| Environment[Heterogeneous] | −.04802 (.00548) | <.001 | −.11321 (.00652) | <.001 | −.04416 (.00540) | <.001 | −.05562 (.00636) | <.001 |
| Imitation[Multi] | −.13010 (.00583) | <.001 | −.05198 (.00543) | <.001 | −.07058 (.00530) | <.001 | −.18877 (.00590) | <.001 |
| Percentage[25] | .30684 (.00535) | <.001 | .56047 (.00638) | <.001 | .21178 (.00540) | <.001 | .39373 (.00614) | <.001 |
| Percentage[50] | .53875 (.00574) | <.001 | .74807 (.00710) | <.001 | .31571 (.00638) | <.001 | .52925 (.00668) | <.001 |
| Intercept | .52479 (.01204) | <.001 | .70713 (.01273) | <.001 | .56035 (.01375) | <.001 | −.03135 (.00904) | <.001 |
| N Observations | 175,477 | | 175,477 | | 175,477 | | 175,477 | |
| N Groups | 9267DateTopic | | 9267DateTopic | | 9267DateTopic | | 9267DateTopic | |
| Marginal $R^2$ / Conditional $R^2$ | .210 / .377 | | .006 / .517 | | .017 / .510 | | .051 / .466 | |
| $\sigma^2$ | .65 | | .56 | | .48 | | .53 | |
| $\tau_{00}$ | .16DateTopic | | .29DateTopic | | .45DateTopic | | .39DateTopic | |
| ICC | .20 | | .34 | | .48 | | .43 | |

**Supplementary Table 3.** Outcomes are the within-cell mean of each textual feature, computed across articles within each DateTopic × environment × imitation strategy × AI prevalence × model cell, and then Z-standardized across cells. Predictors are categorical, with human articles, homogeneous environments, single-source imitation and 10% AI prevalence as reference categories. DateTopic-clustered standard errors are reported in parentheses.

| | Word Count | | Named Entities | | Number Count | | Sentiment | |
|---|---|---|---|---|---|---|---|---|
| | coef. | P | coef. | P | coef. | P | coef. | P |
| Generated[GPT] | −.41798 (.00744) | <.001 | −.59235 (.00601) | <.001 | −.5259 (.00763) | <.001 | .20771 (.00419) | <.001 |
| Generated[Llama] | −.13096 (.00791) | <.001 | −.41512 (.00603) | <.001 | −.5528 (.00874) | <.001 | .16384 (.00375) | <.001 |
| Generated[DeepSeek] | .79833 (.00810) | <.001 | −.50715 (.00636) | <.001 | −.5282 (.00890) | <.001 | .08910 (.00387) | <.001 |
| Environment[Heterogeneous] | −.01724 (.00478) | <.001 | −.01407 (.00520) | .007 | −.00544 (.00461) | .238 | .00744 (.00322) | .021 |
| Imitation[Multi] | .40455 (.00510) | <.001 | .28608 (.00418) | <.001 | .13306 (.00393) | <.001 | −.00113 (.00239) | .637 |
| Percentage[25] | −.08134 (.00322) | <.001 | −.03394 (.00252) | <.001 | −.00768 (.00261) | .003 | .00139 (.00169) | .411 |
| Percentage[50] | −.00411 (.00354) | .245 | −.00905 (.00276) | .001 | .00371 (.00287) | .197 | .00173 (.00186) | .352 |
| Intercept | −.15223 (.00958) | <.001 | .41183 (.01123) | <.001 | .44725 (.01394) | <.001 | −.14956 (.00889) | <.001 |
| N Observations | 175,477 | | 175,477 | | 175,477 | | 175,477 | |
| N Groups | 9267DateTopic | | 9267DateTopic | | 9267DateTopic | | 9267DateTopic | |
| Marginal $R^2$ / Conditional $R^2$ | .329 / .697 | | .006 / .894 | | .005 / .837 | | .003 / .930 | |
| $\sigma^2$ | .33 | | .20 | | .18 | | .07 | |
| $\tau_{00}$ | .38DateTopic | | .78DateTopic | | .80DateTopic | | .92DateTopic | |
| ICC | .53 | | .80 | | .82 | | .93 | |

**Supplementary Table 4**

**(A) Descriptives across homogeneous and heterogeneous environments**

| Measure | Homogeneous | Heterogeneous |
|---|---|---|
| Articles | 104,761 | 51,005 |
| Topics | 9,267 | 5,338 |
| Articles in shared topics | 63,934 | 49,941 |
| Dates in shared topics | 361 | 361 |
| Outlets in shared topics | 130 | 130 |
| Articles per topic, mean (SD) | 11.98 (6.19) | 9.36 (6.07) |
| Outlets per topic, mean (SD) | 8.60 (2.81) | 7.01 (3.36) |
| Baseline cosine similarity, full environment | 0.838 | 0.762 |
| Baseline cosine similarity, shared topics | 0.797 | 0.762 |

**(B) Article characteristics within shared topics between homogeneous and heterogeneous environments**

| Feature | Homogeneous, mean (SD) | Heterogeneous, mean (SD) | Standardized difference |
|---|---|---|---|
| Article word count | 285.43 (117.42) | 284.04 (118.69) | −0.012 |
| Named entities | 19.43 (10.86) | 19.22(10.77) | −0.019 |
| Unique named entities | 11.18 (6.25) | 11.07 (6.16) | −0.018 |
| Numbers | 8.43 (8.98) | 8.72 (9.52) | 0.031 |
| Sentiment | 0.05 (1.16) | 0.06 (1.17) | 0.003 |

**(C) Compositional balance within shared topics between homogeneous and heterogeneous environments**

| Dimension | Total-variation distance |
|---|---|
| Overall outlet distribution | 0.034 |
| Relative distribution across topics | 0.187 |
| Topic-by-outlet distribution | 0.203 |

Note: A topic was considered usable when it contained at least two articles. Standardized differences were calculated as heterogeneous minus homogeneous. Total-variation distance ranges from zero for identical distributions to one for distributions with no overlap. It can be interpreted as the proportion of article share that would need to be redistributed across categories to make two distributions identical. Baseline similarity is mean within-topic cosine similarity.

**Supplementary Table 5.** Linear mixed-effects models of generated-minus-human contrasts in mean within-topic cosine similarity (ΔAvgSim) and within-topic similarity dispersion (ΔStdSim). Outcomes are generated-minus-human contrasts in raw cosine-similarity units. The coefficient for each LLM refers to the homogeneous, single-source, 10% condition. DateTopic-clustered standard errors are reported in parentheses.

| | **ΔAvgSim** | | **ΔStdSim** | |
|---|---|---|---|---|
| | coef. | P | coef. | P |
| Generated[GPT] | −.03227 (.00053) | <.001 | −.00308 (.00030) | <.001 |
| Generated[Llama] | −.02158 (.00052) | <.001 | −.00081 (.00030) | .008 |
| Generated[DeepSeek] | −.04119 (.00054) | <.001 | −.01135 (.00030) | <.001 |
| Generated[GPT] × Environment[Heterogeneous] | .00568 (.00048) | <.001 | .00637 (.00038) | <.001 |
| Generated[Llama] × Environment[Heterogeneous] | .00537 (.00050) | <.001 | .00582 (.00037) | <.001 |
| Generated[DeepSeek] × Environment[Heterogeneous] | .00640 (.00049) | <.001 | .00536 (.00037) | <.001 |
| Generated[GPT] × Imitation[Multi] | .01404 (.00041) | <.001 | −.01819 (.00033) | <.001 |
| Generated[Llama] × Imitation[Multi] | .02047 (.00042) | <.001 | −.01495 (.00034) | <.001 |
| Generated[DeepSeek] × Imitation[Multi] | .02170 (.00038) | <.001 | −.01282 (.00033) | <.001 |
| Generated[GPT] × Percentage [25] | .01018 (.00028) | <.001 | .00018 (.00016) | .285 |
| Generated[Llama] × Percentage [25] | .00658 (.00026) | <.001 | −.00049 (.00016) | .002 |
| Generated[DeepSeek] × Percentage [25] | .01463 (.00025) | <.001 | .00268 (.00016) | <.001 |
| Generated[GPT] × Percentage [50] | .03317 (.00043) | <.001 | .00156 (.00027) | <.001 |
| Generated[Llama] × Percentage [50] | .01962 (.00041) | <.001 | −.00108 (.00026) | <.001 |
| Generated[DeepSeek] × Percentage [50] | .03909 (.00043) | <.001 | .00337 (.00026) | <.001 |
| N Observations | 255,956 | | 251,600 | |
| N Clusters | 9,267DateTopic | | 9,267DateTopic | |
| Marginal $R^2$ / Conditional $R^2$ | .093 / .498 | | .063 / .362 | |
| $\sigma^2$ | .00 | | .00 | |
| $\tau_{00}$ | .00DateTopic | | .00DateTopic | |
| ICC | .45 | | .32 | |

**Supplementary Table 6.** Direction of mixed-LLM and single-LLM dispersion differences. Each cell gives the number of conditions with higher / lower / unresolved dispersion in the mixed world, out of 12 environment × imitation strategy × AI prevalence conditions. Higher and lower mean that the 95% paired t confidence interval for the mixed-minus-single difference lies entirely above or below zero, and unresolved means the interval includes zero and does not establish equivalence. All outcomes are mean within-topic SD. Similarity dispersion is on the main measure, and its per-model counts differ under the two long-context embedding models (Supplementary Text 15).

| Dispersion outcome | Mixed vs GPT-4o | Mixed vs Llama | Mixed vs DeepSeek |
|---|---|---|---|
| Similarity | 5 / 2 / 5 | 6 / 2 / 4 | 2 / 7 / 3 |
| Word count | 12 / 0 / 0 | 11 / 0 / 1 | 0 / 12 / 0 |
| Named entities | 10 / 0 / 2 | 10 / 0 / 2 | 0 / 9 / 3 |
| Numerical references | 8 / 2 / 2 | 0 / 2 / 10 | 2 / 5 / 5 |
| Sentiment | 0 / 8 / 4 | 7 / 1 / 4 | 8 / 0 / 4 |

**Supplementary Table 7.** Similarity by type of article pair at 10% and 25% prevalence under the benchmark task specification. Mean cosine similarity within topics by pair type, per LLM, environment and imitation strategy: human-human; AI-generated article and the article it rewrote (own source); AI-generated article and the other human-written articles; AI-AI; AI-AI pairs sharing a source; and human-human in the human baseline world. Construction as in Extended Data Table 2, which reports the 50% rows. AI-AI pairs are rare at low prevalence (about 0.4% of all pairs at 10% and 5% at 25%; pairs sharing a source at most 0.1% and 2%), so their means rest on few, unevenly distributed topics and are less comparable with the other columns.

| LLM | AI prevalence | Human–human | AI–own source | AI–other human | AI–AI (all) | AI–AI (shared source) | Human–human, baseline |
|---|---|---|---|---|---|---|---|
| Homogeneous environment, single-source imitation | | | | | | | |
| GPT-4o | 10% | 0.774 | 0.914 | 0.736 | 0.697 | 0.925 | 0.774 |
| GPT-4o | 25% | 0.778 | 0.914 | 0.735 | 0.757 | 0.930 | 0.774 |
| Llama | 10% | 0.772 | 0.918 | 0.742 | 0.704 | 0.920 | 0.774 |
| Llama | 25% | 0.772 | 0.914 | 0.739 | 0.758 | 0.926 | 0.774 |
| DeepSeek | 10% | 0.772 | 0.887 | 0.728 | 0.717 | 0.915 | 0.774 |
| DeepSeek | 25% | 0.772 | 0.884 | 0.723 | 0.766 | 0.914 | 0.774 |
| Homogeneous environment, multi-source imitation | | | | | | | |
| GPT-4o | 10% | 0.774 | 0.848 | 0.745 | 0.782 | 0.854 | 0.774 |
| GPT-4o | 25% | 0.778 | 0.847 | 0.744 | 0.824 | 0.875 | 0.774 |
| Llama | 10% | 0.774 | 0.875 | 0.763 | 0.770 | 0.849 | 0.774 |
| Llama | 25% | 0.778 | 0.873 | 0.763 | 0.818 | 0.875 | 0.774 |
| DeepSeek | 10% | 0.772 | 0.845 | 0.748 | 0.793 | 0.858 | 0.774 |
| DeepSeek | 25% | 0.772 | 0.848 | 0.748 | 0.834 | 0.879 | 0.774 |
| Heterogeneous environment, single-source imitation | | | | | | | |
| GPT-4o | 10% | 0.743 | 0.908 | 0.701 | 0.667 | 0.920 | 0.741 |
| GPT-4o | 25% | 0.746 | 0.908 | 0.700 | 0.725 | 0.924 | 0.741 |
| Llama | 10% | 0.741 | 0.909 | 0.708 | 0.661 | 0.917 | 0.741 |
| Llama | 25% | 0.740 | 0.908 | 0.705 | 0.726 | 0.925 | 0.741 |
| DeepSeek | 10% | 0.740 | 0.878 | 0.695 | 0.686 | 0.906 | 0.741 |
| DeepSeek | 25% | 0.740 | 0.877 | 0.691 | 0.739 | 0.909 | 0.741 |
| Heterogeneous environment, multi-source imitation | | | | | | | |
| GPT-4o | 10% | 0.742 | 0.831 | 0.708 | 0.743 | 0.828 | 0.741 |
| GPT-4o | 25% | 0.744 | 0.831 | 0.706 | 0.793 | 0.862 | 0.741 |
| Llama | 10% | 0.742 | 0.857 | 0.723 | 0.714 | 0.809 | 0.741 |
| Llama | 25% | 0.744 | 0.857 | 0.722 | 0.781 | 0.859 | 0.741 |
| DeepSeek | 10% | 0.739 | 0.827 | 0.718 | 0.769 | 0.835 | 0.741 |
| DeepSeek | 25% | 0.739 | 0.832 | 0.716 | 0.811 | 0.867 | 0.741 |

**Supplementary Table 8.** Contribution of each pair type to the change in within-topic similarity under the benchmark task specification. For each topic, the change in mean pairwise cosine similarity from the human baseline is split exactly into pair types: the share of the topic's pairs of each type multiplied by the difference between their mean similarity and the topic's baseline mean. Averaging over topics gives each pair type's contribution, and contributions sum to the total change reported in Fig. 2. Pair types: human-human; AI-generated article and the article it rewrote (own source); AI-generated article and the other human-written articles; AI-AI pairs sharing a source; and other AI-AI pairs. Negative values lower similarity (more diversity), positive values raise it. Main measure; 9,267 topics in the homogeneous and 5,338 in the heterogeneous environment.

| LLM | AI prevalence | Human–human | AI–own source | AI–other human | AI–AI (shared source) | AI–AI (other) | Total change |
|---|---|---|---|---|---|---|---|
| Homogeneous environment, single-source imitation | | | | | | | |
| GPT-4o | 10% | +0.0007 | +0.0015 | −0.0088 | 0.0000 | 0.0000 | −0.0066 |
| GPT-4o | 25% | +0.0027 | +0.0035 | −0.0153 | +0.0007 | −0.0011 | −0.0095 |
| GPT-4o | 50% | −0.0014 | +0.0071 | −0.0212 | +0.0037 | −0.0064 | −0.0181 |
| Llama | 10% | −0.0010 | +0.0016 | −0.0074 | 0.0000 | 0.0000 | −0.0068 |
| Llama | 25% | −0.0013 | +0.0036 | −0.0143 | +0.0006 | −0.0010 | −0.0124 |
| Llama | 50% | −0.0014 | +0.0072 | −0.0186 | +0.0035 | −0.0062 | −0.0155 |
| DeepSeek | 10% | −0.0012 | +0.0008 | −0.0113 | 0.0000 | 0.0000 | −0.0116 |
| DeepSeek | 25% | −0.0012 | +0.0018 | −0.0221 | +0.0006 | −0.0010 | −0.0219 |
| DeepSeek | 50% | −0.0014 | +0.0037 | −0.0281 | +0.0030 | −0.0059 | −0.0287 |
| Homogeneous environment, multi-source imitation | | | | | | | |
| GPT-4o | 10% | +0.0007 | +0.0001 | −0.0062 | +0.0001 | 0.0000 | −0.0053 |
| GPT-4o | 25% | +0.0027 | +0.0006 | −0.0098 | +0.0013 | +0.0004 | −0.0048 |
| GPT-4o | 50% | −0.0014 | +0.0012 | −0.0122 | +0.0066 | +0.0019 | −0.0039 |
| Llama | 10% | +0.0007 | +0.0015 | −0.0025 | +0.0001 | 0.0000 | −0.0002 |
| Llama | 25% | +0.0027 | +0.0036 | −0.0030 | +0.0014 | +0.0004 | +0.0050 |
| Llama | 50% | −0.0014 | +0.0075 | −0.0050 | +0.0072 | +0.0017 | +0.0099 |
| DeepSeek | 10% | −0.0011 | −0.0002 | −0.0064 | +0.0001 | 0.0000 | −0.0075 |
| DeepSeek | 25% | −0.0013 | +0.0006 | −0.0106 | +0.0013 | +0.0005 | −0.0095 |
| DeepSeek | 50% | −0.0014 | +0.0006 | −0.0117 | +0.0068 | +0.0024 | −0.0033 |
| Heterogeneous environment, single-source imitation | | | | | | | |
| GPT-4o | 10% | +0.0007 | +0.0423 | −0.0079 | 0.0000 | 0.0000 | +0.0351 |
| GPT-4o | 25% | +0.0021 | +0.0447 | −0.0133 | +0.0008 | −0.0009 | +0.0335 |
| GPT-4o | 50% | −0.0014 | +0.0572 | −0.0180 | +0.0116 | −0.0053 | +0.0442 |
| Llama | 10% | −0.0007 | +0.0406 | −0.0065 | 0.0000 | 0.0000 | +0.0335 |
| Llama | 25% | −0.0012 | +0.0430 | −0.0120 | +0.0007 | −0.0008 | +0.0296 |
| Llama | 50% | −0.0014 | +0.0550 | −0.0158 | +0.0113 | −0.0049 | +0.0443 |
| DeepSeek | 10% | −0.0006 | +0.0374 | −0.0100 | 0.0000 | 0.0000 | +0.0268 |
| DeepSeek | 25% | −0.0010 | +0.0390 | −0.0184 | +0.0006 | −0.0007 | +0.0195 |
| DeepSeek | 50% | −0.0014 | +0.0487 | −0.0231 | +0.0103 | −0.0044 | +0.0300 |
| Heterogeneous environment, multi-source imitation | | | | | | | |
| GPT-4o | 10% | +0.0003 | +0.0035 | −0.0049 | 0.0000 | 0.0000 | −0.0010 |
| GPT-4o | 25% | +0.0018 | +0.0049 | −0.0078 | +0.0015 | +0.0004 | +0.0007 |
| GPT-4o | 50% | −0.0014 | +0.0094 | −0.0097 | +0.0099 | +0.0015 | +0.0097 |
| Llama | 10% | +0.0003 | +0.0052 | −0.0018 | 0.0000 | 0.0000 | +0.0037 |
| Llama | 25% | +0.0017 | +0.0079 | −0.0023 | +0.0016 | +0.0003 | +0.0092 |
| Llama | 50% | −0.0014 | +0.0154 | −0.0042 | +0.0102 | +0.0013 | +0.0213 |
| DeepSeek | 10% | −0.0009 | +0.0033 | −0.0047 | 0.0000 | 0.0000 | −0.0021 |
| DeepSeek | 25% | −0.0009 | +0.0049 | −0.0078 | +0.0016 | +0.0005 | −0.0017 |
| DeepSeek | 50% | −0.0014 | +0.0090 | −0.0088 | +0.0102 | +0.0021 | +0.0111 |

**References Supplementary Information**

1. Johansen, E. B. & Baumann, O. Platform Reliance: How Usage of Platform Content Shapes Product Similarity. SSRN Scholarly Paper at https://papers.ssrn.com/abstract=5178352 (2025).

2. Grootendorst, M. BERTopic: Neural topic modeling with a class-based TF-IDF procedure. Preprint at https://doi.org/10.48550/arXiv.2203.05794 (2022).